\documentclass[12pt,english,round]{article}
\usepackage[T1]{fontenc}
\usepackage{textcomp}
\usepackage[latin9]{inputenc}
\usepackage{color}
\usepackage{array}
\usepackage{float}
\usepackage{booktabs}
\usepackage{amsmath}
\usepackage{graphicx}
\usepackage[letterpaper]{geometry}
\usepackage{tablefootnote}
\usepackage{setspace}
\usepackage[authoryear]{natbib}
\makeatletter

\providecommand{\tabularnewline}{\\}

\makeatother

\usepackage{babel}
\begin{document}
\title{\textbf{Dynamics of Global Emission Permit Prices and Regional Social
Cost of Carbon under Noncooperation }\thanks{The authors have contributed equally to this work. We are grateful
for comments and suggestions from the participants at PACE 2023, the
Heartland Environmental and Resource Economics Workshop 2023, OSU
Interdisciplinary Research Fall Forum: Computational Approaches for
a Just and Sustainable World 2023, the LSE Environment Camp 2024,
the 2nd China Energy Modeling Youth Forum, Tokyo Workshop of Climate
Finance \& Risk 2024, Climate Change Economics Forum, and seminars
at University of Illinois Urbana-Champaign, Beijing Institute of Technology,
University of Chinese Academy of Sciences, Ohio University, and Peking
University. Cai acknowledges support from the National Science Foundation
grant SES-1739909 and the USDA-NIFA-AFRI grant 2018-68002-27932. Malik
acknowledges funding from the USDA postdoctoral fellowship ``New
Perspectives on Agricultural Economics'' awarded through the National
Bureau of Economic Research. Shin acknowledges the support from the
endowments of The Andersons Program in International Trade and the
Sogang University Research Grant 202510015.01.}}
\author{Yongyang Cai\thanks{Department of Agricultural, Environmental and Development Economics,
The Ohio State University. cai.619@osu.edu}\qquad{}Khyati Malik\thanks{National Bureau of Economic Research (NBER). malikk@nber.org}
\qquad{}Hyeseon Shin\thanks{Corresponding author. Department of Economics, Sogang University.
hsshin@sogang.ac.kr}\qquad{}}
\maketitle
\begin{abstract}
We develop a dynamic multi-region climate--economy model with emissions
trading and solve for the dynamic Nash equilibrium under noncooperation,
where each region follows Paris Agreement--based emissions caps.
The permit price reaches \$923 per ton of carbon by 2050, and global
temperature rises to 1.7°C above pre-industrial levels by 2100. The
regional social cost of carbon equals the difference between regional
marginal abatement cost and the permit price, highlighting complementarity
between carbon taxes and trading. We find substantial heterogeneity
in regional social costs of carbon, show that lax caps can raise emissions,
and demonstrate strong free-rider incentives under partial participation.



\pagebreak{}
\end{abstract}

\section{Introduction}

Climate change has raised concerns of disastrous consequences, ranging
from rising sea levels to increase in the frequency and intensity
of extreme weather events \citep{arias2021climate}. International
efforts to address climate change have evolved over the past three
decades, commencing with the United Nations Framework Convention on
Climate Change (UNFCCC) in 1992, followed by the Kyoto Protocol in
1997, the Paris Agreement in 2015, and the Glasgow Pact in 2021. The
policy outlined in the Paris Agreement codified an objective of ``limiting
global average temperature rise to well below 2 degrees Celsius above
pre-industrial levels, with a pursuit of limiting it to 1.5 degrees
Celsius'' (Article 2). To accomplish this objective, 194 countries
(regions) committed to the Nationally Determined Contributions (NDCs),
which specify their emission mitigation goals. The Glasgow Climate
Pact, adopted at the UN Climate Change Conference in Glasgow (COP26)
in November 2021, revisited the NDCs, reaffirming the previous commitments
made in the Paris Agreement and recognizing the need for more stringent
efforts to attain the global target of 1.5 degrees Celsius.

Various market-based approaches have been proposed to reduce emissions,
among which carbon tax regimes and emission trading systems (ETSs)
are the most prominent.\footnote{It has been documented that a total of 61 carbon pricing policies,
consisting of 30 taxes and 31 ETSs, have been executed or are scheduled
for implementation globally \citep{Stavins2022}.} In a carbon tax regime, a tax is charged to firms for each unit of
carbon emission. Previous studies have argued that the optimal global
carbon tax rate is equal to the global social cost of carbon (SCC),
which is the present value of global climate damages incurred by an
additional unit of atmospheric carbon emissions \citep{nordhaus2017revisiting}.
In an emission trading or a cap-and-trade system, on the other hand,
the maximum amount of permissible emissions is fixed in an economy,
and agents can sell and purchase emission permits (allowances) at
market-determined prices. In the absence of transaction costs and
uncertainty, properly designed carbon tax and emission trading regimes
are argued to yield equivalent outcomes in several key aspects (e.g.,
incentives for emission reductions, aggregate abatement costs, and
carbon leakage) \citep{montgomery1972,goulder2013carbon,Stavins2022}.
In 2005, the European Union (EU) first adopted a legally binding emission
trading system among the EU members,\footnote{In addition to the EU ETS, other national or sub-national ETSs have
been implemented or are in the process of development, including in
Canada, China, Japan, New Zealand, South Korea, Switzerland, and the
United States. For information on the most up-to-date policy in practice,
see World Bank Carbon Pricing Dashboard (https://carbonpricingdashboard.worldbank.org/).} and as of the winter of 2024, the permit price stands at about 70
euros per ton of Carbon Dioxide (CO\textsubscript{2}\textsubscript{}).\footnote{Source of EU permit price: (https://tradingeconomics.com/commodity/carbon).}

In this study, we develop and quantify a dynamic multi-region model
of the climate and economy under an emission trading system (ETS),
while the ETS could be global or partial. The model framework extends
the seminal Regional Integrated Model of Climate and the Economy (RICE)
\citep{nordhaus1996regional,nordhaus2010economic}, which captures
the interactions between economic growth and climate systems in a
multi-regional framework. As in the standard neoclassical growth model
(the so-called Ramsey model), a social planner of each region chooses
investment in capital goods to smoothen consumption over time. Economic
activity generates both output and carbon emissions; the latter induces
climate damages that endogenously reduce regional output. Emission
abatement efforts can reduce carbon emissions but are costly. As emissions
generate global externalities, each region has less incentive to undertake
abatement on its own while benefiting from the abatement efforts of
others. In a noncooperative environment, a set of the optimal strategies
of each regional social planner, involving dynamic choices over consumption
and emission abatement, is characterized as a Nash equilibrium concept
from game theory \citep{nordhaus1996regional}. The introduction of
the ETS imposes exogenous region-specific emissions caps and allows
permit trading across regions. Under the ETS, regional social planners
jointly determine the optimal paths of consumption, emission abatement,
and the amount of permit trading, subject to strictly-enforced emission
cap constraints. Heterogeneity across regions, in abatement technologies,
climate damages, productivity, carbon intensity, and emissions cap
constraints, causes divergence in regional marginal abatement costs
(MAC) and the regional social cost of carbon across regions, shaping
the path of the market equilibrium permit price over time.

In an ETS scheme, the primary forces that shape the equilibrium outcome
include regional climate damages, revenues or costs from permit trading,
and regional abatement costs. Using a simple static framework where
a regional social planner $i$ internalizes climate damages and chooses
optimal net emissions, $E_{i}$, (i.e., emissions after abatement),
we show that the equilibrium condition for optimal abatement can be
expressed as: $\mathrm{SCC}=\mathrm{MAC}(E_{i})-m$, where $m$ is
the market price of emission permits. This relationship between regional
SCC, MAC, and emission price has an important implication: a positive
regional SCC implies $\mathrm{MAC}(E_{i})>m$, suggesting that a region
experiencing climate damages has an incentive to abate emissions up
to a level where the MAC exceeds the market price of emission permits.
We emphasize that this equilibrium condition represents the necessary
condition for achieving the optimal level of net emissions \emph{from
the perspective of each regional social planner}.

In the literature, the equality between $\mathrm{MAC}$ and permit
price is used as a competitive equilibrium condition under an ETS
regime \emph{in a decentralized economy}, where individual firms internalize
the emission price. This condition is derived from the perspective
of individual firms in the ETS when they face no additional carbon
tax. If a carbon tax $\tau_{i}$ is levied on firms on top of the
permit price, the competitive equilibrium condition becomes $\tau_{i}=\mathrm{MAC}(E_{i}')-m'$
in the decentralized economy, where $E_{i}'$ and $m'$ are the corresponding
net emissions and permit price under the decentralized equilibrium
with tax $\tau_{i}$. If the carbon tax is set equal to the regional
SCC in each region, then $E_{i}'$ and $m'$ are equal to the regional
social planner's optimal net emissions and permit price. Moreover,
because climate damages---and hence the regional SCC---differ across
regions, a uniform global carbon price alone may not achieve regionally
optimal level of emission pricing. The appropriate equilibrium condition
in this context, $\mathrm{SCC}=\mathrm{MAC}(E_{i})-m$, further implies
that carbon taxation and cap-and-trade are not competing instruments
but rather complementary.

We bring the model to data by aggregating 152 countries into 12 world
regions and simulating the model at annual time steps. The model is
calibrated by fitting it to historical data and the recent predicted
trends of regional climate damage\textcolor{black}{{} \citep{burke2018large}},
regional total factor productivity\textcolor{black}{{} \citep{burke2018large}},
regional abatement costs \textcolor{black}{\citep{ueckerdt2019economically}},
and population projections based on the Shared Socioeconomic Pathway
2 (SSP2; \citealp{samir2017human}), also known as the ``Middle-of-the-Road''
scenario. Besides this calibration, our model uses a stylized but
stable climate system, called the Transient Climate Response to Emissions
(TCRE; \citealp{matthews2009}), which assumes that increases in the
global averaged atmospheric temperature have a nearly linear functional
dependence on the cumulative carbon emissions. We demonstrate that
this temperature system can be calibrated to match closely with the
various Representative Concentration Pathways (RCPs; \citealp{meinshausen2011rcp}).
Due to its simplicity and effectiveness, the TCRE scheme has found
applications in recent economic analyses, as evidenced by studies
such as \citet{brock2017climate}, \citet{dietz2019cumulative}, \citet{mattauch_steering_2020},
and \citet{Barnett2020}. Additionally, \citet{dietz2021} show that
the TCRE scheme does not lead to a large difference in economic analysis
with the seminal DICE framework \citep{nordhaus2017revisiting}, compared
to other more complicated climate systems.

For regional emission cap scenarios, we construct emission cap pathways
reflecting the latest emission targets for 2030 and net-zero pledges
for 2050-2070, contributing to the scant literature on evaluating
the economic and environmental implications of these commitments (e.g.,
\citealp{ven2023}; \citealp{meinshausen2022}; \citealp{elzen2022}).
For each region, emissions cap trajectories are constructed to align
with the latest nationally determined contributions (NDCs) under the
Paris Agreement and the Glasgow Climate Pact. These trajectories reflect
near-term targets for 2025 and 2030, as well as long-term net-zero
commitments spanning 2050 to 2070, and are treated as exogenous constraints.

Our numerical simulations, based on newly calibrated parameters, provide
comprehensive results on the potential economic and environmental
outcomes under the ETS. Assuming full compliance with these emission
caps and no future revisions,\footnote{This paper assumes that regions do not revise their NDC limits (emission
cap constraints) in the future. While primarily introduced to address
computational challenges, this assumption is also supported by economic
and institutional considerations: (i) relaxing NDC limits could result
in penalties from other regions or damage to the region's reputation;
(ii) reducing NDC limits would lead to a loss of benefits from selling
emission permits or more cost from purchasing; and (iii) renegotiating
NDC limits is often time-consuming and infrequent.} the emission permit price is endogenously determined by annual supply
and demand in the global permit market. Under the baseline emission
cap scenario based on the Paris Agreement and Glasgow Pact commitments,
our simulation results show that the oversupply of global permits
in the initial years results in zero permit prices, but by 2050, the
emission permit price can reach up to \$923 per ton of carbon. The
corresponding global average temperature is expected to reach 1.7
degree Celsius above the pre-industrial level by the end of this century.
Furthermore, our simulation results numerically confirm that the regional
SCC exactly equals the difference between the regional MAC and the
permit price, as explored in the predictions of the static framework.
For instance, in 2050, the MAC for the United States is estimated
at \$1,159 per ton of carbon, the permit price at \$923, and the resulting
difference of \$236 matches the regional SCC calculated independently
from the model.

A comparative analysis of the roles of emission caps and the ETS further
illustrates several key findings. First, the findings indicate that
the ETS with the baseline emission caps leads to higher emissions
under noncooperation, as regions with binding emission caps can purchase
permits from less constrained regions, exploiting the surplus permits
in the initial years. This highlights the necessity of maintaining
stringent global emission caps to ensure the efficient functioning
of the ETS. Second, our welfare analysis reveals that when the global
emission cap is sufficiently tight, the ETS can lead to welfare improvements
for all participating regions. These welfare improvements reflect
efficiency gains achieved by reallocating abatement efforts across
regions through market mechanisms.

Additionally, leveraging the flexibility of our model, we evaluate
two additional set of policy scenarios, including a partial ETS participation
and alternative emission caps consistent with different net-zero targets.
Recognizing the challenges of achieving international cooperation
on climate policies, we consider a scenario in which one major region
(the United States) opts out of the global cap-and-trade system. The
results show that the United States experiences a notable welfare
gain. This outcome underscores strong free-rider incentives, as the
opt-out region faces a lighter abatement burden while benefiting from
the emission reductions undertaken by participating regions. In a
second policy experiment, we find that tightening the global emission
cap leads to higher permit prices across future periods.  For instance,
if all regions achieve net-zero emissions by 2050, the permit price
could reach \$1,621 per ton of carbon by 2049. However, even with
such strict emissions restrictions, the global temperature is projected
to rise by 1.62°C by the end of the century. These results suggest
that meeting the global target of limiting warming to 1.5°C will require
even more stringent emission reduction commitments than those set
out in the Glasgow Pact, albeit at the cost of significantly higher
permit prices. Our results are qualitatively consistent with the findings
of other IAMs \citep{ven2023,meinshausen2022,elzen2022}, which indicate
that the strengthened post-Glasgow NDC and net-zero targets, covering
both near-term and long-term goals, are insufficient to limit warming
to 1.5°C above the pre-industrial level, though full implementation
could restrict warming to below 2°C. Finally, to ensure the robustness
of our results, we conduct sensitivity analyses using alternative
parameter values for climate damages and abatement costs, calibrated
to empirical estimates reported in the key existing literature.

\section{Related Literature}

This paper contributes to the literature on macroeconomic modeling
of climate change. Our model framework is closely related to the RICE
model \citep{nordhaus1996regional,nordhaus2010economic,yang_model_2023},
an extension of the global DICE model \citep{nordhaus2014} in a multi-region
framework, capturing interactions between economic growth and climate
systems. However, \citet{nordhaus1996regional} and \citet{yang_model_2023}
do not investigate an ETS, and \citet{nordhaus2010economic} does
not solve a problem with an ETS under noncooperation. Following the
RICE model, a group of studies have explored climate policy in a dynamic
multi-regional framework under cooperation and noncooperation. For
instance, \citet{luderer2012economics} and \citet{jakob2012time}
compare the long-term predictions of three-region energy-economy models
under specific environmental targets, such as stabilizing the atmospheric
CO\textsubscript{2} concentrations at 450 ppm. However, these studies
do not incorporate actual NDCs or regional net-zero targets to assess
the region-specific emissions and economic pathways. Other studies,
such as \citet{van2016non} and \citet{jaakkola2019non}, focus on
stochastic events, including climate tipping points and technological
breakthroughs, comparing the results under different levels of cooperation.
\citet{cai2023climate} build a dynamic IAM for two economic regions
(North and Tropic/South) to compute regional SCCs under cooperation
and noncooperation. \citet{hambel2021social} extend the RICE framework
by integrating endogenous international trade under noncooperation
and provide a closed-form analytical solution for the regional SCCs
under certain model assumptions. \citet{iverson2021carbon} study
a Markov perfect equilibrium in a dynamic game with a social planner
deciding climate policies and non-constant discount rates. For a comprehensive
review of macroeconomic models of climate change, see \citet{fernandezvillaverde2025}.
Nonetheless, the literature has yet to examine the dynamics of an
ETS with an endogenous market price of emission permits and regional
SCCs under a noncooperative setting. This paper addresses this gap
by providing a comprehensive analysis of an ETS as well as identifying
the relationship between permit prices and regional SCCs under a multi-region
dynamic noncooperative framework.

This paper also adds to the extensive literature on environmental
economic policy, with a particular focus on carbon pricing. ETSs,
alongside carbon taxes, have garnered attention as promising mechanisms
for reducing global emissions. However, comprehensive analyses of
ETSs as instruments for global climate policy remain relatively limited.
Among the most closely related studies are extensions of the WITCH
model \citep{Bosetti2006} that incorporate an ETS, including the
analyses of regional ETSs for Asian countries \citep{massetti2012},
endogenous technological change \citep{DeCian2012}, and banking of
emission permits \citep{bosetti2009}.  The WITCH model considers
alternative emission permit allocation schemes in which permits are
distributed according to population or current emission shares, whereas
our study applies emission caps consistent with the Paris Agreement
and the Glasgow Climate Pact. In addition, our study departs from
previous WITCH analyses by examining regional SCCs and their economic
policy implications under an ETS, which were not explored in earlier
studies. \citet{carbone2009case} present another relevant study,
constructing a computable general equilibrium model incorporating
countries' endogenous participation in an ETS and allocation of emission
permits, and solving for the equilibrium permit price. Yet, \citet{carbone2009case}
examine the ETS in a static setting, missing the crucial dynamic aspects
of climate change and its connection with emission abatement decisions.
\citet{fischer2011emissions} develop a dynamic stochastic general
equilibrium model to compare outcomes of emission caps and tax policies,
but they do not allow for emission trading between regions. Another
group of studies examines the potential efficiency gains from integrating
regional ETSs into a global system \citep{habla2018strategic,doda2019linking,holtsmark2020effects,holtsmark2021dynamics,mehling2018linking}.
Some studies highlight that differentiated emission pricing under
ETSs based on location-specific damages can be welfare improving \citep{muller2009efficient,holland2015optimal,fowlie2019market}.

More broadly, a substantial body of literature has examined ETSs from
various perspectives. Several studies have analyzed how the initial
allocation of permits affects the equilibrium outcomes \citep{Hahn2011},
or have compared allocation methods between auctioning and free allocation
\citep{goulder2008instrument,goulder2010impacts}. Other studies have
compared the relative efficiency of carbon policies under cost uncertainty,
showing that carbon tax can be more efficient than ETS under certain
conditions, and vice versa \citep{weitzman1974,Stavins1996,Karp2024}.
In another strand of literature, a number of studies have focused
on empirically analyzing the regional emission trading markets currently
in practice, such as the EU ETS \citep{hitzemann2018equilibrium,fuss2018,perino2022european},
China ETS \citep{goulder2022china}, or California ETS \citep{borenstein2019}.
An important issue with a regional ETS is carbon leakage, which refers
to the shift of emission intensive production to regions outside the
jurisdiction of the ETS. Previous studies have compared different
policy instruments aimed at mitigating this problem and providing
a level-playing field to the firms operating within the ETS \citep{ambec2024economics,bohringer2014cost,farrokhi2021can,fowlie2022mitigating,levinson2023}.

While there is no general agreement among economists on whether carbon
taxation or ETS is better \citep{Stavins2022}, previous works have
often considered these two pricing instruments (trading versus taxes)
as policy substitutes. For instance, some studies support carbon taxation
over ETS, highlighting concerns about price volatility in ETS \citep{Nordhaus2007},
or the presence of uncertainty regarding emission abatement costs
\citep{newell2003regulating}. For instance, \citet{newell2003regulating}
find that carbon taxation can yield higher welfare benefits than ETS
under such uncertainties. Conversely, other studies advocate ETS over
carbon taxation because allocation of permits in ETS allows for more
flexibility, and ETS faces less uncertainty in controlling the cumulative
amount of carbon emissions compared to taxes \citep{Keohane2009}.
\citet{harstad2010trading}, \citet{Hahn2011}, and \citet{Stavins2022}
argue that from a practical perspective, ETS may be a preferred instrument
over tax, as free allowances could be negotiated among participating
agents to redistribute burden and, thus be used to gain political
support. Moreover, in a global ETS, emission permits will be traded
globally with one price for all nations based on market forces, which
may alleviate the problem of carbon leakage \citep{Fowlie_JPE2016}.
However, \citet{harstad2010trading} argue that although an ETS in
a perfect market is the first-best system, frequent government interventions
to redistribute allocations among firms may result in distortions
in the market allocation.\footnote{Environmental pollution has also been studied from the context of
fiscal federalism, which considers what level of government should
regulate pollution \citep{oates1999,ogawa2009think,banzhaf2012,williams2012growing}.} For a comprehensive comparison on carbon taxation and trading regimes,
see e.g., \citet{strand2013strategic,schmalensee2017lessons,Cai2021_uncertainty,Stavins2022}.
This study contributes to the literature by showing that regional
carbon tax is complementary to ETS under noncooperation.

\section{A Static Framework of Climate and the Economy with an ETS\label{subsec:StaticModel}}

This section presents a simplified static model framework with a global
ETS to provide intuition on the relationship between the regional
SCC and regional MAC. In a global ETS regime, the main forces determining
the equilibrium are climate damages, costs (or revenue) from emissions
trading, and costs from regional emission abatement efforts. In this
simplified static framework, it can be easily shown that the regional
SCC equals the regional MAC minus the market price of emission permits.
Building on these insights, the next section presents the multi-region
dynamic general equilibrium model, followed by the quantification
strategies and numerical analysis.

Consider a world economy with multiple regions under a global ETS
regime, and let $\mathcal{I}$ denote the set of regions. Each region
$i\in\mathcal{I}$ is allocated an exogenous emission cap $\bar{E}_{i}$,
which represents the region's maximum allowable emissions and is strictly
enforced. Economic activity generates emissions, with $E_{i}^{\mathrm{Gross}}$
denoting a region's \emph{gross emissions}, which are treated as exogenous
in this simplified framework.\footnote{It is innocuous to assume that gross emissions are exogenous in this
static framework. In the full dynamic model introduced later, regional
gross emissions are proportional to GDP, which is determined by capital
and the exogenous population growth. Given that the current level
of capital is determined in the previous period, gross emissions can
be considered exogenous for each period in a static equilibrium.} To comply with the emission cap constraint, each region's social
planner can either (i) purchase or sell emission permits in the global
emission market, or (ii) undertake costly abatement efforts. Let $E_{i}$
denote its \emph{emissions net of abatement} (henceforth referred
to as \emph{net emissions}), and $E_{i}^{P}$ the quantity of \emph{emission
permits purchased} in the market. A positive $E_{i}^{P}$ indicates
that the region purchases permits, while a negative $E_{i}^{P}$ indicates
that the region sells permits. The emission cap constraint for each
region is then given by $E_{i}\leq\overline{E}_{i}+E_{i}^{P}$.

In this economy, the regional social planner aims to minimize the
total economic costs associated with its emissions, including (i)
climate damages due to global emissions, (ii) costs (or revenue) from
emission trading, (iii) and costs from own regional emission abatement.
Formally, a social planner of each region solves the following minimization
problem:

\begin{equation}
\min_{0\leq E_{i}\leq\overline{E}_{i}+E_{i}^{P}}\quad\underbrace{D_{i}\left(\sum_{j\in\mathcal{I}}E_{j}\right)}_{\text{Climate Damages}}+\underbrace{\vphantom{{D_{i}\left(\sum_{j\in\mathcal{I}}E_{j}\right)\}}}mE_{i}^{P}}_{\text{ETS Costs/Revenue}}+\underbrace{\vphantom{{D_{i}\left(\sum_{j\in\mathcal{I}}E_{j}\right)\}}}\int_{E_{i}}^{E_{i}^{\mathrm{Gross}}}\mathrm{MAC}_{i}(E)dE}_{\text{Emission Abatement Cost}}.\label{eq:total_cost_SP}
\end{equation}
The first term of the objective function represents regional climate
damages, where $D_{i}(\cdot)$ is a function that captures the regional
climate damages from global net emissions, $\sum_{j\in\mathcal{I}}E_{j}$.
Here $D_{i}(\cdot)$ can also be considered as the present value of
future climate damages from the current global net emissions in the
corresponding dynamic framework. The second term accounts for the
cost (or revenue) from purchasing (or selling) permits at price $m$,
which is the Nash equilibrium price satisfying the market clearing
condition $\sum_{j\in\mathcal{I}}E_{j}^{P}=0$. The quantity of emission
permits purchased may influence the global permit price, making the
market price depend on their emission permit purchase choices, which
in turn are contingent on the trade decisions of other regions. Lastly,
the third term reflects the total abatement cost incurred to reduce
emissions by $(E_{i}^{\mathrm{Gross}}-E_{i})$, where $\mathrm{MAC}_{i}(\cdot)$
denotes the region's marginal abatement cost and is assumed to be
monotonically increasing over $E_{i}$, with $\mathrm{MAC}_{i}(E_{i}^{\mathrm{Gross}})=0$.
For simplicity, it is assumed that net emissions are nonnegative throughout
the paper.\footnote{This assumption ensures that no region can abate more than its gross
emissions solely to sell permits to other regions.}

In the static Nash equilibrium of the world economy, every regional
social planner simultaneously solves equation (\ref{eq:total_cost_SP})
while satisfying the market-clearing condition $\sum_{j\in\mathcal{I}}E_{j}^{P}=0$,
which implies that $\sum_{j\in\mathcal{I}}E_{j}\leq\sum_{j\in\mathcal{I}}\overline{E}_{j}$.
We assume that each region is a price taker for permits, as a large
number of firms in each region participate in the permit trading market
in practice. Thus, the Karush-Kuhn-Tucker conditions of equation (\ref{eq:total_cost_SP})
imply that the solution to the region's optimal net emissions leads
to the following relationship between the marginal regional damage,
the MAC, and the permit price:

\begin{equation}
\underbrace{\frac{\partial D_{i}(\sum_{j\in\mathcal{I}}E_{j})}{\partial E_{i}}}_{\text{Regional SCC}}=\underbrace{\mathrm{\mathrm{MAC}}_{i}(E_{i})-m^ {}\vphantom{{D_{i}\left(\sum_{j\in\mathcal{I}}E_{j}\right)\}}}}_{\text{Deviation from Market Equilibrium}},\label{eq:Lambda-condition-SP}
\end{equation}
when $E_{i}>0$. The left hand side of this equation represents the
marginal regional damage, which corresponds to the regional SCC. This
term captures the additional regional economic cost imposed by a unit
increase in global emissions, reflecting the region's contribution
to global climate change. The right hand side captures the difference
between $\mathrm{MAC}_{i}(E_{i})$ and $m$, reflecting the deviation
of regional abatement efforts from the market equilibrium. The equation
(\ref{eq:Lambda-condition-SP}) simplifies to $\mathrm{SCC}_{i}=\mathrm{MAC}_{i}(E_{i})-m^ {}$,
where $\mathrm{SCC}_{i}$ denotes the regional SCC. A positive regional
SCC implies $\mathrm{MAC}_{i}(E_{i})>m^ {}$, suggesting that a region
experiencing climate damages has an incentive to abate emissions to
a level where the marginal abatement cost exceeds the market price
of emission permits.

In the environmental economics literature, it is well established
that when production generates emissions as a negative externality
in an economy, a regulator can achieve the socially efficient level
of abatement by setting an emission cap at that level and issuing
tradable permits to firms. The resulting market equilibrium under
cap-and-trade is equivalent to the outcome of an optimal emission
tax equal to the MAC at the efficient abatement level, in the absence
of cost uncertainty or transaction costs. However, in a multi-region
framework where climate damages are region-specific, the decentralized
equilibrium outcome under the global ETS does not necessarily coincide
with the regional social planner\textquoteright s optimal outcome.
The underlying reason is that, because climate damages are heterogeneous
across regions, a uniform global carbon price alone may not yield
regionally optimal levels of emission abatement or the corresponding
emission prices.

The intuition behind $\mathrm{SCC}_{i}=\mathrm{MAC}_{i}(E_{i})-m$
can be further illustrated through an extreme case. If emission caps
are sufficiently large such that they are not binding for any region---effectively
a scenario without emission constraints---permit prices and trading
volumes would be zero. Imposing the equilibrium condition $\mathrm{MAC}=m$
in this situation would imply no abatement (i.e., $\mathrm{MAC}=0$
when $m=0$). However, it is well established that even in the absence
of binding caps, the optimal level of emission abatement from the
perspective of a regional social planner remains strictly positive
(i.e., $\mathrm{SCC}=\mathrm{MAC}>0=m$) under noncooperative Nash
equilibrium (see, e.g., \citealp{nordhaus1996regional,cai2023climate}).
This example further highlights the gap between the regional social
planner\textquoteright s optimal abatement level and the decentralized
market outcome, as firms undertake no abatement in the absence of
an emission price. The basic intuition explored here carries over
to the dynamic general equilibrium model introduced in the next section,
where we further confirm numerically that $\mathrm{SCC}_{i}=\mathrm{MAC}_{i}(E_{i})-m$
continues to hold.

An additional important implication of this relationship is that the
two pricing instruments---emission taxes and cap-and-trade---are
not competing policies but rather complementary under the global ETS
framework. Although our model does not explicitly introduce firms,
the regional MAC is effectively obtained as an aggregation of individual
firms\textquoteright{} marginal abatement quantities at a given price
level within a region \citep{keohane2016}. To achieve the optimal
level of net emissions, $E_{i}$, \emph{from each regional social
planners perspective} under the global ETS, the optimal policy for
each regional planner is to impose, within its own economy, a carbon
tax equal to its regional $\mathrm{SCC}_{i}$ on top of the permit
price from the global ETS. This ensures that individual firms internalize
both the regional SCC and the global permit price. With the regional
tax incorporated, each individual firm chooses net emissions such
that its own MAC equals the sum of the permit price and the regional
carbon tax, i.e., $\mathrm{MAC_{i}}(E_{i}')=\mathrm{\tau}_{i}+m'$,
where the optimal carbon tax is $\mathrm{\tau}_{i}=\mathrm{SCC}_{i}$
from the perspective of the regional social planner.

For a partial ETS, if region $i$ participates in the ETS, then the
relation $\mathrm{SCC}_{i}=\mathrm{MAC}_{i}(E_{i})-m$ still holds,
where $m$ is the equilibrium permit price between the participating
regions; if region $i$ does not participate, then the relation is
changed to $\mathrm{SCC}_{i}=\mathrm{MAC}_{i}(E_{i})$, which is a
standard relation under no ETS.

\section{A Dynamic Regional Model of Climate and the Economy with an ETS\label{sec:Dyn-Model}}

We now introduce a dynamic regional model of climate and economy that
integrates a global ETS across multiple regions. \textcolor{black}{Our
model framework extends the RICE model \citep{nordhaus1996regional,nordhaus2010economic}
by incorporating the global ETS in a dynamic setting and the TCRE
climate system with annual time steps. }One of the main focuses of
the model is to characterize the equilibrium path of carbon prices
in a noncooperative environment, where a social planner of each region
maximizes its own social welfare by optimally choosing emissions abatement,
permit purchases in the global carbon market, and consumption over
time. Future regional emissions are constrained by emission caps based
on commitments established under the Paris Agreement, later updated
by the Glasgow Pact, and by net-zero targets.

The macroeconomic framework of our model employs a multi-regional
representation of the Ramsey growth model. Each region is indexed
by $i\in\mathcal{I}$, where $\mathcal{I}$ is the set of regions.
Time is discrete and infinite, with annual time steps indexed by $t=0,1,2,\cdots$.
All agents are forward-looking with complete information. The model
presented here abstracts from uncertainty, excludes international
trade of goods other than emission permits, and assumes frictionless
trading in the ETS.

\subsection{The Economic System}

Each region consists of a representative household with a population
size of\textcolor{black}{{} $L_{i,t}$, and utility of the representative
consumer is given by}
\begin{equation}
u(c_{i,t})=\frac{c_{i,t}^{1-\gamma}}{1-\gamma},\label{eq:preference}
\end{equation}
where $c_{i,t}$ is per capita consumption and $\gamma$ is the inverse
of intertemporal elasticity of substitution. Following DICE-2016R
\citep{nordhaus2017revisiting}, $\gamma$ is set to 1.45.

There is a representative firm in each region that employs a Cobb-Douglas
production technology using capital and labor as inputs, and produces
a numeraire good whose price is normalized to 1. The representative
household owns all input factors and the firm of the regional economy.
The gross output, or pre-damage output, $Q_{i,t}$, is given by

\textcolor{black}{
\begin{equation}
Q_{i,t}=A_{i,t}K_{i,t}^{\alpha}L_{i,t}^{1-\alpha},\label{eq:production function}
\end{equation}
}where\textcolor{black}{{} $K_{i,t}$ is capital stock, and $\alpha=0.3$
is the elasticity of gross output with respect to capital, as in \citet{nordhaus2017revisiting}.}
Consistent with the standard neoclassical growth model, capital depreciates
over time, and the firm invests to replenish and accumulate capital
stock. The evolution of capital follows\textcolor{black}{
\begin{equation}
K_{i,t+1}=(1-\delta)K_{i,t}+I_{i,t},\label{eq: capital transition}
\end{equation}
}w\textcolor{black}{here $I_{i,t}$ is investment and $\delta=0.1$
is the} rate of depreciation of capital stock.

\textcolor{black}{Regions experience climate damages resulting from
the externalities of global emissions, with the extent of these damages
varying across regions. The output net of climate-induced damages,
denoted as }$Y_{i,t}$\textcolor{black}{, is given by:}

\textcolor{black}{
\begin{equation}
Y_{i,t}=\left(\frac{1}{1+\pi_{1,i}T_{t}+\pi_{2,i}T_{t}^{2}}\right)Q_{i,t},\label{eq:net output}
\end{equation}
where $\pi_{1,i}$ and $\pi_{2,i}$ are region-specific climate damage
parameters, and $T_{t}$ is the global average temperature increase
in }degrees\textcolor{black}{{} Celsius above the pre-industrial level.
The specification captures the adverse (or potentially beneficial
for some regions) effects of rising global average temperature, where
local damages increase nonlinearly with temperature, following the
quadratic form commonly used in the climate economics literature \citep{nordhaus2014,nordhaus2017revisiting}.}

\subsection{Emissions}

Each region's economic activity produces carbon emissions, and the
representative firm faces emission constraints of the region. The
\emph{gross emissions} before abatement, in gigatonnes of carbon (GtC),
is assumed to be proportional to the gross output for region $i$
at time $t$:\textcolor{black}{
\begin{equation}
E_{i}^{\mathrm{Gross}}=\sigma_{i,t}Q_{i,t},\label{eq: emissions}
\end{equation}
where} $Q_{i,t}$ is the gross output, and\textcolor{black}{{} $\sigma_{i,t}>0$
is the exogenous carbon intensity. The firm may choose to reduce a
fraction of gross emissions, with its efforts represented by emission
control rate $\mu_{i,t}\in[0,1]$.}\footnote{Alternatively, abatement can also be modeled through the reduced use
of fossil fuel energy inputs in the production function (see, e.g.,
\citealp{Bosetti2006}; \citealp{bauer_remind-r_2012}; \citealp{golosov2014optimal};
\citealp{BaldwinCaiKuralbayeva}). However, such models often require
disaggregation of fossil fuel energy firms, renewable energy firms,
final-goods producers, and other sectors, increasing the complexity
of the model. The emission control rate approach simplifies this by
offering a more streamlined representation.}\textcolor{black}{{} The }\textcolor{black}{\emph{amount of emissions
abated}}\textcolor{black}{{} by each region, $E_{i,t}^{A}$, is then
expressed as follows.}

\textcolor{black}{
\begin{equation}
E_{i,t}^{A}=\mu_{i,t}\sigma_{i,t}Q_{i,t}.\label{eq:abatement of emissions}
\end{equation}
The emission abatement efforts incur costs to the firm and are heterogeneous
across regions due to technological differences. The emission abatement
cost, $\Phi_{i,t}$, is specified as
\begin{equation}
\Phi_{i,t}=b_{1,i,t}\mu_{i,t}^{b_{2,i}}Q_{i,t},
\end{equation}
where $b_{1,i,t}=(b_{1,i}+b_{3,i}\exp(-b_{4,i}t))\sigma_{i,t}$}.
The parameters $b_{1,i}$, $b_{2,i}$, $b_{3,i},$ and $b_{4,i}$
govern the cost structure of emission abatement cost, $\Phi_{i,t}$,
which depends on both the gross output $Q_{i,t}$ and the exponential
function of the emission control rate, $\mu_{i,t}$. This specification
captures the dynamic nature of abatement costs, reflecting potential
technological advancements that reduce the cost of emissions abatement
over time.

In a global ETS, or a cap-and-trade system, each region is provided
with an emission allowance and can trade emission permits with other
regions. The representative firm in each region that emits beyond
its cap can purchase permits, while those emitting below their allowance
can sell excess permits. Denoting the \emph{emissions net of abatement}
(\emph{net emissions}) as $E_{i,t}=E_{i}^{\mathrm{Gross}}-E_{i,t}^{A}$,
the emission cap constraint is represented by\textcolor{black}{
\begin{equation}
E_{i,t}-E_{i,t}^{P}\leq\overline{E}_{i,t},\label{eq:emissions cap constraint}
\end{equation}
where $\overline{E}_{i,t}$ is the }\textcolor{black}{\emph{emission
cap}}\textcolor{black}{{} assigned to region $i$ at time $t$, and
$E_{i,t}^{P}$ denotes the amount of }\textcolor{black}{\emph{emissions
purchased}}\textcolor{black}{{} from other regions. Note that $E_{i,t}^{P}>0$
indicates that region $i$ is a net buyer of emission permits at time
$t$, while $E_{i,t}^{P}<0$ implies region $i$ is a net seller of
emission permits at time $t$. In a model that does not consider emission
permit trade, the emissions purchase is simply set at $E_{i,t}^{P}=0$
for all $i$ and $t$.}

\subsection{The Climate System}

The global average temperature rises as carbon emissions accumulate
in the atmosphere. Adopting the TCRE climate system representation
\citep{matthews2009}, it is assumed that the global average temperature
increase above the pre-industrial level is approximately linear to
cumulative global emissions $\mathcal{E}_{t}$, i.e.,
\begin{equation}
T_{t}=\zeta\mathcal{E}_{t},\label{eq:TCRE}
\end{equation}
where\textcolor{black}{{} $\zeta$ represents the c}ontribution rate
of cumulative global emissions to temperature. The cumulative global
emissions evolve according to\textcolor{black}{
\begin{equation}
\mathcal{E}_{t+1}=\mathcal{E}_{t}+\sum_{i\in\mathcal{I}}E_{i,t}.\label{eq:cumEmission}
\end{equation}
This dynamic process captures the accumulation of emissions in the
atmosphere over time, with each region contributing to the global
emissions stock and, consequently to the increase in global temperature.}

\subsection{Market Clearing}

The goods market clearing implies that the total consumption of each
region under the ETS is constrained by\textcolor{black}{
\begin{equation}
c_{i,t}L_{i,t}=Y_{i,t}-I_{i,t}-\Phi_{i,t}-m_{t}E_{i,t}^{P},\label{eq:consumption}
\end{equation}
where $m_{t}$ is the market equilibrium price of emission permits.
As in the static framework, the last term on the right-hand side reflects
the cost or revenue generated from the emission permit trade.}

Finally, the emission trading market clears each period, given by:\footnote{The current model assumes that intertemporal lending or borrowing
of emission permits is not allowed.}\textcolor{black}{
\begin{equation}
\sum_{i\in\mathcal{I}}E_{i,t}^{P}=0.\label{eq:market clearing condition}
\end{equation}
If it is a partial ETS and region $i$ does not participate in the
partial ETS, then we set $E_{i,t}^{P}$ to be fixed at zero.}

\section{Solving for the Equilibrium}

Building on the model components outlined in Section \ref{sec:Dyn-Model},
we now define the noncooperative equilibrium of our multi-region dynamic
model with the ETS and present an algorithm used to obtain the Nash
Equilibrium solution.

\subsection{The Noncooperative Equilibrium}

In the noncooperative model, the regional social planner of each region
maximizes the region's own lifetime social welfare. The maximization
problem for each region $i$ is defined as\textcolor{black}{
\begin{equation}
\max_{c_{i,t},E_{i,t}^{P},\mu_{i,t}}\sum_{t=0}^{\infty}\beta^{t}u(c_{i,t})L_{i,t},\label{eq:non-cooperative model}
\end{equation}
where $\beta$ is the discount factor. We follow DICE-2016R \citep{nordhaus2017revisiting}
and set $\beta=0.985$.} Since one region's emissions will influence
the global average temperature, and therefore other regions' output,
the maximization problems of all regions have to be solved simultaneously
as a dynamic game. Then we define the dynamic Nash equilibrium of
the economy as follows.
\begin{description}
\item [{DEFINITION:}] Given the initial capital and cumulative global emissions,
$\{K_{i,0},\mathcal{E}_{0}:\ i\in\mathcal{I}\}$, and the exogenous
paths of emission caps $\{\overline{E}_{i,t}:\ i\in\mathcal{I},t\geq0\}$,
the dynamic Nash equilibrium for the noncooperative model is a sequence
of quantities $\{c_{i,t},E_{i,t}^{P},\mu_{i,t},K_{i,t},\mathcal{E}_{t},T_{t}:\ i\in\mathcal{I},t\geq0\}$
and prices $\{m_{t}:\ t\geq0\}$ that simultaneously solve the maximization
problem (\ref{eq:non-cooperative model}) for all regions subject
to equations (\ref{eq:preference})-(\ref{eq:market clearing condition}).
\end{description}
The optimal solution for this dynamic multi-region model involves
three choice problems. First, as in the standard Ramsey-type growth
model, each region faces an intertemporal choice problem in which
there is a trade-off between current consumption and future consumption.
Each region may sacrifice present consumption to make investments,
which can contribute to higher consumption in the future. Second,
the intertemporal choice problem is further compounded by climate
damages. Current production increases the global temperature, which
subsequently lowers future productivity. Since emissions abatement
has positive externalities, a region's returns from abatement efforts
may not be large enough to offset the cost of abatement. Therefore,
the optimal solution of each region is highly dependent on the choices
made by other regions. Lastly, the ETS allows each region to choose
between purchasing emission permits from the market and undertaking
further abatement. The ETS promotes efficient abatement globally by
encouraging regions with better abatement technology or capacity (thus,
with lower abatement cost) to conduct more abatement, and regions
with less efficient abatement technology to purchase permits from
other regions.

Note that the equilibrium concept in our noncooperative model is an
Open-Loop Nash equilibrium (OLNE), which provides a solution path
over time depending on the initial state. In an OLNE, regions commit
to the strategies over time for their decision variables---consumption
($c_{i,t}$), emission purchase ($E_{i,t}^{P}$), and emission control
rate ($\mu_{i,t}$)--- at the initial period and cannot change their
behavior over time. This concept contrasts with the Markov Perfect
Equilibrium (MPE), where regions may make multiple decisions over
time, allowing adaptation in their strategies. While the OLNE concept
may be less satisfactory than the MPE concept (since OLNE is not subgame
perfect), it has the computational advantages of solving open-loop
versus feedback, particularly when the dimension of the state space
is large and there are occasionally binding constraints as in our
case.

\subsection{The Algorithm for the Noncooperative Model}

Obtaining the optimal solution of the dynamic model involving multiple
regions under noncooperation is challenging. In particular, finding
an equilibrium solution for the emission permit prices that satisfies
the optimality conditions for each region as well as the market clearing
condition poses significant computational challenges.

Here we outline the algorithm we develop to obtain the optimal solution
for our model. With the discount factor $\beta=0.985$, the discounted
utilities after 300 years are nearly zero and have little impact on
the solution in the first 100 years. Let 
\[
V_{i,300}(K_{1,300},...,K_{12,300},\mathcal{E}_{300})=u\left(\frac{0.75Y_{i,300}}{L_{i,300}}\right)\frac{L_{i,300}}{1-\beta}
\]
be a terminal value function at the terminal year 300, which approximates
the present value of utilities after 300 years, assuming that consumption
at any $t\geq300$ is equal to 75 percent of the output $Y_{i,300}$
at $t=300$ and that the exogenous population after 300 years stays
at its value at $t=300$. Note that $Y_{i,300}$ is computed with
a function of the terminal state $K_{i,300}$ and $\mathcal{E}_{300}$.
Thus, we can transform the infinite horizon models to finite horizon
models, where region $i$'s social welfare is rewritten as
\[
\sum_{t=0}^{299}\beta^{t}u(c_{i,t})L_{i,t}+\beta^{300}V_{i,300}(K_{1,300},...,K_{12,300},\mathcal{E}_{300}),
\]
We also numerically verify that this time horizon truncation at 300
years has little impact on the OLNE solution in the first 100 years,
by solving the same model but with a time horizon truncation at 400
years. This time horizon truncation method is common in solving infinite-horizon
non-stationary models. For example, DICE-2016 truncates its model's
infinite horizon to 500 years, with a terminal value function being
zero everywhere, for obtaining its numerical solution.

The algorithm to solve the noncooperative model is as follows:
\begin{description}
\item [{Step}] 1. \textit{Initialization}. Set an initial guess of permit
prices $\{m_{t}^{0}:\ t\geq0\}$, and emissions $\{E_{i,t}^{0}:\ t\geq0\}$.
Iterate through steps 2, 3 and 4 for $j=1,2,...$, until convergence.
\item [{Step}] 2\emph{. Maximization Step }at iteration $j$. For each
region $i$, solve the maximization problem (\ref{eq:non-cooperative model})
without the market clearing condition (\ref{eq:market clearing condition}),
assuming the permit prices \emph{$\{m_{t}^{j-1}:\ t\geq0\}$ }and
other regions' emissions $\{E_{i',t}^{j-1}:\ i'\neq i,\ t\geq0\}$
are given from the initialization step when $j=1$ or Step 3 at iteration
$j-1$ when $j>1$. The optimal emissions and permits purchased for
region $i$ are denoted $\{E_{i,t}^{*,j},E_{i,t}^{P,j}:\ t\geq0\}$.
\item [{Step}] 3\emph{. Update Step }at iteration $j$.\emph{ }After solving
for the optimization problem of all regions respectively in Step 2,
update the permit prices and emissions as
\begin{eqnarray*}
m_{t}^{j} & = & m_{t}^{j-1}\exp\left(\omega\sum_{i\in\mathcal{I}}E_{i,t}^{P,j}\right),\\
E_{i,t}^{j} & = & \omega E_{i,t}^{*,j}+(1-\omega)E_{i,t}^{j-1},\ \ \forall i\in\mathcal{I},
\end{eqnarray*}
where $\omega=0.1$ is a weight parameter and $\sum_{i\in\mathcal{I}}E_{i,t}^{P,j}$
is the net quantity of traded emission permits.
\item [{Step}] 4\emph{. Check the convergence criterion.} Check if $m_{t}^{j}\simeq m_{t}^{j-1}$,
$E_{i,t}^{j}\simeq E_{i,t}^{j-1}$, and $E_{i,t}^{P,j}\simeq E_{i,t}^{P,j-1}$
for every region $i$ and $t\geq0$. If so, stop the iteration, otherwise
go to Step 2 by increasing $j$ with 1. Note that $m_{t}^{j}=m_{t}^{j-1}$
implies that the market clearing condition (\ref{eq:market clearing condition})
holds at the solution.
\end{description}
This algorithm embodies the concept of market equilibrium. When there
is a net positive quantity of traded emission permits in the market,
indicating excess demand, we increase the permit prices. Conversely,
when there is a net negative quantity of traded emission permits,
indicating excess supply, we lower the permit prices. This mechanism
ensures that the market reaches a balance between supply and demand,
and thus the market clears. Furthermore, this algorithm guarantees
that each region obtains its optimal solution and reaches an equilibrium
state. In other words, no region has the incentive to deviate from
the Nash equilibrium for the noncooperative model solution.

\section{Data and Calibration}

When taking our model to data, two key objectives are pursued. The
first objective is to generate regional emission cap pathways ($\overline{E}_{i,t}$)
for future periods, constraining the constituent nations of each region
to meet their emissions commitments under the Glasgow Pact and the
long-term net zero emission targets. The second objective is to determine
parameters for total factor productivity (\textcolor{black}{$A_{i,t}$}),
carbon intensity ($\sigma_{i,t}$), abatement cost (\textcolor{black}{$b_{1,i}$,
$b_{2,i}$, $b_{3,i}$, $b_{4,i}$}), and climate damage (\textcolor{black}{$\pi_{1,i},\pi_{2,i}$)}
that reflect the future projections provided in recent studies \citep{burke2018large,ueckerdt2019economically,kahn2021long}.
We obtain these region-specific parameters, which capture regional
heterogeneities in GDP growth, emissions, technologies, and climate
damages.

We assume that the world is divided into 12 aggregated regions: \textcolor{black}{the
United States (US), the EU, Japan, Russia, Eurasia, China, India,
Middle East (MidEast), Africa, Latin America (LatAm), Other High-Income
countries (OHI) and other non-OECD Asia (OthAs).} These 12 regions
are formed by aggregating 152 countries around the world, following
the regional classification in the RICE model \citep{nordhaus1996regional,nordhaus2010economic}.\footnote{\textcolor{black}{See Appendix 1 \ref{sec:constituent countries of multi-country regions}
for the full list of countries and regional aggregation.}} Note that, while we present our simulation results at a regionally
aggregated level for computational tractability, our methods are applicable
to a larger number of regions and can be readily extended to obtain
country-level outcomes. For example, the regional SCC for Africa can
be disaggregated to approximate country-level SCCs, since the regional
SCC is equal to the sum of the SCCs of the individual countries within
the region. \footnote{Regional SCC is the present value of future climate damages in a region
resulting from an additional unit of global emissions released in
the current period.} The initial year is set to 2020, with country-level historical data
on population \textcolor{black}{(billions)}, capital (\$ trillions,
2020), GDP (\$ trillions, 2020), and emissions (CO\textsubscript{2}\textcolor{black}{{}
equival}ent, GtC) sourced from the World Bank. \footnote{To obtain initial capital stock, we used capital formation data from
the World Bank. Specifically, we computed $K_{i,2020}=\sum_{g=0}^{20}(0.9)^{g}\text{CF}_{i.2020-g}$,
where $\text{CF}_{i,t}$denotes capital formation adjusted to 2020
values in trillions of USD.} For future population pathways, projections from the SSP2 scenario
\citep{samir2017human} are employed.

\subsection{Regional Emission Cap Pathways\label{subsec:Regional-Emission-Cap}}

Since there is no global cap-and-trade system or ETS currently in
place, we consider the emission commitments outlined in the Paris
Agreement and the COP26 Glasgow Climate Pact. Under the Paris Agreement,
195 countries or regions set Nationally Determined Contributions (NDCs)
that specify their near-term targets for 2025 or 2030, along with
long-term net-zero commitments for 2050 to 2070. These near-term targets
were further strengthened during the Glasgow Climate Pact in 2021,
with most countries and regions submitting updated or new NDCs.\footnote{Individual NDC documents were obtained from the following source:
UNFCCC NDC Registry (https://unfccc.int/NDCREG)} We collect reports of the most updated NDCs after the Glasgow Climate
Pact and obtain the target years to reach net zero emissions of different
countries from Climate Action Tracker.\footnote{See https://climateactiontracker.org/.}
Based on these datasets, we create the baseline regional emission
cap pathways for future periods following the strategy detailed below.

As the first step, we obtain the near-term emission targets for each
country. In their NDC reports, most countries express targets as a
specific percentage reduction in emissions by 2030 (or, in some cases,
2025) compared to their Business As Usual (BAU) emission level at
some base year. Some countries, including China, Chile, Malaysia,
Singapore, and Tunisia specified their targets as a percentage reduction
in carbon intensity instead of a percentage reduction in emissions.
For the countries that did not make a specific emissions reduction
pledge, we assume that their carbon intensity reduction and emission
reduction percentages are the same as those of the most populous country
in that region.

Next, we generate annual emission cap pathways based on the regions'
historical emission levels \citep{world_bank_open_data_2020}, their
emission targets for 2030 (or 2025), and net zero emission target
years. We use five-year emissions data (2014 - 2018) from the World
Bank and the INDC emission targets in 2030 (or 2025) to fit a quadratic
function and use this fitted function to project the emission pathways
for the periods between 2018 and 2030. Emission projections for the
years between 2030 (or 2025) and the net zero emission target year
are obtained by linearly interpolating the emissions. After obtaining
the country-level emission pathways, we aggregate them to find the
regional emission cap pathways.\footnote{We also find that our aggregated regional emission cap pathways are
close to those used in \citet{nordhaus2010economic}.} To assess the implications of varying stringency in the emission
caps, we also create additional emission cap pathways by choosing
alternative net zero scenarios, wherein we assume that all regions
achieve net zero emissions by 2050 (most stringent), 2070, or 2090
(most lax). See Figures \ref{fig:Emission-caps-of-regions} and \ref{fig:Global-emission-caps}
in Appendix \ref{sec:Emission-Cap-Pathways-Appendix} for the regional
emission cap pathways in the baseline cap scenario and the different
emission cap pathways at the global level.

\subsection{Total Factor Productivity and Climate Damage\label{subsec:Total-Factor-Productivity}}

We follow \citet{cai2023climate} to calibrate total factor productivity
(TFP), \textcolor{black}{$A_{i,t}$, and }the climate damage parameters
\textcolor{black}{$\pi_{1,i}$ and $\pi_{2,i}$, based on \citet{burke2018large},
who provide the projected GDP of 165 countries till 2099 assuming
no climate-related impacts, and their GDP assuming the }contemporaneous\textcolor{black}{{}
climate impact of the RCP4.5 temperature $T_{t}^{\mathrm{RCP4.5}}$}\footnote{\textcolor{black}{We choose the RCP4.5 scenario instead of the other
RCP scenarios, because the RCP4.5 scenario is the closest which covers
the range of temperature in our solution.}}\textcolor{black}{{} until 2049, under the SSP2 population pathway.}\footnote{\textcolor{black}{It is nontrivial to use historical data to calibrate
future TFP, particularly when we need to isolate the climate impacts
from the data. For simplicity, we use the projected GDP for the calibration
in this paper. We also use the TFP growth values in RICE to do sensitivity
analysis, and find our results are still robust (see Appendix \ref{subsec:Sensitivity-over-TFPgrowth}).}}\textcolor{black}{{} We aggregate these projections according to our
12 regions and employ the SSP2 population scenario to obtain regional
GDP per capita estimates, $y_{i,t}^{\mathrm{BDD,NoCC}}$ under no
climate impact and $y_{i,t}^{\mathrm{BDD}}$ under climate impact,
which are used to calibrate the regional TFP, $A_{i,t}$, and }the
climate damage parameters, \textcolor{black}{$\pi_{1,i}$ and $\pi_{2,i}$.
For each region $i$, the dynamic path of the TFP is modeled by the
relationship $A_{i,t+1}=A_{i,t}\exp\left(g_{i,t}\right)$, where $g_{i,t}$
is the growth rate of $A_{i,t}$ at time $t$. When $t<80$ (i.e.,
within this century), we assume
\begin{equation}
g_{i,t}=g_{i,0}\exp\left(-d_{i}t\right).\label{eq:TFP_growth_rate}
\end{equation}
For $t\geq80$ (i.e., beyond this century), since the cumulative effect
is huge for a long horizon, it is often inappropriate to simply extrapolate
TFP growth rate using the formula (\ref{eq:TFP_growth_rate}). Therefore,
we follow RICE to generate $g_{t,i}$ for $t\geq80$. See Appendix
\ref{sec:GDP-Growth-Rate} for the details.}

\textcolor{black}{In our structural estimation, we obtain $(g_{i,0},d_{i},\pi_{1,i},\pi_{2,i})$
by solving the following minimization problem:
\begin{equation}
\min_{g_{i,0},d_{i},\pi_{1,i},\pi_{2,i}}\ \sum_{t=0}^{79}\left(\frac{y_{i,t}^{\mathrm{NoCC}}}{y_{i,0}^{\mathrm{NoCC}}}-\frac{y_{i,t}^{\mathrm{BDD,NoCC}}}{y_{i,0}^{\mathrm{BDD,NoCC}}}\right)^{2}+\sum_{t=0}^{29}\left(\frac{y_{i,t}}{y_{i,0}}-\frac{y_{i,t}^{\mathrm{BDD}}}{y_{i,0}^{\mathrm{BDD}}}\right)^{2}.\label{eq:StrEstObj-noDam}
\end{equation}
Here $y_{i,t}^{\mathrm{NoCC}}$ is GDP per capita obtained under no
climate impact by solving the following optimal growth model with
a choice of $(g_{i,0},d_{i},\pi_{1,i},\pi_{2,i})$ and its associated
TFP $A_{it}$:
\begin{eqnarray}
\max_{c_{i,t}} &  & \sum_{t=0}^{\infty}\beta^{t}u(c_{i,t})L_{i,t},\label{eq:ModelForCalibrateTFP}\\
\mathrm{s.t.} &  & K_{i,t+1}=(1-\delta)K_{i,t}+\left(y_{i,t}^{\mathrm{NoCC}}-c_{i,t}\right)L_{i,t},\nonumber 
\end{eqnarray}
where $y_{i,t}^{\mathrm{NoCC}}=A_{i,t}K_{i,t}^{\alpha}L_{i,t}^{-\alpha}$
and $u(c_{i,t})$ is defined as in equation (\ref{eq:preference});
and $y_{i,t}$ is GDP per capita under climate impact by solving the
following optimal growth model:
\begin{eqnarray}
\max_{c_{i,t}} &  & \sum_{t=0}^{\infty}\beta^{t}u(c_{i,t})L_{i,t},\label{eq:ModelForCalibrateDam-1}\\
\mathrm{s.t.} &  & K_{i,t+1}=(1-\delta)K_{i,t}+\left(y_{i,t}-c_{i,t}\right)L_{i,t},\nonumber 
\end{eqnarray}
where 
\[
y_{i,t}=\frac{1}{1+\pi_{1,i}T_{t}^{\mathrm{RCP4.5}}+\pi_{2,i}\left(T_{t}^{\mathrm{RCP4.5}}\right)^{2}}A_{i,t}K_{i,t}^{\alpha}L_{i,t}^{-\alpha}.
\]
The initial TFP $A_{i,0}$ is chosen such that $A_{i,0}K_{i,0}^{\alpha}L_{i,0}^{-\alpha}/\left(1+\pi_{1,i}T_{0}+\pi_{2,i}T_{0}^{2}\right)$
is equal to the observed GDP per capita in 2020.} Figures \ref{fig:GDP_fitted_noDam}
and \textcolor{black}{\ref{fig:GDPdam_fitted_RCP45}} in Appendix
\ref{sec:Calibration-of-TFP-appendix} shows that with our calibrated
\textcolor{black}{$(g_{i,0},d_{i},\pi_{1,i},\pi_{2,i})$,} the GDP
per capita \textcolor{black}{$y_{i,t}^{\mathrm{NoCC}}$ or $y_{i,t}$
matches well with the projected data }$y_{i,t}^{\mathrm{BDD,NoCC}}$\textcolor{black}{{}
or }$y_{i,t}^{\mathrm{BDD}}$\textcolor{black}{{} from \citet{burke2018large},
respectively, for all regions.}

\subsection{Carbon Intensity{\normalsize{} \label{subsec:Carbon-Intensity}}}

To obtain the time-varying and region-specific carbon intensities
$\sigma_{i,t}$, we use the projections of GDP and emissions in \textcolor{black}{\citet{ueckerdt2019economically},
who report simulation results of future emissions and GDP under different
scenarios based on climate policy regimes, technology portfolios,
and }carbon\textcolor{black}{{} tax implementation. As the carbon intensity
in our model reflects the zero-carbon tax regime, we employ results
from }the scenario defined as `FFrun111' in \citet{ueckerdt2019economically}.
Specifically, this FFrun111 scenario corresponds to climate action
from 2010 with full technology portfolio and no carbon tax. . Based
on the equation (\ref{eq: emissions}), we calculate the carbon intensities
as $\sigma_{i,t}=E_{i,t,\mathrm{FFrun111}}^{U}/Q_{i,t,\mathrm{FFrun111}}$,
where $E_{i,t,\mathrm{FFrun111}}^{U}$ and $Q_{i,t,\mathrm{FFrun111}}$
are the projected regional emissions and GDP under the FFrun111 scenario
for region $i$ at time $t$. \footnote{\citet{ueckerdt2019economically} provide data for 11 regions. Upon
comparison, we find that the countries constituting the \textquoteleft Rest
of the World (ROW)\textquoteright{} region are the ones that are in
the \textquoteleft Other High Income (OHI)\textquoteright{} and \textquoteleft Eurasia\textquoteright{}
region in our study. Therefore, the carbon intensity obtained from
the ROW in \citet{ueckerdt2019economically} corresponds to that of
OHI and Eurasia regions in our work.}

\subsection{Abatement Cost}

\textcolor{black}{Our estimation of }the abatement cost parameters
\textcolor{black}{$b_{1,i}$, $b_{2,i}$, $b_{3,i}$, and $b_{4,i}$
relies on} \textcolor{black}{the simulation results under ten different
levels of carbon taxes in \citet{ueckerdt2019economically}. }For
each scenario $j$ with associated carbon taxes $\tau_{t,j}^{U}$
for every region $i$, \citet{ueckerdt2019economically} report the
projected regional emissions net of abatement ($E_{i,t,j}^{U}$),
for region $i$ at time $t$. Since\textcolor{black}{{} \citet{ueckerdt2019economically}}
do not consider an ETS, according to the discussion in Section \ref{subsec:StaticModel},
\textcolor{black}{we can assume that their carbon taxes} $\tau_{t,j}^{U}$
are equal to the marginal abatement costs when emissions are strictly
positive. That is,
\begin{equation}
\tau_{t,j}^{U}=\frac{1,000b_{2,i}\mu_{i,t,j}^{b_{2,i}-1}b_{1,i,t}}{\sigma_{i,t}}=1,000b_{2,i}\mu_{i,t,j}^{b_{2,i}-1}(b_{1,i}+b_{3,i}\exp(-b_{4,i}t)),\label{eq:tax_abatement}
\end{equation}
for $\mu_{i,t,j}\in(0,1)$. Therefore, we can use their \textcolor{black}{simulation
results }under different carbon tax levels to estimate \textcolor{black}{$b_{1,i}$,
$b_{2,i}$, $b_{3,i}$}, and $b_{4,i}$. At first, we estimate the
associated emission control rate $\mu_{i,t,j}^{U}$ using the equations
(\ref{eq: emissions})-(\ref{eq:abatement of emissions}) as follows:
\begin{equation}
\mu_{i,t,j}^{U}=1-\frac{E_{i,t,j}^{U}}{E_{i,t,\mathrm{FFrun111}}^{U}},\label{eq:mu}
\end{equation}
where $E_{i,t,\mathrm{FFrun111}}^{U}$ is the regional emission from
the FFrun111 scenario with zero carbon tax (see\textcolor{black}{{}
\citet{ueckerdt2019economically}} for details). Then we use the computed
$\mu_{i,t,j}^{U}$ and the associated carbon tax $\tau_{t,j}^{U}$
to find the abatement cost coefficients---\textcolor{black}{$b_{1,i}$,
$b_{2,i}$, $b_{3,i}$, and $b_{4,i}$---}such that the equality
(\ref{eq:tax_abatement}) can hold in an approximate manner for every
scenario $j$ and time $t$.

\subsection{Climate System: Transient Climate Response to Emissions}

In the TCRE climate system, $\mathcal{E}_{0}$, is chosen such that
the initial global mean temperature $T_{0}$ in year 2020 is 1.2 degrees
Celsius above the pre-industrial level. \textcolor{black}{The c}ontribution
rate of cumulative global emissions to temperature\textcolor{black}{{}
}is calibrated at\textcolor{black}{{} $\zeta=0.0021$ by using the projections}
of emissions and temperatures across the four Representative Carbon
Concentration Pathways (RCPs) scenarios: RCP 2.6, RCP 4.5, RCP 6.0,
and RCP 8.5 \citep{meinshausen2011rcp}. \textcolor{black}{That is,
we solve the following minimization problem: 
\begin{eqnarray*}
\min_{\zeta} &  & \sum_{t}\sum_{j}\left\Vert T_{t}^{j}-\zeta\mathcal{E}_{t}^{j}\right\Vert \\
s.t. &  & \mathcal{E}_{t+1}^{j}=\mathcal{E}_{t}^{j}+E_{t}^{j},\ \forall t,j,
\end{eqnarray*}
where $j$ represents one of four RCP scenarios, $E_{t}^{j}$ and
$T_{t}^{j}$ are the exogenous projections of emissions and temperatures,
respectively, at time $t$ for RCP scenario $j$. Figure \ref{fig:TCRE}
in Appendix \ref{sec: Atmospheric Temperature Anomaly } demonstrates
that }the calibrated TCRE climate system matches well all four RCP
temperature pathways using their associated RCP emission pathways.

\section{Baseline Model Results}

This section presents the numerical simulation results under the assumption
that all regions participate in the global ETS. The baseline emission
cap scenario is defined in accordance with the NDCs in the near term
and the net-zero targets in the mid- to long term. We report the results
through the end of the century (2020--2100), focusing on both economic
and environmental outcomes across key variables. Additionally, we
provide an in-depth analysis of the numerical simulation results to
evaluate the effects of implementing a global cap-and-trade system.

\subsection{{\Large Emissions, Temperature, and Permit Prices \label{subsec:Noncooperative-Model-baseline}}}

\begin{figure}[h]
\centering{}\includegraphics[width=1\textwidth]{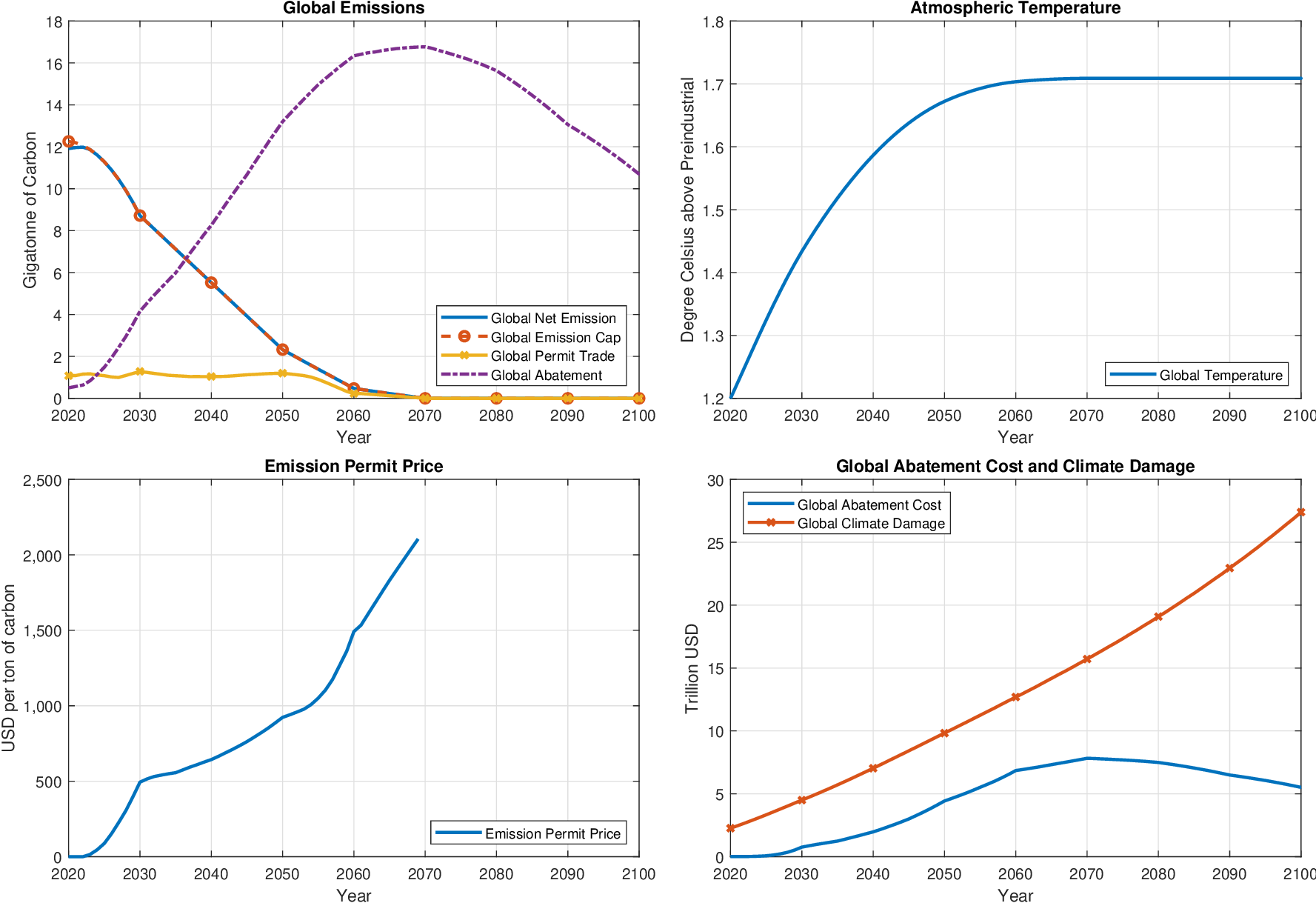}\caption{\label{fig:Non_Cooperation_Key_Figures} Simulation results at the
global scale under the baseline emission cap scenario.}
\end{figure}

Figure \ref{fig:Non_Cooperation_Key_Figures} displays key simulation
results at the global level, from the noncooperative model with the
ETS under the baseline emission cap scenario. Until 2100, we can delineate
three periods according to the global emission trading patterns. Prior
to 2024, global emissions do not reach the global emission caps and
therefore the permit price remains zero, implying an excess supply
of emission permits in the initial years (Figure \ref{fig:Non_Cooperation_Key_Figures}
top-left and bottom-left), as the emission caps in the initial years
are large while they become much smaller over time (Figure \ref{fig:Emission-caps-of-regions}).
This result is not surprising, considering that an oversupply of emission
permits has been observed in the EU ETS \citep{fuss2018}, resulting
in zero or very low permit prices. During this period, the volume
of global emission abatement and its associated abatement cost remain
at low levels (Figure \ref{fig:Non_Cooperation_Key_Figures} top-left
and bottom-right). This result implies that the global emission cap
should be set at a level such that there is no over-supply of emission
permits, so that permit prices are strictly positive under noncooperation
and the ETS.

Global emissions are constrained by the global emission caps starting
from 2024. To comply with the monotonically decreasing emission caps,
the regions undertake additional abatement efforts and/or purchase
emission permits. This results in a substantial increase in the volume
of emission abatement, along with the abatement cost and permit price.
The decrease in global emissions is mainly achieved by the concomitant
increase in global abatement, which peaks by around 2070 (Figure \ref{fig:Non_Cooperation_Key_Figures}
top-left). The global abatement cost also increases steeply to reach
its maximum value of \$7.82 trillion by 2070, and then decreases as
the world reaches net zero emissions by 2070. The emission permit
trade also gradually decreases to zero by 2070 (Figure \ref{fig:Non_Cooperation_Key_Figures}
top-left). As the emission cap becomes tighter over the years, the
permit price increases to \$923 per ton of carbon in 2050 and \$2,105
in 2069 (Figure \ref{fig:Non_Cooperation_Key_Figures} bottom-left).
The kinks in the permit price path in 2030 result from a kink in emission
cap pathways (see Figure \ref{fig:Global-emission-caps}), while those
in 2050 and 2060 are a result of some regions achieving net zero emissions,
as shown in Figure \ref{fig:Emission-caps-of-regions}. Essentially,
binding emission caps lead to a rise in the overall emission abatement,
while the steep increase in the permit price limits the trading of
emission permits.

Post-2070 is the period when global emissions are at net zero. For
this period, emission permits are no longer traded and all emissions
are abated in each region.\footnote{In the bottom-left panel of Figure \ref{fig:Non_Cooperation_Key_Figures},
we plot the emission permit price only when the traded volume is positive.} By the end of this century, the atmospheric temperature is projected
to reach 1.7 degrees Celsius above the pre-industrial level, which
is driven by net positive emissions before 2070 (Figure \ref{fig:Non_Cooperation_Key_Figures}
top-right), restricted by the emission caps.\footnote{If there is no emission cap, then the temperature anomaly will be
much higher. This can be seen in a later discussion with alternative
emission caps.} Lastly, the global climate damage rises almost linearly over the
entire period (Figure \ref{fig:Non_Cooperation_Key_Figures} bottom-right).
Overall, our noncooperative model simulation predicts that the global
emission caps set by the Paris Agreement and Glasgow Pact are not
restrictive enough to achieve the global target of limiting the temperature
rise to 1.5 degrees Celsius.

\begin{figure}[h]
\begin{centering}
\includegraphics[width=1\textwidth]{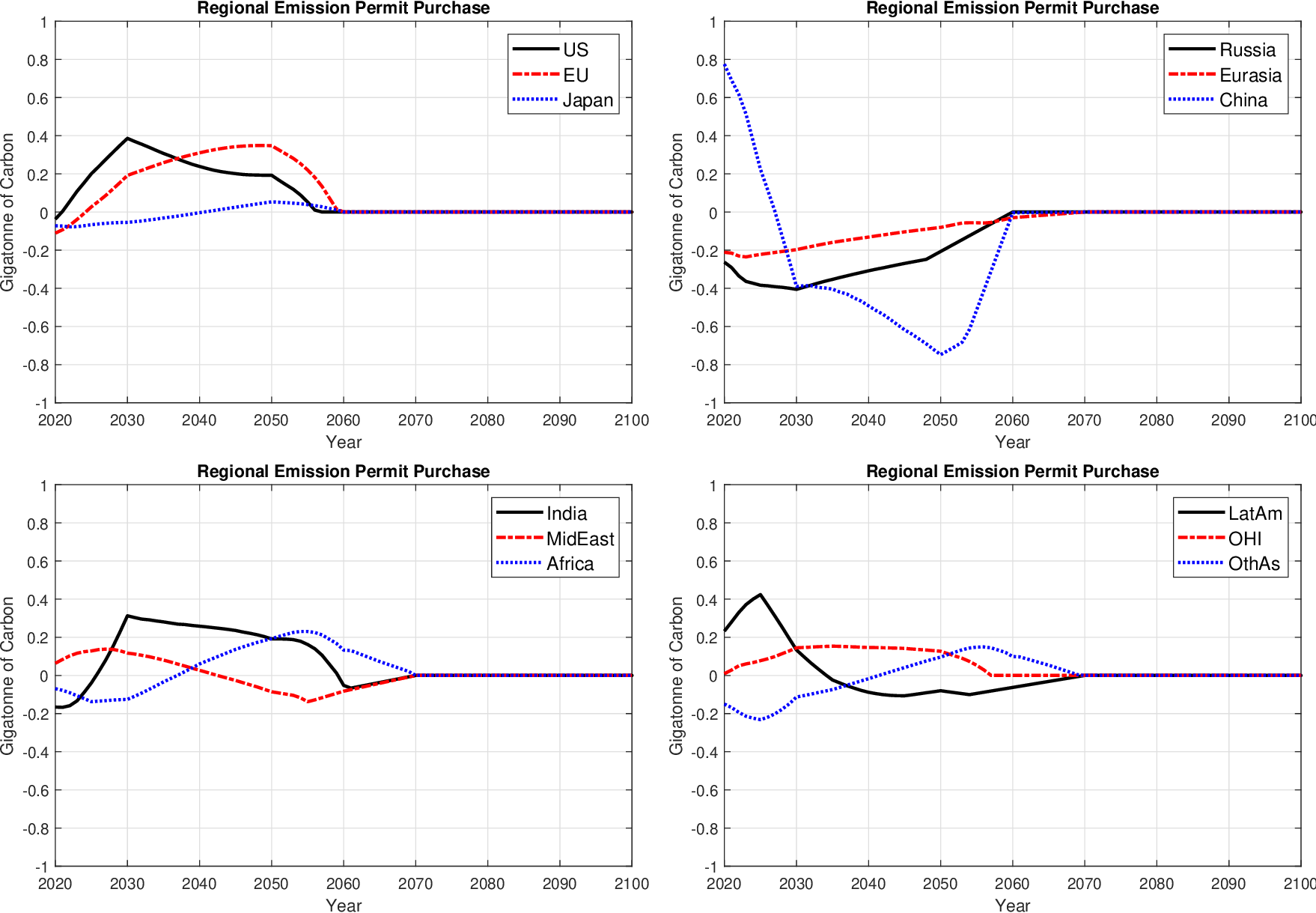}
\par\end{centering}
\centering{}\caption{\label{fig:Non_Cooperation_Emission_Purchases} Simulation results
of regional emission permit purchase under the baseline emission cap
scenario.}
\end{figure}

It is instructive to examine the trading patterns of different regions
in the emission permit market over the years to identify the permit
buyers and sellers. Figure \ref{fig:Non_Cooperation_Emission_Purchases}
displays the volumes of traded emission permits for each region over
time, with a positive value denoting permit purchase and a negative
value indicating permit sales. Before 2024, although there is an excess
permit supply at the global level, the emission cap constraint is
effective for some regions, such as China and Latin America, as these
regions become permit buyers at this early stage. After 2024, when
the global emission cap constraint becomes binding, the group of permit
buyers consists of the US, EU, India, and OHI regions; the group of
permit sellers consists of Russia, China, and Eurasia, with China
being the largest permit provider after 2032. Japan is expected to
be involved in a relatively small volume of permit trading, and the
other regions (Africa, MidEast, Latin America, and OthAs) change their
status of permit suppliers to buyers or vice versa over time.

\subsection{Regional MAC and SCC}

\begin{figure}[h]
\begin{centering}
\includegraphics[width=1\textwidth]{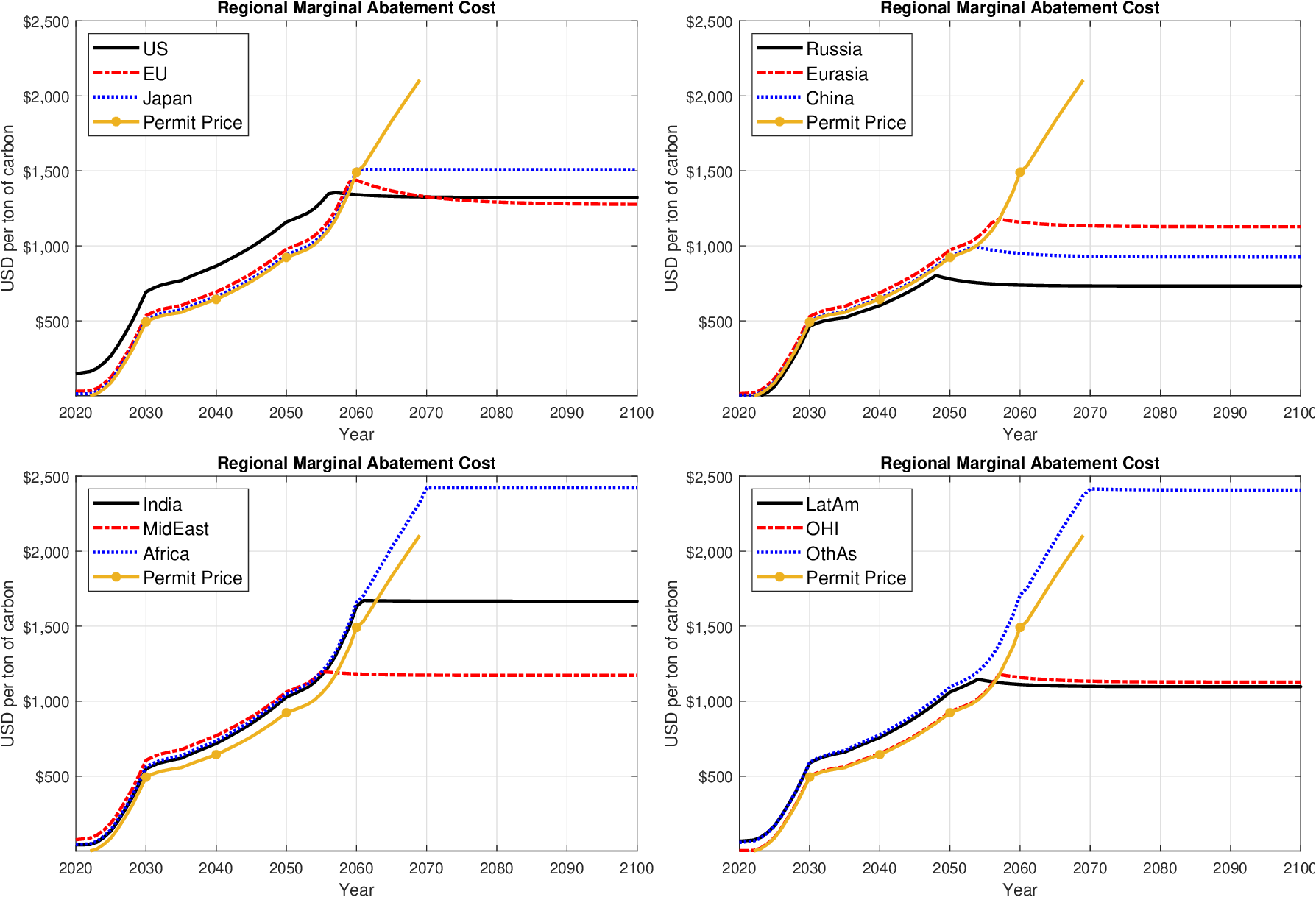}
\par\end{centering}
\centering{}\caption{\label{fig:MAC-regional}Simulation results of regional MAC under
the baseline emission cap scenario.}
\end{figure}

Based on the simulation results under the baseline emission cap scenario,
this section examines the relationship between the regional MAC, SCC,
and the market equilibrium emissions permit price under the global
ETS regime. The MAC captures the additional cost incurred due to an
increase in emission abatement. From our model equation (\ref{eq:abatement of emissions}),
the total abatement cost is $\Phi_{i,t}=b_{1,i,t}\mu_{i,t}^{b_{2,i}}Q_{i,t},$
in trillions of USD. Thus, the MAC, in 2020 USD per ton of carbon,
is obtained as follows:
\begin{equation}
\mathrm{MAC}{}_{i,t}=1,000\left(\frac{\partial\Phi_{i,t}}{\partial E_{i,t}^{A}}\right)=1,000\left(\frac{b_{1,i,t}b_{2,i}\mu_{i,t}^{b_{2,i}-1}}{\sigma_{i,t}}\right).\label{eq:MAC}
\end{equation}
Note that, with $b_{1,i,t}>0$ and $b_{2,i}>2,$ MAC is strictly increasing
on \textcolor{black}{emission control rate $\mu_{i,t}\in[0,1].$}\footnote{\textcolor{black}{See Table \ref{tab:Abatement-cost-parameters_Ueckerdt}
}in Appendix\textcolor{black}{{} \ref{sec:List-of-Parameters} for the
list of our calibrated abatement cost parameters.}}\textcolor{black}{{} }Figure \ref{fig:MAC-regional} displays the regional
MAC along with the market equilibrium price of emission permits. The
simulation result shows that, for most regions, the MAC increases
rapidly and remains strictly greater than the permit price until the
emissions hit zero by 2050s or 2060s. \footnote{For regional optimal emissions, see Figure \ref{fig:Net_emission}
in Appendix \ref{subsec:Regional-Net-Emissions}.} Russia is an exception, where the MAC falls below the permit price
starting in 2021, which we will elaborate shortly in the next paragraph.
The MAC of each region gradually decreases below the permit price
after its net zero emission level is achieved (i.e., $E_{i,t}^{*}=0$).
Russia is the first region to achieve net zero emissions in 2048 (as
shown in Figure \ref{fig:Net_emission} in Appendix \ref{subsec:Regional-Net-Emissions}).
However, Russia continues to sell permits afterwards, as shown in
Figure \ref{fig:Non_Cooperation_Emission_Purchases}, because its
net zero emission target year is 2060, under the baseline emission
cap scenario. After 2048, Russia's MAC begins to decline slightly
and stabilizes over time, as its emission control rate reaches its
upper limit.\footnote{When emission control rates are one, the MACs are $1,000b_{2,i}(b_{1,i}+b_{3,i}\exp(-b_{4,i}t))$,
and they are nearly constant when $t$ is large.} Similar trends are observed in other regions as they attain net zero
emissions in the 2050s (China, Latin America, MidEast, the US, the
OHI, Eurasia) and the 2060s (the EU, Japan and India). Africa and
OthAs are the last group of regions to achieve net zero emissions
(before trading of permits) by 2070, after which emission permits
are no longer traded, and the MACs of all regions decline slowly over
time. Note that if a region's after-permit-trade zero emission cap
constraint is binding (i.e., the inequality (\ref{eq:emissions cap constraint})
is binding with $\overline{E}_{i,t}=0$), then even when its regional
MAC is smaller than the permit price, the region will not be able
to sell emission permits, otherwise it will violate the constraint.

The SCC, a central concept in the climate change literature, is widely
used to quantify the present value of climate damages induced by an
additional unit of carbon emissions. While the SCC is often calculated
in a global social planner's problem (e.g., the DICE model), we consider
the SCC for each region. Similar to \citet{van2016non} and \citet{cai2023climate},
we define the noncooperative SCC of a region as the marginal rate
of substitution between global emissions and regional capital as follows:

\begin{equation}
\mathrm{SCC}_{i,t}=\frac{-1,000(\partial V_{i,t}/\partial\mathcal{E}_{t})}{\partial V_{i,t}/\partial K_{i,t}},\label{eq:SCC_OLNE}
\end{equation}
where $V_{i,t}$ is the value function of the noncooperative model
at time $t$ for region $i$, depending on the state variables $\{K_{i,t},\mathcal{E}_{t}:\ i\in\mathcal{I}\}$;
that is,
\[
V_{i,t}(K_{1,t},...,K_{12,t},\mathcal{E}_{t})=\max_{c_{i,s},E_{i,s}^{P},\mu_{i,s}}\sum_{s=t}^{\infty}\beta^{s-t}u(c_{i,s})L_{i,s},
\]
for each region $i$ under the open loop Nash equilibrium.\footnote{To compute the regional SCC in the noncooperative model, it is equivalent
to replace the numerator in equation (\ref{eq:SCC_OLNE}) with the
shadow price of the transition equation of cumulative global emissions
(\ref{eq:cumEmission}) at time $t$, and replace the denominator
with the shadow price of the regional capital transition equation
(\ref{eq: capital transition}) for each region.} Since our cumulative global emissions are measured in gigatonnes
of carbon (GtC) and capital is measured in trillions of USD, our SCC
is measured in monetary unit of 2020 USD per ton of carbon. \footnote{The choice of discount rates has been a critical factor contributing
to the gap in social cost of carbon estimates across different studies
\citep{guo2006,weitzman2013}. Our concept of regional SCC differs
from the definition in \citet{nordhaus2017revisiting}, where the
regional SCC is obtained by assuming that it is a fraction of the
global SCC, with the share determined by the discounted value of regional
output using an exogenous, constant discount rate. We verified numerically
that our regional noncooperative SCC is equal to the present value
of future climate damages in the region resulting from an additional
unit of global emissions released in the current period, with our
social discount rates $r_{i,t+1}$ defined endogenously with the following
formula: 
\[
r_{i,t+1}=\frac{u'(c_{i,t})}{\beta u'(c_{i,t+1})}-1,
\]
where $c_{i,t}$ are the optimal per-capita consumptions. }

\begin{figure}[h]
\begin{centering}
\includegraphics[width=1\textwidth]{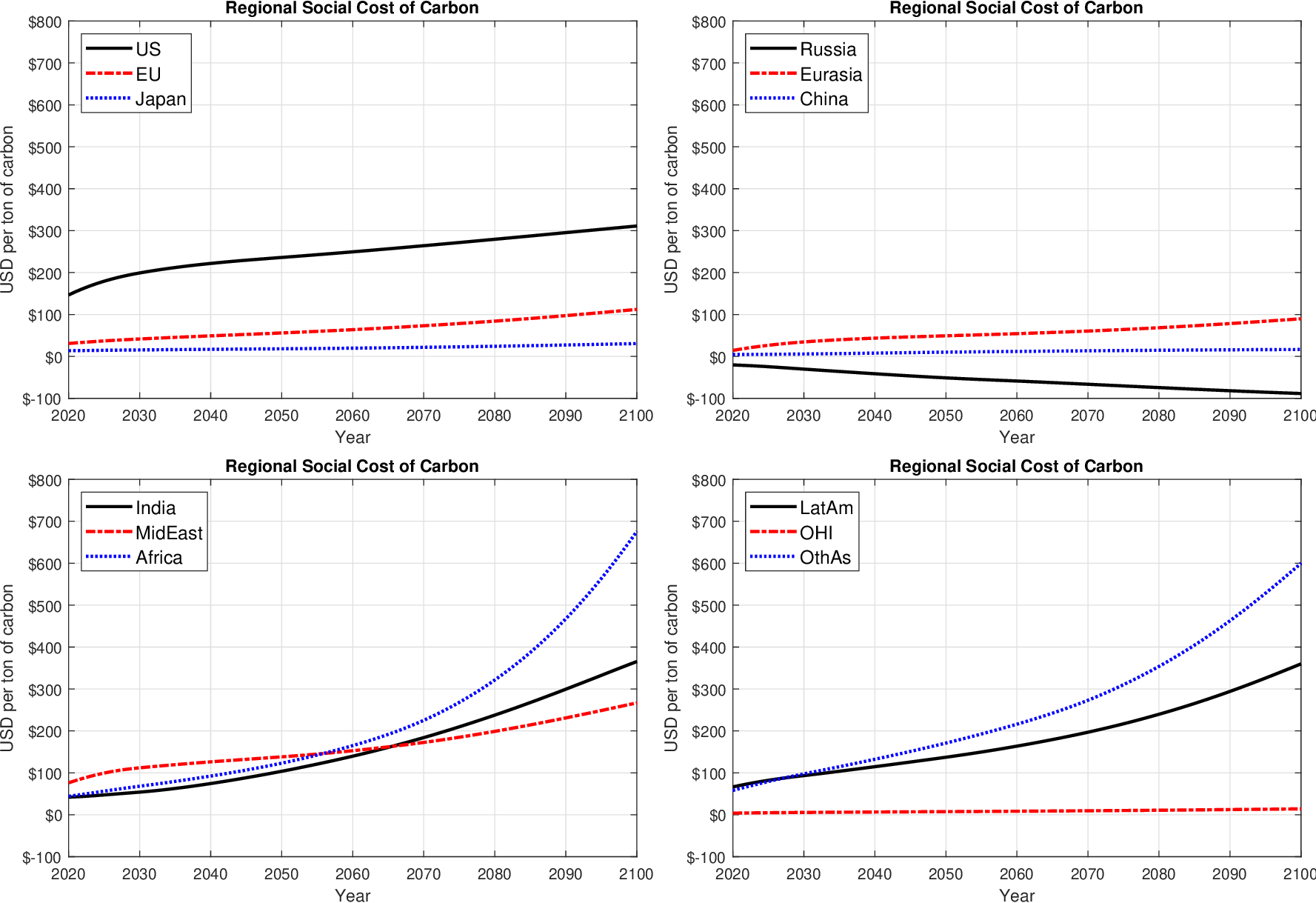}
\par\end{centering}
\centering{}\caption{\label{fig:Regional_SCC} Simulation results of regional SCC under
the baseline emission cap scenario.}
\end{figure}

The simulation results confirm the relationship between the regional
SCC, MAC, and permit prices in a multi-region economy with a global
ETS, as explored in Section \ref{subsec:StaticModel}. Figure \ref{fig:Regional_SCC}
demonstrates that, when a region's net emissions are strictly positive,
the regional SCC is exactly equal to the gap between the regional
MAC and the permit price shown in Figure \ref{fig:MAC-regional}.\footnote{When regional net emissions are at zero, the regional SCC can be larger
than the gap, $\left(\mathrm{MAC}_{i}(E_{i})-m^{*}\right)$, which
could be negative, as shown in Figure \ref{fig:Regional_SCC}.} For example, in 2050, the MAC for the US is \$1,159 per ton of carbon,
the permit price is \$923, and their difference is exactly equal to
the regional SCC of the US, \$236. Russia is an exceptional case with
a negative SCC in our simulation, suggesting that global warming creates
benefits rather than causing climate damages in Russia, a result heavily
influenced by the climate damage parameters calibrated using projections
from \citet{burke2018large}. Nevertheless, the relationship between
the SCC, MAC, and permit price remains intact; for example, in 2040,
the SCC for Russia is $-\$41$, which corresponds to the difference
between the MAC (\$602) and the permit price (\$643). This explains
why the MAC falls below the permit price starting in 2021, despite
Russia having nonzero emissions until 2050. Among all the regions,
the US has the highest SCC in the near term, at around \$199 per ton
of carbon in 2030. However, the SCCs of Africa and non-OECD Asia are
expected to experience substantial increases, reaching \$675 for Africa
and \$599 for non-OECD Asia by the end of the century. Russia, with
the lowest SCC, experiences a steady negative SCC, reaching (-\$88)
per ton of carbon by the end of the century, indicating that it benefits
from global warming throughout the entire period. Our results show
considerable heterogeneity in the SCC across regions.

\subsection{Effects of ETS Implementation}

\begin{figure}[h]
\begin{centering}
\includegraphics[width=1\textwidth]{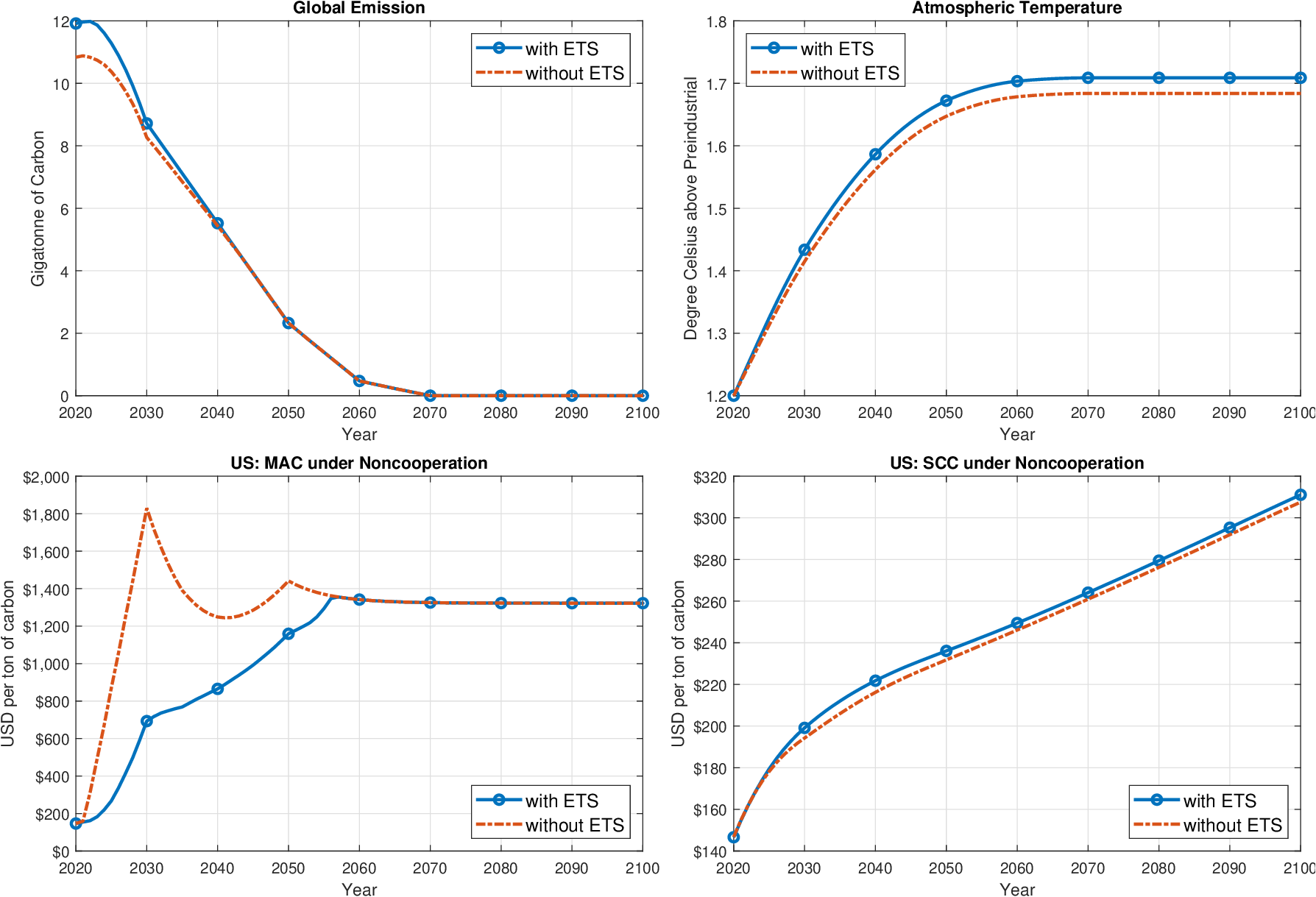}
\par\end{centering}
\centering{}\caption{\label{fig:compare_ETS} Comparative analysis: effects of the ETS
implementation.}
\end{figure}

Next, we examine the economic and climate implications of the ETS
implementation by comparing the global economy with and without the
ETS regime, while maintaining the baseline emission caps in both economies.
The top two panels of Figure \ref{fig:compare_ETS} show that the
ETS implementation results in slightly higher emissions and temperature
increases compared to the case without the ETS. This pattern persists
until 2043. This occurs because, under the ETS, regions with binding
regional emission cap constraints now have the option to purchase
permits from regions with less restrictive emission caps, fully exploiting
the total amount of permits allowed at the global level. In contrast,
without the ETS, regions with binding regional emission caps cannot
utilize surplus emission permits from other regions, resulting in
global net emissions that are lower than or equal to those in the
ETS scenario over time. In the bottom two panels of Figure \ref{fig:compare_ETS},
we compare the MAC and the SCC using the US as an illustrative example.
The MAC of the US economy without the ETS is higher until 2056, reflecting
that the US becomes a permit buyer under the ETS regime as MAC increases
with increasing emission abatement.\footnote{Before 2030, the emission control rate grows fast so the MAC of the
US increases rapidly along time, but after 2030, the emission control
rate grows slowly such that the improvement of emission abatement
technology makes the MAC decline along time until 2042.} The SCC comparison demonstrates that the ETS has little impact on
the regional SCC. This pattern holds for the other regions as well,
as shown in Figure \ref{fig:SCC_Noncoop}. Since the regional noncooperative
SCC is the present value of future climate damages in the region resulting
from an additional unit of global emissions released in the current
period, these results indicate that the ETS has little impact on either
the marginal climate damages or our endogenous social discount rates.

What are the welfare implications of implementing the global ETS?
To quantify these effects, we compute a compensating variation (CV)
per capita associated with the ETS implementation. Specifically, the
CV per capita for region $i$ is computed by numerically solving the
following equation:

\textcolor{black}{
\begin{equation}
W_{i,0}(\mathbf{c}_{i}^{1}-CV)=W_{i,0}(\mathbf{c}_{i}^{0}),\label{eq:CV}
\end{equation}
}where $\mathbf{c}_{i}^{1}=(c_{i,0}^{1},\cdots,c_{i,t}^{1},\cdots,c_{i,T}^{1})$
represents the vector of optimal consumption per capita under the
ETS implementation, $\mathbf{c}_{i}^{0}=(c_{i,0}^{0},\cdots,c_{i,t}^{0},\cdots,c_{i,T}^{0})$
is the vector of optimal consumption per capita without the ETS implementation,
and 
\[
W_{i,0}(\mathbf{c}_{i})=\sum_{t=0}^{T}\beta^{t}u(c_{i,t})L_{i,t}
\]
is the social welfare associated with the vector of consumption per
capita $\mathbf{c}_{i}=(c_{i,0},\cdots,c_{i,t},\cdots,c_{i,T})$ with
the terminal time $T=299$.

Table \ref{tab:welfare_ETS-pre} shows welfare effects of the global
ETS, measured by the CV per capita in 2020 USD, and its share (\%)
out of per capita consumption in each region at $t=0$. The results
reveal significant heterogeneity in welfare effects across regions:
some regions experience welfare gains, with Russia benefiting the
most, showing a CV per capita of \$552.75, equivalent to 6.699\% of
its per capita consumption. In contrast, other regions, including
Africa, the Middle East, and non-OECD Asia, experience negative welfare
effects. This outcome may seem counterintuitive, as one might expect
an additional market mechanism (ETS) to enhance welfare by providing
greater flexibility in managing emissions. It is important to note
that a direct comparison of the noncooperative model with and without
the ETS does not provide a clear picture of the welfare effects of
the ETS under the current baseline emission cap scenario, as the global
net emissions and resulting climate damages differ across the two
economies. Recall that, global average temperature is higher under
ETS implementation in the baseline emission cap scenario, as shown
in Figure \ref{fig:compare_ETS}. Therefore, the differences between
$\mathbf{c}_{i}^{1}$ and $\mathbf{c}_{i}^{0}$ depend on both the
global net emissions (and resulting climate damages) and the ETS implementation.
As a result, some regions may experience additional climate damages
due to higher temperatures, outweighing the benefits of lower abatement
costs from permit purchases or additional profits from permit sales.
This explains why some regions experience negative welfare effects
from ETS implementation.

\begin{table}[H]
\caption{\textcolor{black}{\label{tab:welfare_ETS-pre}}Welfare effects of
the ETS implementation: pre-emission cap adjustment.}

\begin{centering}
\begin{tabular}{ccccccc}
\toprule 
 & {\small US} & {\small EU} & {\small Japan} & {\small Russia} & {\small Eurasia} & {\small China}\tabularnewline
\midrule 
CV per capita (\$) & 38.94 & 83.04 & -0.03 & 552.75 & 53.18 & 42.90\tabularnewline
\midrule 
CV per capita (\%) & 0.089 & 0.330 & 0.000 & 6.699 & 1.062 & 0.429\tabularnewline
\end{tabular}
\par\end{centering}
\centering{}%
\begin{tabular}{ccccccc}
\toprule 
 & {\small India} & {\small MidEast} & {\small Africa} & {\small LatAm} & {\small OHI} & {\small OthAs}\tabularnewline
\midrule 
CV per capita (\$) & 22.94 & -36.03 & -5.40 & 66.68 & 109.69 & -18.93\tabularnewline
\midrule 
CV per capita (\%) & 1.469 & -0.387 & -0.354 & 1.001 & 0.317 & -0.806\tabularnewline
\bottomrule
\end{tabular}
\end{table}

To isolate the impact of the ETS, we adjust the emission cap $\overline{E}_{i,t}$
in the noncooperative model with the ETS to $\overline{E}'_{i,t}$,
which is the optimal level of net emissions obtained from the noncooperative
model with the emission caps $\overline{E}_{i,t}$ but without the
ETS. We then compare economies with and without the ETS under the
emission cap $\overline{E}'_{i,t},$ ensuring two economies have the
same pathways of global net emissions and temperature. Table \ref{tab:welfare_ETS-post}
shows that the welfare impacts of the global ETS are strictly positive
for all regions, irrespective of whether they are permit sellers or
buyers, with considerable heterogeneity in the magnitude of these
effects across the regions. For example, in the United States, the
CV per capita is \$116.18, which is a relatively small fraction of
per capita consumption (0.267\%). Among all regions, Russia still
experiences the largest welfare improvement, with a CV per capita
of \$235.17, equivalent to 2.850\% of its per capita consumption,
though its welfare gains are smaller than in the pre-adjustment analysis.
This is because Russia, a country with a negative SCC, benefits from
higher temperatures, meaning that its welfare gains in the pre-adjustment
analysis reflect both climate-induced benefits and ETS implementation.
Conversely, Africa and non-OECD Asia experience the smallest welfare
gains from the global ETS, indicating that additional climate damages
from higher temperatures contributed to the negative welfare effects
observed in the pre-adjustment analysis. Overall, the results in the
post-adjustment analysis highlight that the principle of gains from
trade applies to emissions trading as well, driven by efficiency gains
achieved through the reallocation of emission abatement efforts across
regions.

\begin{table}[H]
\caption{\textcolor{black}{\label{tab:welfare_ETS-post}}Welfare effects of
the ETS implementation: post-emission cap adjustment.}

\begin{centering}
\begin{tabular}{ccccccc}
\toprule 
 & {\small US} & {\small EU} & {\small Japan} & {\small Russia} & {\small Eurasia} & {\small China}\tabularnewline
\midrule 
CV per capita (\$) & 116.18 & 100.26 & 38.60 & 235.17 & 57.49 & 34.62\tabularnewline
\midrule 
CV per capita (\%) & 0.267 & 0.399 & 0.136 & 2.850 & 1.148 & 0.346\tabularnewline
\end{tabular}
\par\end{centering}
\centering{}%
\begin{tabular}{ccccccc}
\toprule 
 & {\small India} & {\small MidEast} & {\small Africa} & {\small LatAm} & {\small OHI} & {\small OthAs}\tabularnewline
\midrule 
CV per capita (\$) & 33.34 & 42.73 & 16.82 & 75.30 & 100.63 & 12.40\tabularnewline
\midrule 
CV per capita (\%) & 2.135 & 0.459 & 1.103 & 1.130 & 0.291 & 0.529\tabularnewline
\bottomrule
\end{tabular}
\end{table}

\section{Alternative Policy Simulations}

Our model framework allows flexibility to explore alternative policy
simulations. Recognizing that the actual implementation of a global
ETS requires substantial international commitment, we consider two
alternative policy scenarios to gain additional insights. First, we
examine a partial ETS in which a major player in the world economy---the
United States---does not participate, and the ETS operates among
the remaining regions. Second, we analyze alternative net-zero pathways
that vary in the stringency of global emission reduction commitments.

\subsection{Partial ETS: US Non-participation}

Recent political developments under the Trump administration have
indicated that the United States is unlikely to participate in international
cooperation aimed at mitigating global emissions in the near term.
In light of this potential fragmentation of global climate cooperation,
it is useful to numerically examine the economic and environmental
consequences of a partial ETS. To this end, we simulate a scenario
in which the United States refrains from any form of climate action---neither
enforcing its emission cap nor participating in the ETS---while the
remaining 11 regions continue to operate a cap-and-trade system.

\begin{figure}[h]
\begin{centering}
\includegraphics[width=1\textwidth]{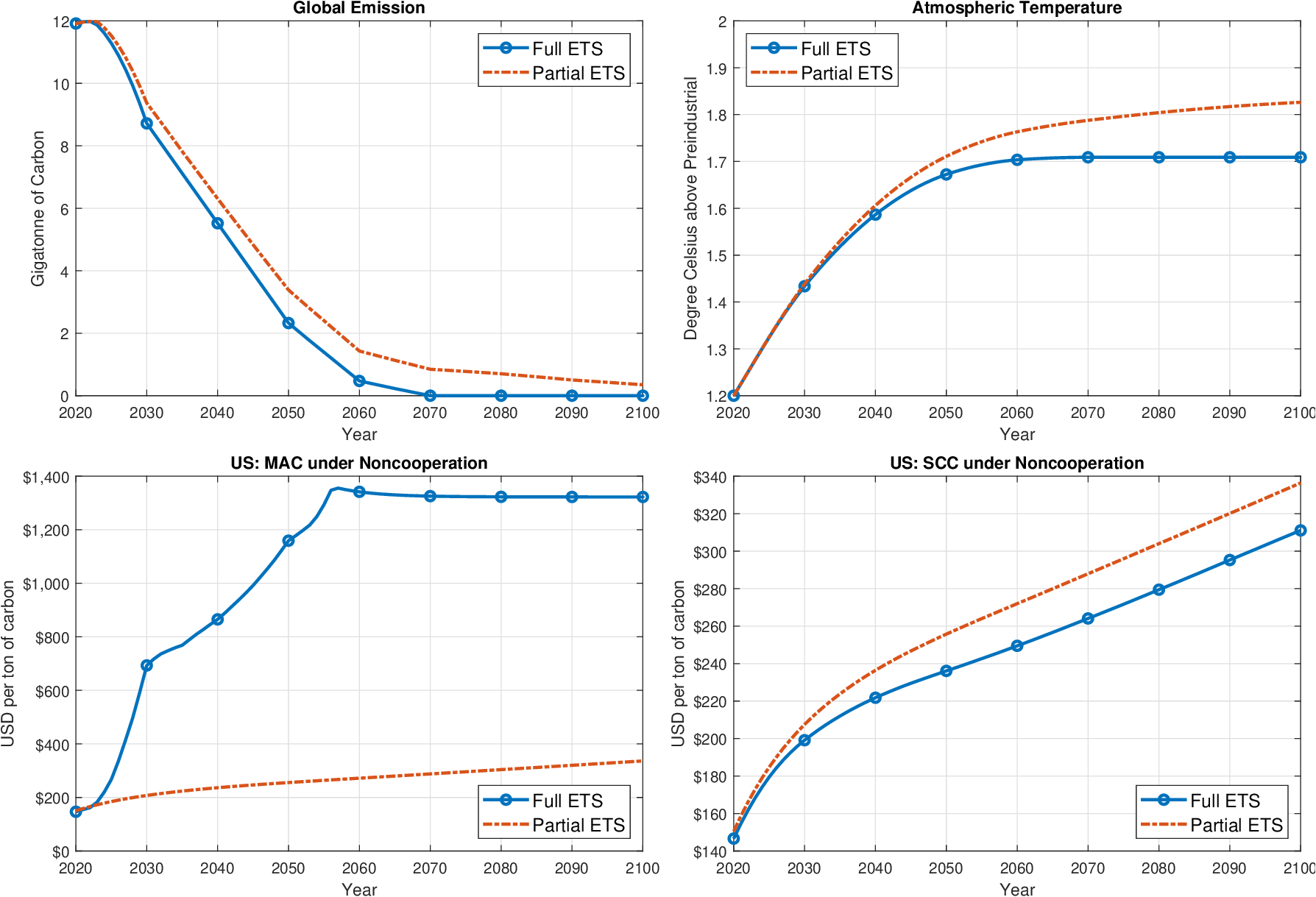}
\par\end{centering}
\centering{}\caption{\label{fig:PartialETS} Comparative analysis: effects of US non-participation
in the global ETS}
\end{figure}

Figure \ref{fig:PartialETS} compares the key outcomes under the baseline
(\textquotedblleft full ETS\textquotedblright ) and the partial ETS
scenarios. As intuitively expected, US non-participation leads to
a substantial increase in global emissions and a corresponding rise
in global temperature, reaching 1.83 degrees Celsius---about 0.12
degrees Celsius higher than in the baseline. Under the partial ETS
scenario, the US MAC is equal to its SCC which rises to \$336 by 2100
due to higher global average temperatures.

Table \ref{tab:welfare_PartialETS} shows welfare effects of the US
non-participation by comparing the partial ETS to the full ETS, further
illustrating the distributional effects of fragmented climate cooperation.
The United States experiences a significant welfare gain, with CV
per capita of \$565---approximately 1.3\% of per capita consumption---reflecting
its reduced abatement burden and the low level of MACs as shown in
the figure above. On the other hand, the other regions (except for
Russia) all experience significant welfare losses, ranging from -3.8\%
(MidEast) to -0.01\% (OHI) of per capita consumption, as they bear
the costs of both higher climate damages and continued emission constraints
under the ETS. Russia is the only exception among the participating
regions, experiencing welfare gains because higher global average
temperatures are expected to benefit its economy, as noted earlier.
Overall, this exercise underscores the critical importance of sustained
and collective international commitment to climate mitigation, given
the strong incentives for free-riding under noncooperative settings.

\begin{table}[H]
\caption{\textcolor{black}{\label{tab:welfare_PartialETS}}Welfare effects
of US non-participation in the global ETS}

\begin{centering}
\begin{tabular}{ccccccc}
\toprule 
 & {\small US} & {\small EU} & {\small Japan} & {\small Russia} & {\small Eurasia} & {\small China}\tabularnewline
\midrule 
CV per capita (\$) & 564.96 & -54.00 & -115.07 & 105.06 & -129.73 & -10.15\tabularnewline
\midrule 
CV per capita (\%) & 1.309 & -0.215 & -0.404 & 1.226 & -2.561 & -0.103\tabularnewline
\end{tabular}
\par\end{centering}
\centering{}%
\begin{tabular}{ccccccc}
\toprule 
 & {\small India} & {\small MidEast} & {\small Africa} & {\small LatAm} & {\small OHI} & {\small OthAs}\tabularnewline
\midrule 
CV per capita (\$) & -28.94 & -327.27 & -52.56 & -121.99 & -3.23 & -79.91\tabularnewline
\midrule 
CV per capita (\%) & -1.867 & -3.793 & -3.409 & -1.966 & -0.009 & -3.378\tabularnewline
\bottomrule
\end{tabular}
\end{table}

\subsection{Net Zero Emission Scenarios}

\begin{figure}[h]
\begin{centering}
\includegraphics[width=1\textwidth]{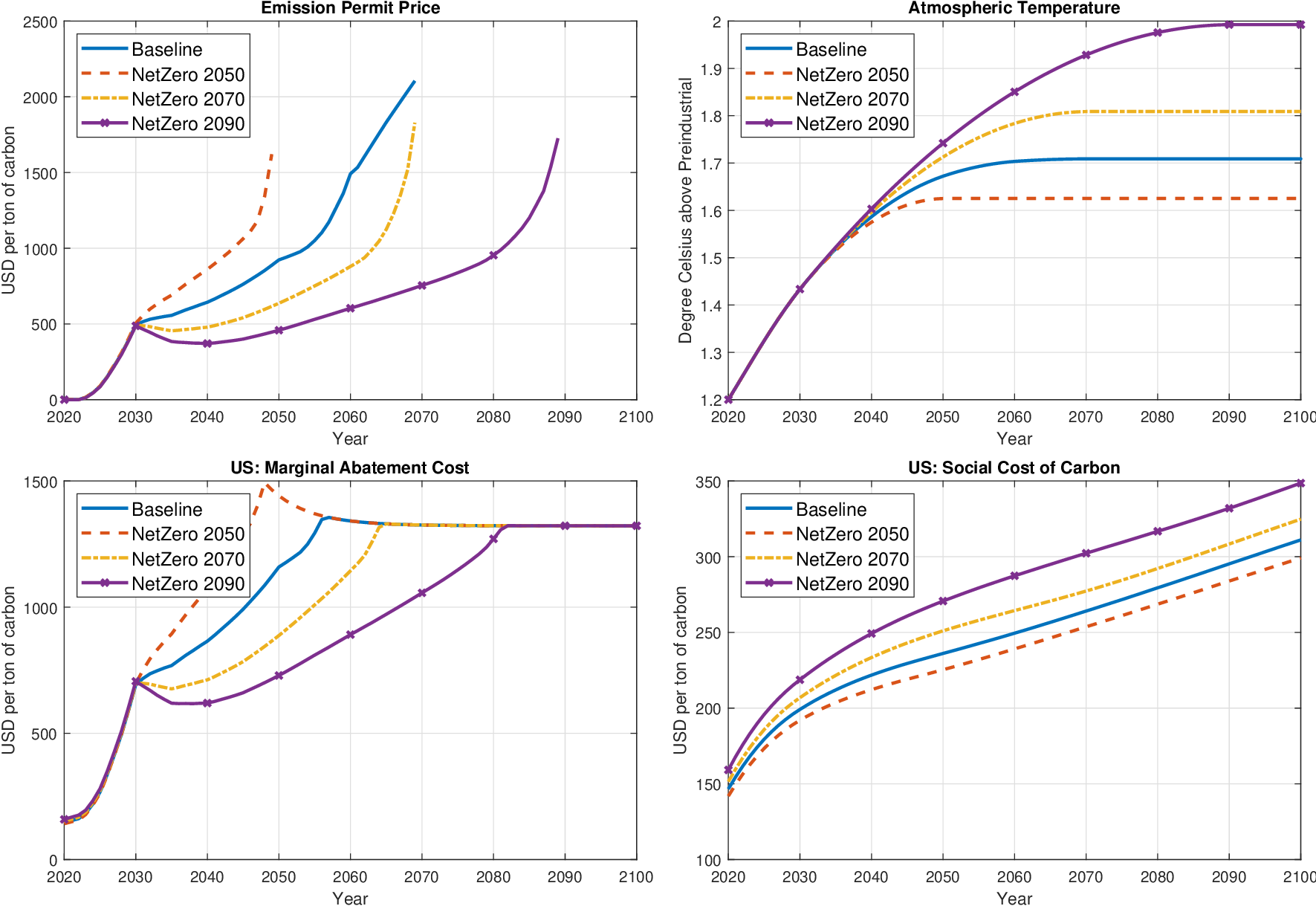}
\par\end{centering}
\centering{}\caption{\label{fig:Netzero_price_n_temp} Sensitivity analysis: alternative
emission cap scenarios.}
\end{figure}

Along with the baseline emission cap scenario, we analyze simulation
results for the noncooperative model with alternative emission cap
paths, defined by the net zero emission targets in 2050, 2070, and
2090, where all regions achieve net zero emissions by the specified
year. The top-left and top-right panels in Figure \ref{fig:Netzero_price_n_temp}
display the emission permit prices and expected temperature increases
under different emission cap scenarios. Under the net zero 2050 scenario,
which is the most strict emission cap schedule, the emission permit
price reaches \$1,621 per ton of carbon in 2049, and the temperature
rise is restricted to 1.62 degrees Celsius by the end of this century.
Net zero 2070 and net zero 2090 are more relaxed scenarios, leading
to permit prices at \$616 and \$446 per ton of carbon in 2049, respectively.
In these scenarios, the temperature rise by the end of the century
is expected to reach 1.80 degrees Celsius and 1.99 degrees Celsius
above the pre-industrial level, respectively. This result shows that
the global target of restricting the temperature rise to 1.5 degrees
Celsius is unattainable in a noncooperative world, even under the
most restrictive net zero 2050 scenario, suggesting that stronger
measures are needed to effectively regulate global emissions.

The bottom-left and bottom-right panels in Figure \ref{fig:Netzero_price_n_temp}
show the regional MAC and SCC, taking the US as an example. The comparison
of the regional SCC and MAC for all other regions are available in
Appendix \ref{subsec:sensitivity_ECAP}, which show the same patterns
as the US. Our results show that stricter emission caps lead to higher
MACs. Specifically, the MAC of the US under the net zero 2050 scenario
can reach up to \$1,493 per ton of carbon in 2048, compared to the
peak of \$1,322 in 2082 under the net zero 2090 scenario. This is
because more rigorous emission caps imposed on each region entail
additional abatement efforts, resulting in a higher MAC. We also find
that more stringent emission caps result in a smaller SCC: the SCC
of the US is \$225 per ton of carbon in 2050 in the net zero 2050
scenario, while it is \$270 in 2050 in the net zero 2090 scenario.
This is because the permit price in the net zero 2050 scenario grows
more quickly over time and even faster than the MAC.

\section{Sensitivity Analysis}

Lastly, recognizing that climate damages and emission abatement costs
are key drivers of our model outcomes, including permit prices, MAC,
and the SCC, we conduct sensitivity analyses on the parameters for
climate damages ({\small$\pi_{1,i}$, $\pi_{2,i}$}) and emission
abatement ({\small$b_{1,i}$, $b_{2,i}$, $b_{3,i},$ $b_{4,i}$}).

\begin{figure}[h]
\begin{centering}
\includegraphics[width=1\textwidth]{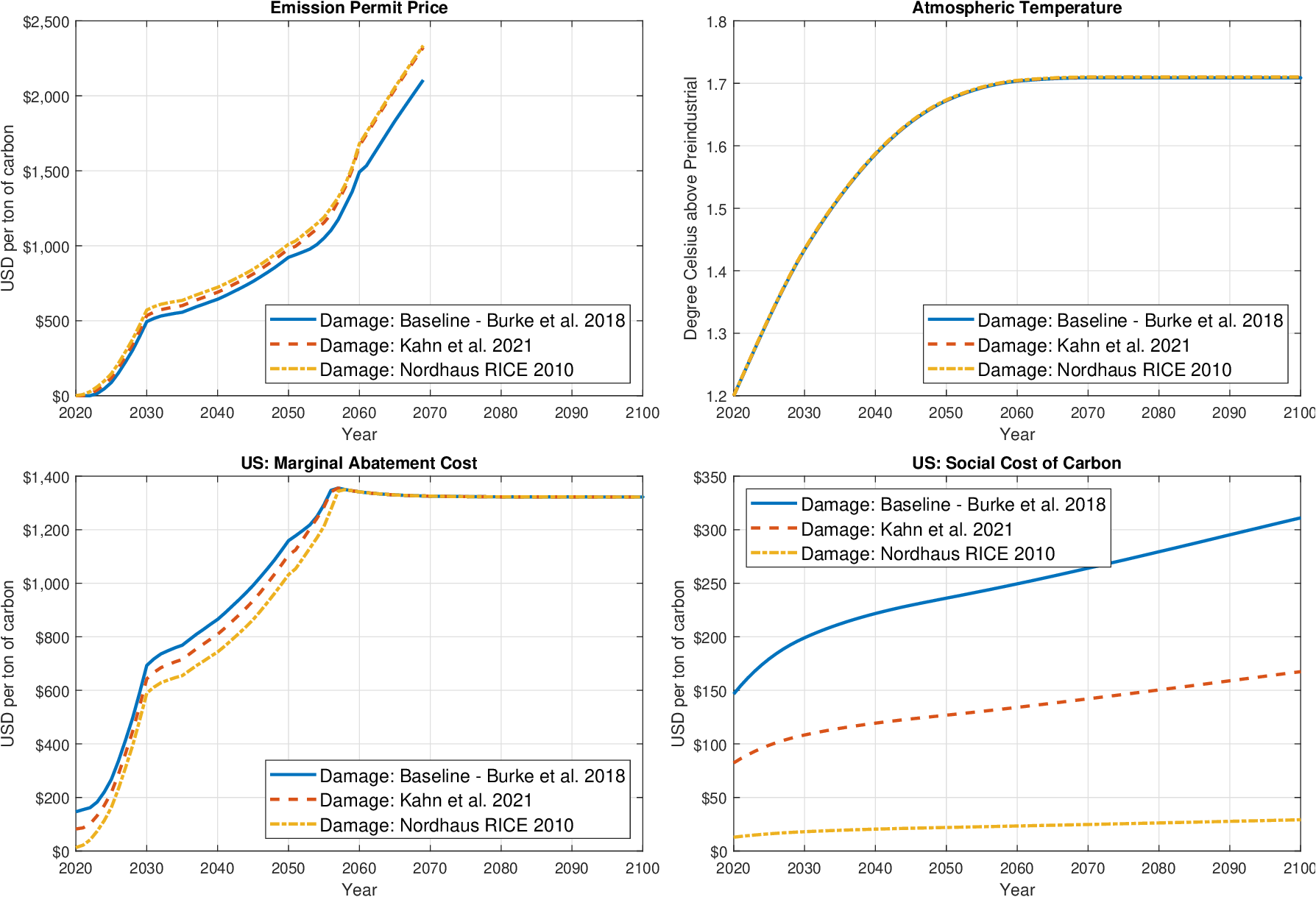}
\par\end{centering}
\centering{}\caption{\label{fig:sensitivity_Damage} Sensitivity analysis: alternative
climate damage parameters.}
\end{figure}

\subsection{Climate Damage Parameters\label{subsec:Alternative-Climate-Damages}}

. While we incorporate climate damage projections from \citet{burke2018large}
as our baseline model simulation, we additionally consider projections
from \citet{kahn2021long} and \citet{nordhaus2010excel}. Specifically,
we calibrate the climate damage parameters to match the projections
from \citet{kahn2021long} and directly adopt the parameters from
\citet{nordhaus2010excel}. Figure \ref{fig:sensitivity_Damage} demonstrates
the key economic and climate outcomes under different climate damage
assumptions. The baseline climate damage parameters from \citet{burke2018large}
result in slightly lower emission permit prices, reaching \$2,105
by 2069, compared to \$2,322 under the damage parameters estimated
from \citet{kahn2021long} and \$2,334 under the damage parameters
from \citet{nordhaus2010excel}. With the same emission cap constraints
imposed, the global temperature outcomes remain unaffected despite
variations in the damage parameters. For the US, the baseline climate
damage estimation from \citet{burke2018large} leads to higher MAC
and higher SCC, indicating that the baseline marginal climate damages
are relatively higher than those projected by \citet{kahn2021long}
and \citet{nordhaus2010excel}. However, regional heterogeneity exists;
for example, Russia experiences negative SCC under the baseline parameter
values, while experiencing small but positive SCC under the parameter
values estimated from \citet{kahn2021long} and \citet{nordhaus2010excel}.
See Appendix \ref{subsec:sensitivity_Damage} for a comparison across
all 12 regions.

\subsection{Abatement Cost Parameters}

\begin{figure}[h]
\begin{centering}
\includegraphics[width=1\textwidth]{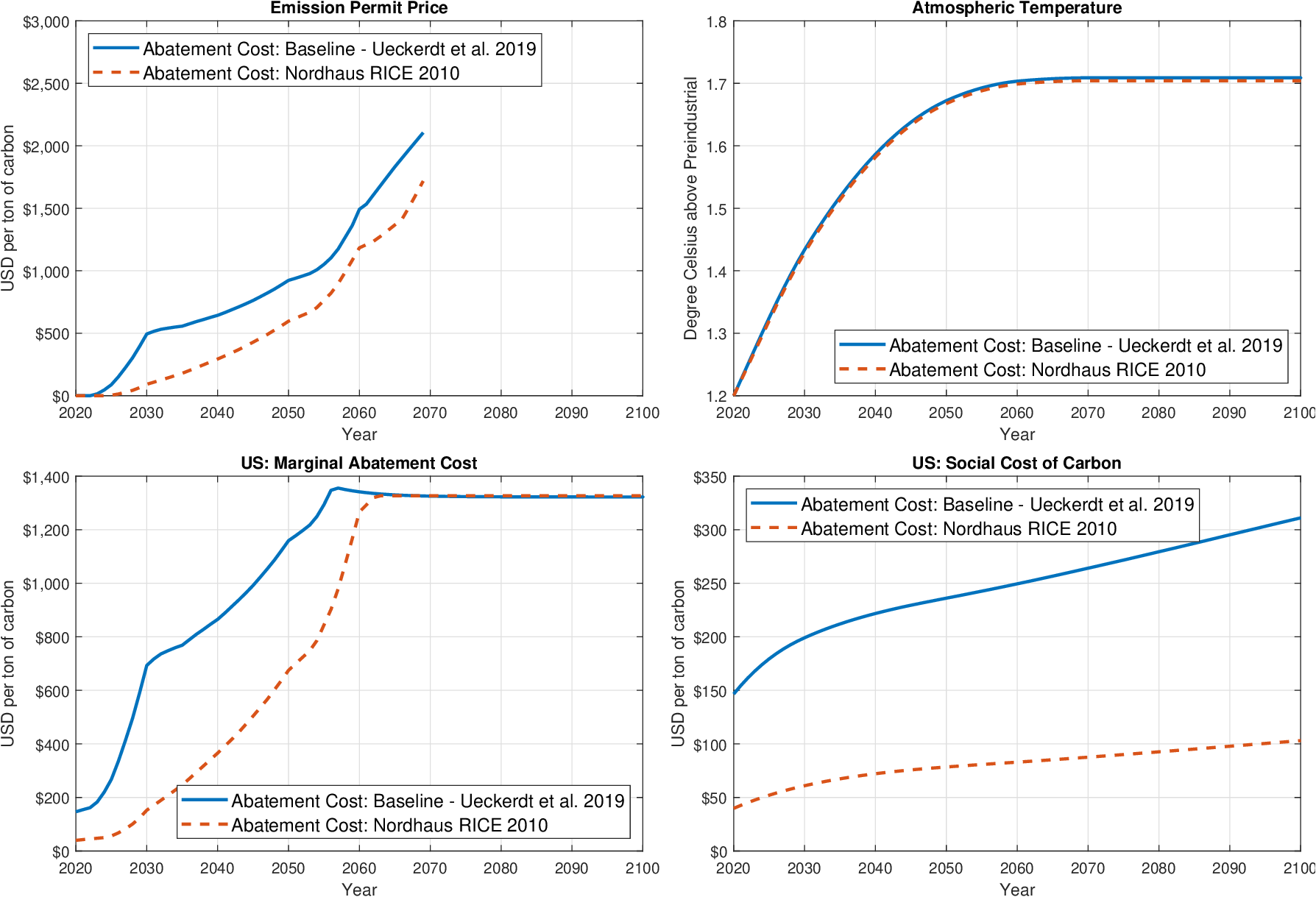}
\par\end{centering}
\centering{}\caption{\label{fig:sensitivity_MAC} Sensitivity analysis: alternative abatement
cost parameters}
\end{figure}

For our sensitivity analysis of the emission abatement cost parameters,
we consider the parameter values from \citet{nordhaus2010excel} in
addition to the baseline parameter values calibrated from \citet{ueckerdt2019economically},
both of which share the same functional form of abatement cost. As
shown in Figure \ref{fig:sensitivity_MAC}, the permit price rises
to \$1,717 by 2069 under the abatement cost estimate of \citet{nordhaus2010excel},
approximately 81\% of the permit price projected under the baseline
scenario. Despite the lower permit price, the global temperature increase
remains similar to the baseline simulation, reaching 1.70 degrees
Celsius by the end of the century, due to the emission cap constraints.
The lower emission permit price under the abatement cost estimate
of \citet{nordhaus2010excel} reflects lower MAC, as illustrated with
the US case in Figure \ref{fig:sensitivity_MAC}, with similar patterns
observed across most regions (see Appendix \ref{subsec:sensitivity_abatement}).
Lastly, the SCC is also lower under \citet{nordhaus2010excel} parameters,
with the SCC of the US reaching \$102 by 2100---just 33\% of the
baseline scenario.

\section{Conclusion}

In this work, we build a dynamic multi-region model of climate and
the economy with a global emission cap-and-trade system. In our integrated
assessment framework, regions participating in a global or partial
ETS are allocated emission caps in line with the emission targets
of the NDCs and net zero commitments, as established under the Paris
Agreement and the Glasgow Pact. We solve for the market prices of
emission permits under the dynamic Nash equilibrium in a noncooperative
setting among the participating regions. The permit prices are endogenously
determined by demand and supply of emission permits in the permits
market, reflecting regional heterogeneity in future productivity growth,
abatement technologies, climate damage, and population growth. For
strictly positive emissions, we show both theoretically and numerically
that the regional SCC is equal to the difference between the regional
MAC and the market price of permits for regions participating in the
ETS.

This work has several policy implications. First, our results indicate
that the current global target of restricting the global temperature
rise to 1.5 degrees Celsius above pre-industrial level by 2100 is
unattainable under noncooperation, with the current emission commitments
outlined in the Paris Agreement and the Glasgow Pact. Our findings
suggest that more stringent emission reduction targets and global
cooperation are needed to curb the trend of rising global temperature.
Second, our baseline simulation shows that the current emission commitments
lead to excess emission permit supply in the initial years, resulting
in permit prices of zero. This finding suggests that effective implementation
of the global ETS requires stricter emission caps so that the global
supply of permits does not exceed the global demand of permits. Third,
we demonstrate that an ETS is not a perfect substitute for a carbon
tax; rather, the two instruments are complementary and can be jointly
employed to enhance policy efficiency. Finally, our numerical analysis
highlights the strong free-rider incentives in a partial ETS, illustrating
why achieving sustained global cooperation in carbon mitigation remains
a major challenge.

\section{Declaration of generative AI and AI-assisted technologies in the
writing process}

During the preparation of this work, the authors used generative AI
(ChatGPT) in order to improve the manuscript\textquoteright s readability
during the revision process. After using this tool/service, the authors
reviewed and edited the content as needed and take full responsibility
for the content of the published article.

\newpage{}

\bibliographystyle{apalike}
\bibliography{References}

\newpage{}

\global\long\def\thefigure{A.\arabic{figure}}%
\setcounter{figure}{0} 
\global\long\def\thesection{A.\arabic{section}}%
\setcounter{section}{0} 
\global\long\def\thetable{A.\arabic{table}}%
\setcounter{table}{0} 
\global\long\def\theequation{A.\arabic{equation}}%
\setcounter{equation}{0} 
\global\long\def\thepage{A.\arabic{page}}%
\setcounter{page}{1}

\part*{Appendix for Online Publication}

\section{List of Parameters\label{sec:List-of-Parameters}}

Table \ref{tab:Key-parameters.} lists the key parameters and their
values.

\begin{table}[H]
\caption{{\small\textcolor{black}{\label{tab:Key-parameters.}Key parameters.}}}

\centering{}{\small{}%
\begin{tabular}{>{\centering}p{27mm}>{\centering}p{15mm}p{0.4\textwidth}}
\toprule 
\textbf{\textcolor{black}{Parameter}} & \textbf{\textcolor{black}{Value}} & \textbf{\textcolor{black}{Description}}\tabularnewline
\midrule
\multicolumn{3}{l}{{\small 1) Economic system parameters (from \citet{nordhaus2017revisiting})}}\tabularnewline
\midrule
{\small\textcolor{black}{$\beta$}} & {\small\textcolor{black}{$0.985$}} & {\small\textcolor{black}{Annual discount factor}}\tabularnewline
{\small$\gamma$} & {\small$1.45$} & {\small Elasticity of marginal utility}\tabularnewline
{\small\textcolor{black}{$\alpha$}} & {\small\textcolor{black}{$0.3$}} & {\small\textcolor{black}{Output elasticity of capital}}\tabularnewline
{\small\textcolor{black}{$\delta$}} & {\small\textcolor{black}{$0.1$}} & {\small\textcolor{black}{Annual depreciation rate of capital}}\tabularnewline
\midrule 
\multicolumn{3}{l}{{\small 2) Climate system parameters}}\tabularnewline
\midrule
$\zeta$ & $0.0021$ & {\small Contribution rate of carbon emissions to temperature}\tabularnewline
\bottomrule
\end{tabular}}{\small\par}
\end{table}

Table \ref{tab:Abatement-cost-parameters_Ueckerdt} lists the values
of the baseline abatement cost parameters calibrated from\textcolor{black}{{}
}\citet{ueckerdt2019economically}. The values of carbon intensity
at annual time steps will be provided upon request.

\begin{table}[H]
\caption{\textcolor{black}{\label{tab:Abatement-cost-parameters_Ueckerdt}Abatement
cost parameters (baseline) calibrated from }\citet{ueckerdt2019economically}}

\begin{centering}
\begin{tabular}{>{\centering}m{0.9cm}>{\centering}m{2cm}>{\centering}m{2cm}>{\centering}m{2cm}>{\centering}m{2cm}>{\centering}m{2cm}>{\centering}m{2cm}}
\toprule 
 & {\small US} & {\small EU} & {\small Japan} & {\small Russia} & {\small Eurasia} & {\small China}\tabularnewline
\midrule 
{\small$b_{1,i}$} & {\small 0.462} & {\small 0.477} & {\small 0.750} & {\small 0.292} & {\small 0.347} & {\small 0.328}\tabularnewline
\midrule 
{\small$b_{2,i}$} & {\small 2.859} & {\small 2.670} & {\small 2.011} & {\small 2.499} & {\small 3.243} & {\small 2.822}\tabularnewline
\midrule 
{\small$b_{3,i}$} & {\small 9.920} & {\small 5.832} & {\small 2.492} & {\small 7.625} & {\small 7.966} & {\small 7.189}\tabularnewline
\midrule 
{\small$b_{4,i}$} & {\small 0.182} & {\small 0.114} & {\small 0.2} & {\small 0.2} & {\small 0.168} & {\small 0.168}\tabularnewline
\end{tabular}
\par\end{centering}
\centering{}%
\begin{tabular}{>{\centering}m{0.9cm}>{\centering}m{2cm}>{\centering}m{2cm}>{\centering}m{2cm}>{\centering}m{2cm}>{\centering}m{2cm}>{\centering}m{2cm}}
\toprule 
 & {\small India} & {\small MidEast} & {\small Africa} & {\small LatAm} & {\small OHI} & {\small OthAs}\tabularnewline
\midrule 
{\small$b_{1,i}$} & {\small 0.594} & {\small 0.455} & {\small 0.665} & {\small 0.286} & {\small 0.347} & {\small 0.602}\tabularnewline
\midrule 
{\small$b_{2,i}$} & {\small 2.802} & {\small 2.574} & {\small 3.636} & {\small 3.828} & {\small 3.243} & {\small 3.995}\tabularnewline
\midrule 
{\small$b_{3,i}$} & {\small 6.336} & {\small 11.205} & {\small 6.558} & {\small 11.496} & {\small 7.966} & {\small 6.518}\tabularnewline
\midrule 
{\small$b_{4,i}$} & {\small 0.2} & {\small 0.2} & {\small 0.2} & {\small 0.2} & {\small 0.168} & {\small 0.163}\tabularnewline
\bottomrule
\end{tabular}
\end{table}

Table \ref{tab:Climate-damage-parameters} lists the calibrated values
of the climate damage parameters used in the baseline analysis.

\begin{table}[H]
\caption{\textcolor{black}{\label{tab:Climate-damage-parameters}Climate damage
parameters (baseline) calibrated from \citet{burke2018large}}}

\begin{centering}
\begin{tabular}{>{\centering}m{0.9cm}>{\centering}m{2cm}>{\centering}m{2cm}>{\centering}m{2cm}>{\centering}m{2cm}>{\centering}m{2cm}>{\centering}m{2cm}}
\toprule 
 & {\small US} & {\small EU} & {\small Japan} & {\small Russia} & {\small Eurasia} & {\small China}\tabularnewline
\midrule 
{\small$\pi_{1,i}$} & 0.0842 & 0.0489 & 0.0090 & -0.4169 & 0.2678 & 0.0003\tabularnewline
\midrule 
{\small$\pi_{2,i}$} & 0.0096 & 0.0011 & 0.0748 & 0.3094 & 0.0002 & 0.0008\tabularnewline
\end{tabular}
\par\end{centering}
\centering{}%
\begin{tabular}{>{\centering}m{0.9cm}>{\centering}m{2cm}>{\centering}m{2cm}>{\centering}m{2cm}>{\centering}m{2cm}>{\centering}m{2cm}>{\centering}m{2cm}}
\toprule 
 & {\small India} & {\small MidEast} & {\small Africa} & {\small LatAm} & {\small OHI} & {\small OthAs}\tabularnewline
\midrule 
{\small$\pi_{1,i}$} & 0.0017 & 0.3595 & 0.1886 & 0.1801 & 0.0123 & 0.2161\tabularnewline
\midrule 
{\small$\pi_{2,i}$} & 0.3276 & 0.0088 & 0.0764 & 0.0030 & 0.0044 & 0.0224\tabularnewline
\bottomrule
\end{tabular}
\end{table}

Table \ref{tab:TFP-parameters} lists the values of the TFP parameters
calibrated from \textcolor{black}{\citet{burke2018large}}. 
\begin{table}[H]
\caption{\textcolor{black}{\label{tab:TFP-parameters}TFP parameters calibrated
from \citet{burke2018large}}}

\begin{centering}
\begin{tabular}{>{\centering}m{0.9cm}>{\centering}m{2cm}>{\centering}m{2cm}>{\centering}m{2cm}>{\centering}m{2cm}>{\centering}m{2cm}>{\centering}m{2cm}}
\toprule 
 & {\small US} & {\small EU} & {\small Japan} & {\small Russia} & {\small Eurasia} & {\small China}\tabularnewline
\midrule 
{\small$g_{i,0}$} & 0.0033 & 0.0089 & 0.0085 & 0.0170 & 0.0094 & 0.0345\tabularnewline
\midrule 
{\small$d_{i}$} & 0.0011 & 0.0010 & 0.0010 & 0.0154 & 0.0010 & 0.0308\tabularnewline
\end{tabular}
\par\end{centering}
\centering{}%
\begin{tabular}{>{\centering}m{0.9cm}>{\centering}m{2cm}>{\centering}m{2cm}>{\centering}m{2cm}>{\centering}m{2cm}>{\centering}m{2cm}>{\centering}m{2cm}}
\toprule 
 & {\small India} & {\small MidEast} & {\small Africa} & {\small LatAm} & {\small OHI} & {\small OthAs}\tabularnewline
\midrule 
{\small$g_{i,0}$} & 0.0332 & 0.0093 & 0.0218 & 0.0134 & 0.0076 & 0.0221\tabularnewline
\midrule 
{\small$d_{i}$} & 0.0151 & 0.0010 & 0.0013 & 0.0010 & 0.0010 & 0.0062\tabularnewline
\bottomrule
\end{tabular}
\end{table}

\section{List of Countries\label{sec:constituent countries of multi-country regions}}

\begin{table}[H]
\begin{centering}
\caption{List of countries for regional aggregation\label{tab:Countries-in-the-multi-country-regions}}
\par\end{centering}
\centering{}%
\begin{tabular}{>{\centering}m{0.13\textwidth}m{0.85\textwidth}}
\toprule 
\textbf{Region} & \textbf{Constituent Countries}\tabularnewline
\midrule 
Africa & Algeria, Angola, Benin, Botswana, Burkina Faso, Burundi, Cameroon,
Cape Verde, Central African Republic, Chad, Comoros, Democratic Republic
of the Congo, Republic of the Congo, Cote d'Ivoire, Djibouti, Arab
Republic of Egypt, Ethiopia, Gabon, Gambia, The Ghana, Guinea, Guinea-Bissau,
Kenya, Lesotho, Libya, Madagascar, Mali, Mauritania, Mauritius, Morocco,
Mozambique, Namibia, Niger, Rwanda, Senegal, Sierra Leone, South Africa,
Sudan, Swaziland, Tanzania, Togo, Tunisia, Uganda, Zambia, Zimbabwe.\tabularnewline
\midrule 
European Union\tablefootnote{The current EU does not contain {\small the United Kingdom, but in
this paper we still assume the United Kingdom is in the EU for convenience.}}{\small .} & Austria, Belgium, Bulgaria, Croatia, Czech Republic, Denmark, Finland,
France, Germany, Greece, Hungary, Ireland, Italy, Latvia, Lithuania,
Luxembourg, Netherlands, Poland, Portugal, Romania, Slovak Republic,
Spain, Sweden, United Kingdom.\tabularnewline
\midrule 
Eurasia & Albania, Armenia, Azerbaijan, Belarus, Bosnia and Herzegovina, Estonia,
Georgia, Kazakhstan, Kyrgyz Republic, FYR Macedonia, Moldova, Serbia,
Slovenia, Tajikistan, Turkey, Ukraine, Uzbekistan.\tabularnewline
\midrule 
Latin America & Argentina, Bahamas, The Belize, Bolivia, Brazil, Chile, Colombia,
Costa Rica, Cuba, Dominican Republic, Ecuador, El Salvador, Guatemala,
Haiti, Honduras, Mexico, Nicaragua, Panama, Paraguay, Peru, Puerto
Rico, Trinidad and Tobago, Uruguay.\tabularnewline
\midrule 
Middle East & Cyprus, Islamic Republic of Iran, Iraq, Israel, Jordan, Lebanon, Oman,
Qatar, Saudi Arabia, United Arab Emirates\tabularnewline
\midrule 
Other Non-OECD Asia & Afghanistan, Bangladesh, Bhutan, Brunei Darussalam, Cambodia, Fiji,
Indonesia, Malaysia, Mongolia, Nepal, Pakistan, Philippines, Samoa,
Solomon Islands, Sri Lanka, Thailand, Vanuatu, Vietnam.\tabularnewline
\midrule 
Other High-Income & Australia, Canada, Iceland, Republic of Korea, New Zealand, Norway,
Switzerland.\tabularnewline
\bottomrule
\end{tabular}
\end{table}

\pagebreak{}

\section{Details about Calibration and Data}

\subsection{Calibration of the TCRE Climate System {\Large\label{sec: Atmospheric Temperature Anomaly }}}

\begin{figure}[h]
\centering{}\includegraphics[width=0.65\textwidth]{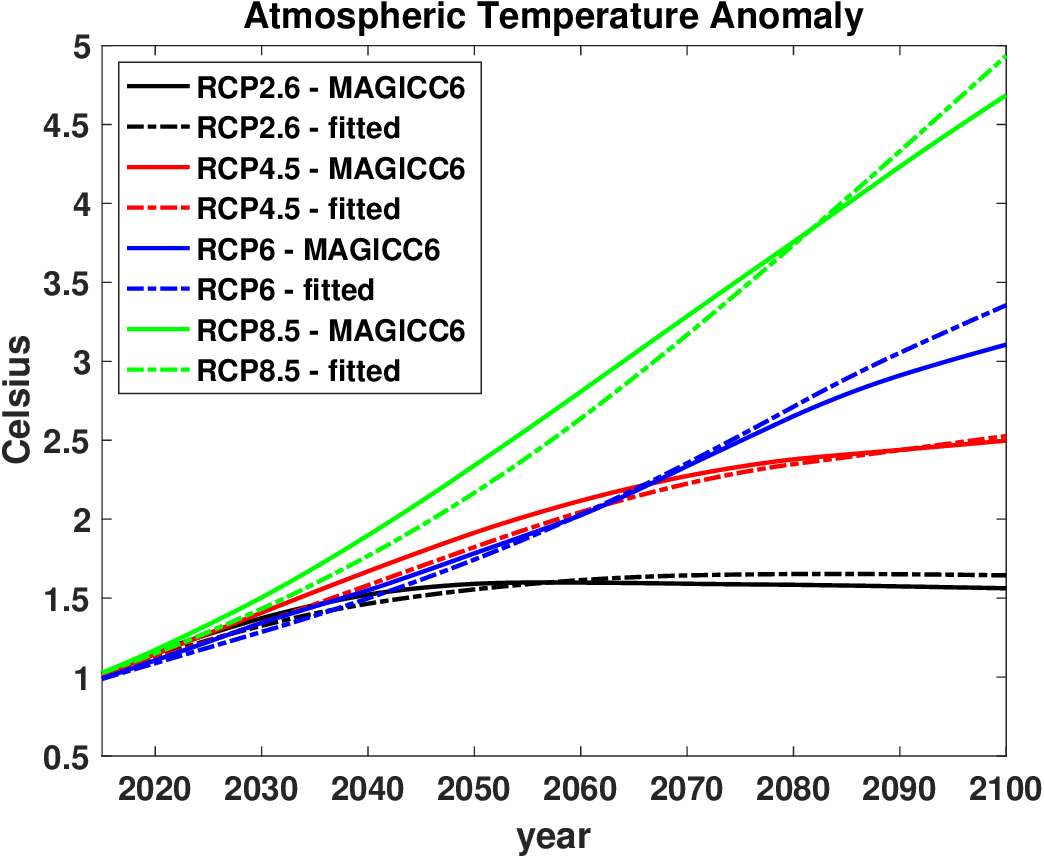}\caption{\label{fig:TCRE}Calibration of the TCRE climate system.}
\end{figure}

Each of the four RCP scenarios \citep{meinshausen2011rcp} --- RCP
2.6, RCP 4.5, RCP 6, and RCP 8.5 --- provide their pathways of emissions,
atmospheric carbon concentration, radiative forcing, and atmospheric
temperature anomaly. When we calibrate the contribution rate of carbon
emissions on temperature, $\zeta$, in a climate system, we use the
pathways of emissions and atmospheric temperature anomaly of the four
RCP scenarios. Figure \ref{fig:TCRE} shows that our calibrated TCRE
climate system provides a very good projection of the atmospheric
temperature anomaly (increase relative to pre-industrial levels) based
on cumulative emissions only.

\pagebreak{}

\subsection{Calibration of Total Factor Productivity and Climate Damage \label{sec:Calibration-of-TFP-appendix}}

\begin{figure}[H]
\centering{}\includegraphics[width=0.9\textwidth]{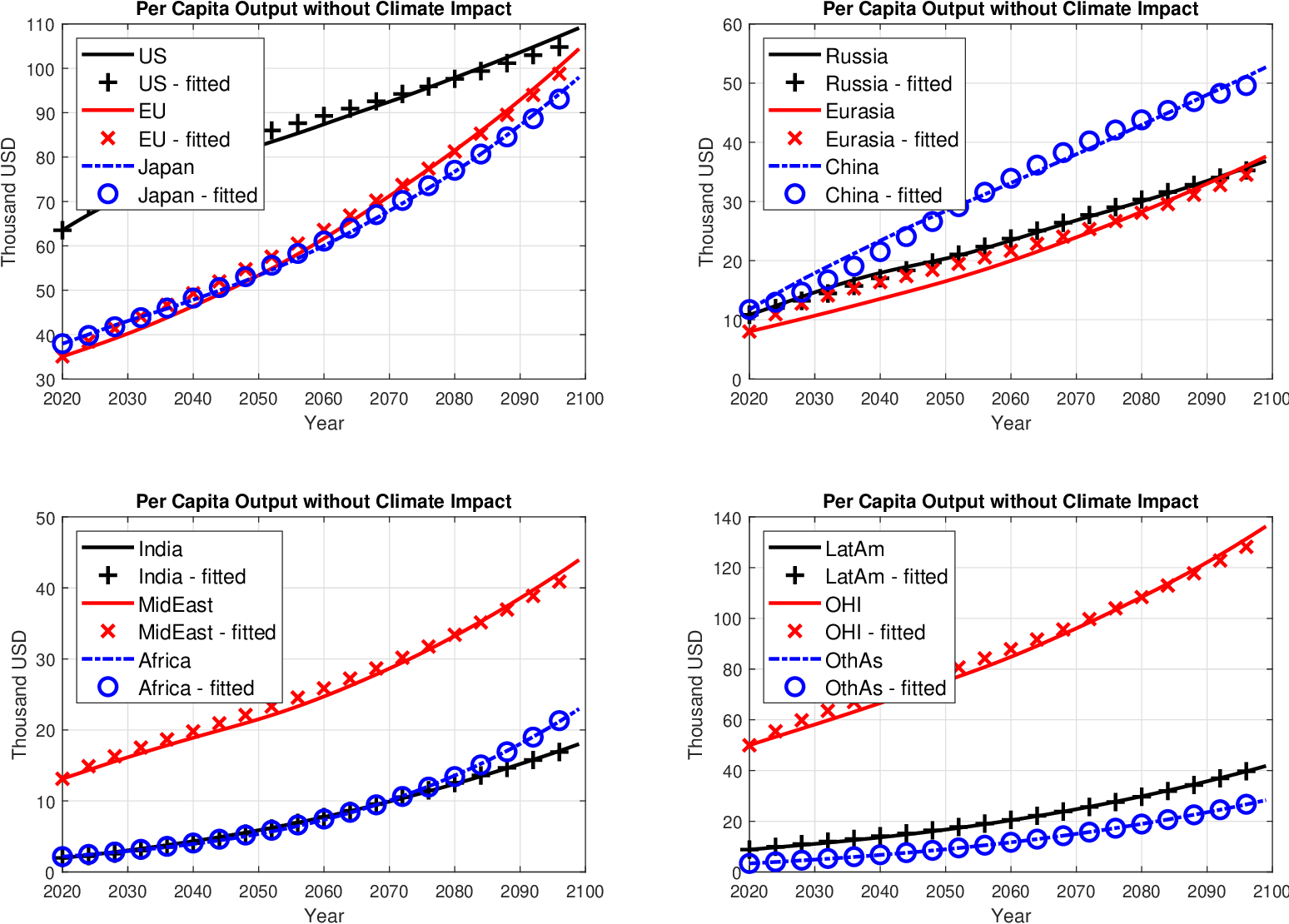}\caption{\label{fig:GDP_fitted_noDam}Fitting GDP per capita under no climate
impact. Lines represent GDP per capita under no climate impact from
\citet{burke2018large}; marks represent fitted GDP per capita.}
\end{figure}

\begin{figure}[H]
\centering{}\includegraphics[width=0.9\textwidth]{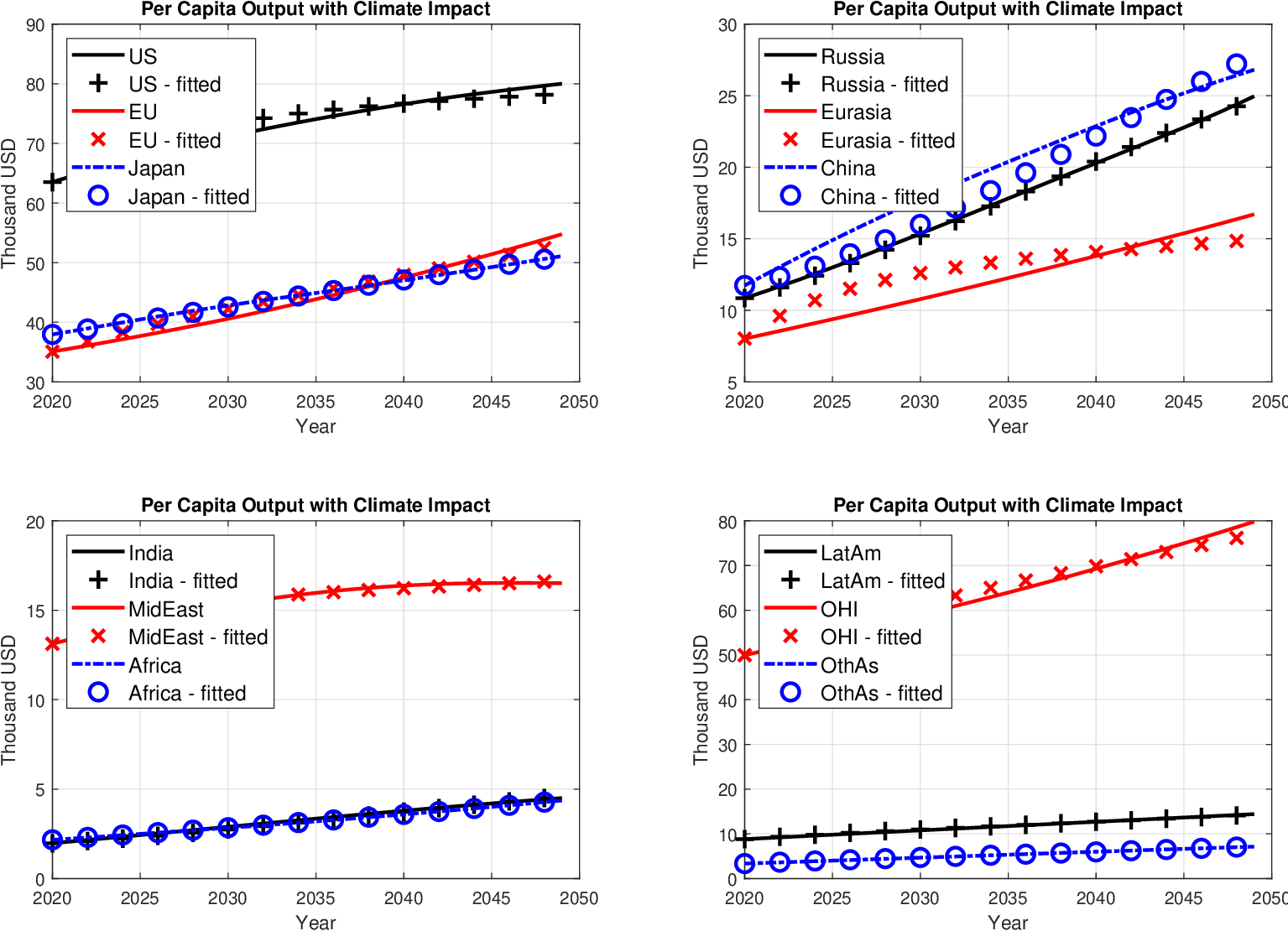}\caption{\label{fig:GDPdam_fitted_RCP45} Fitting GDP under climate impact.
Lines represent the GDP per capita \textcolor{black}{under the climate
impact of RCP 4.5 }from\textcolor{black}{{} \citet{burke2018large}};
marks represent fitted values.}
\end{figure}

Figures \ref{fig:GDP_fitted_noDam} and \textcolor{black}{\ref{fig:GDPdam_fitted_RCP45}}
show that with our calibrated \textcolor{black}{TFP and climate damage
coefficients,} the GDP per capita \textcolor{black}{$y_{i,t}^{\mathrm{NoCC}}$
or $y_{i,t}$ matches well with the projected data }$y_{i,t}^{\mathrm{BDD,NoCC}}$
or $y_{i,t}^{\mathrm{BDD}}$\textcolor{black}{{} from \citet{burke2018large},
respectively, for all regions.}


\pagebreak{}

\subsection{Calibration of Climate Damage from \citet{kahn2021long}}

For sensitivity analysis on climate damage parameters in Section \ref{subsec:Alternative-Climate-Damages},
we calibrate the climate damage parameters \textcolor{black}{$\pi_{1,i}$
and $\pi_{2,i}$ by considering projections on GDP loss across different
climate scenarios in \citet{kahn2021long}, which shows the percentage
loss in GDP per capita by 2030, 2050, and 2100 under the RCP 2.6 and
RCP 8.5 scenarios for China, EU, India, Russia, and the US. We use
their method and data to project the percentage loss in GDP per capita
($\Delta_{i,t}^{\mathrm{RCP26}}$ and $\Delta_{i,t}^{\mathrm{RCP85}}$)
every year from 2020 to 2114 under the RCP 2.6 and RCP 8.5 scenarios
for each of our 12 regions, employing the baseline setup in \citet{kahn2021long}.
Specifically, $\Delta_{i,t}^{\mathrm{RCP26}}=1-y_{i,t}^{\mathrm{RCP26}}/y_{i,t}^{\mathrm{base}}$
and $\Delta_{i,t}^{\mathrm{RCP85}}=1-y_{i,t}^{\mathrm{RCP85}}/y_{i,t}^{\mathrm{base}}$,
where $y_{i,t}^{\mathrm{RCP26}}$, $y_{i,t}^{\mathrm{RCP85}}$, and
$y_{i,t}^{\mathrm{base}}$ are GDP per capita under RCP 2.6, RCP 8.5,
and the baseline scenario, respectively. Thus from equation (\ref{eq:net output})
we obtain $(\pi_{1,i},\pi_{2,i})$ by solving the following minimization
problem for each region $i$:
\begin{equation}
\min_{\pi_{1,i},\pi_{2,i}}\ \sum_{t=0}^{94}\left(\frac{1+\pi_{1,i}T_{t}^{\mathrm{RCP85}}+\pi_{2,i}\left(T_{t}^{\mathrm{RCP85}}\right)^{2}}{1+\pi_{1,i}T_{t}^{\mathrm{RCP26}}+\pi_{2,i}\left(T_{t}^{\mathrm{RCP26}}\right)^{2}}-\frac{1-\Delta_{t,i}^{\mathrm{RCP26}}}{1-\Delta_{t,i}^{\mathrm{RCP85}}}\right)^{2}.\label{eq:StrEstObj_Dam-1}
\end{equation}
Here $T_{t}^{\mathrm{RCP26}}$ and $T_{t}^{\mathrm{RCP85}}$ are the
global average temperature anomalies at time $t$ (deviation from
the pre-industrial temperature) under the RCP 2.6 and RCP 8.5 scenarios.
Figure \ref{fig:GDPdam_fitted_RCP45} shows, with our calibrated climate
damage coefficients, the ratios of GDP per capita between RCP 2.6
and RCP8.5 from our model, matches well with the ratios in \citet{kahn2021long}.}

\begin{figure}[H]
\centering{}\includegraphics[width=0.75\textwidth]{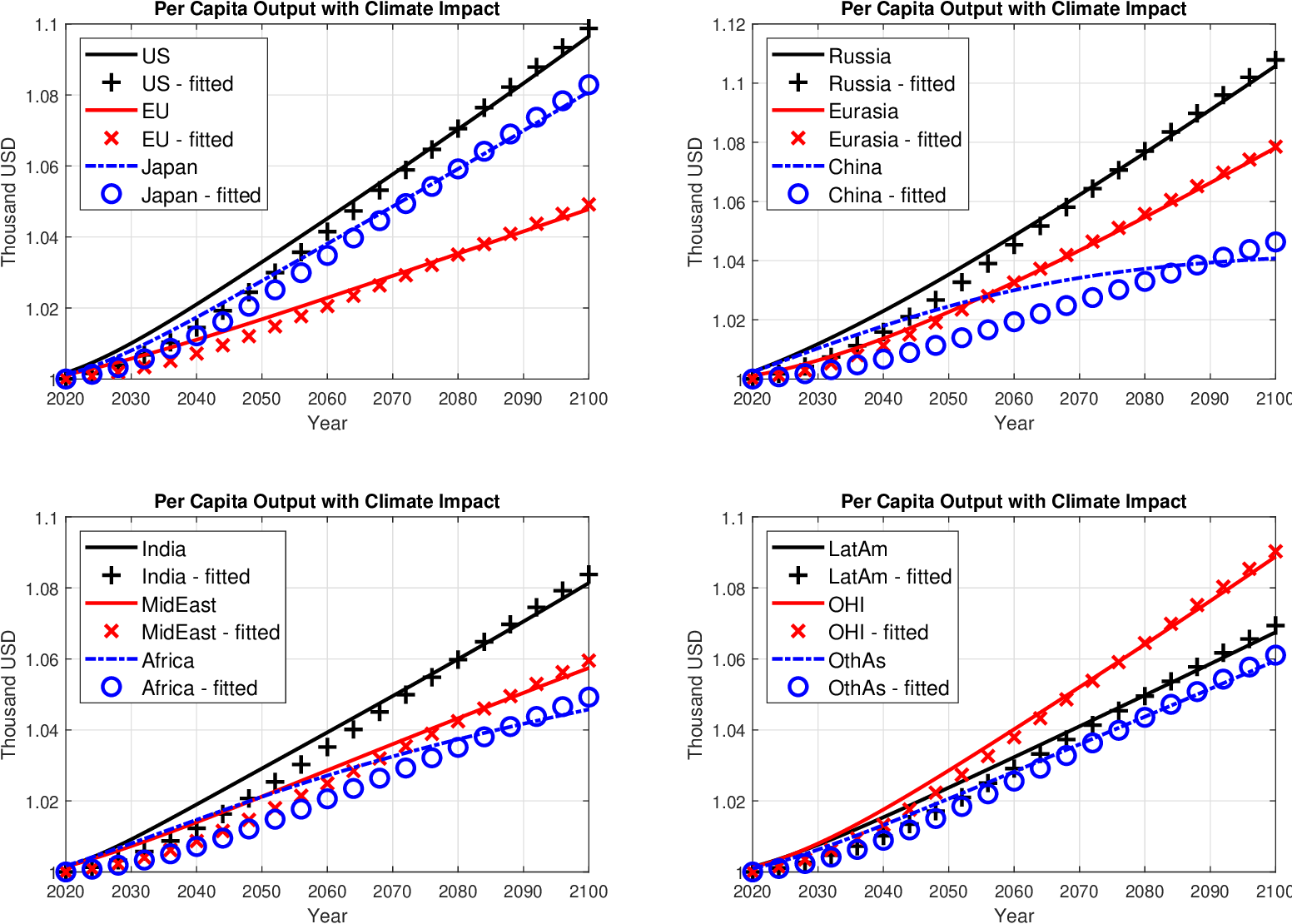}\caption{\label{fig:GDPloss_fitted_RCP85-1} Fitting climate damage parameters.
Lines represent the ratios of GDP per capita \textcolor{black}{between
RCP 2.6 and RCP8.5} from\textcolor{black}{{} \citet{kahn2021long}};
marks represent fitted ratios.}
\end{figure}

\subsection{Regional Emission Cap Pathways\label{sec:Emission-Cap-Pathways-Appendix}}

\begin{figure}[h]
\centering{}\includegraphics[width=0.8\textwidth]{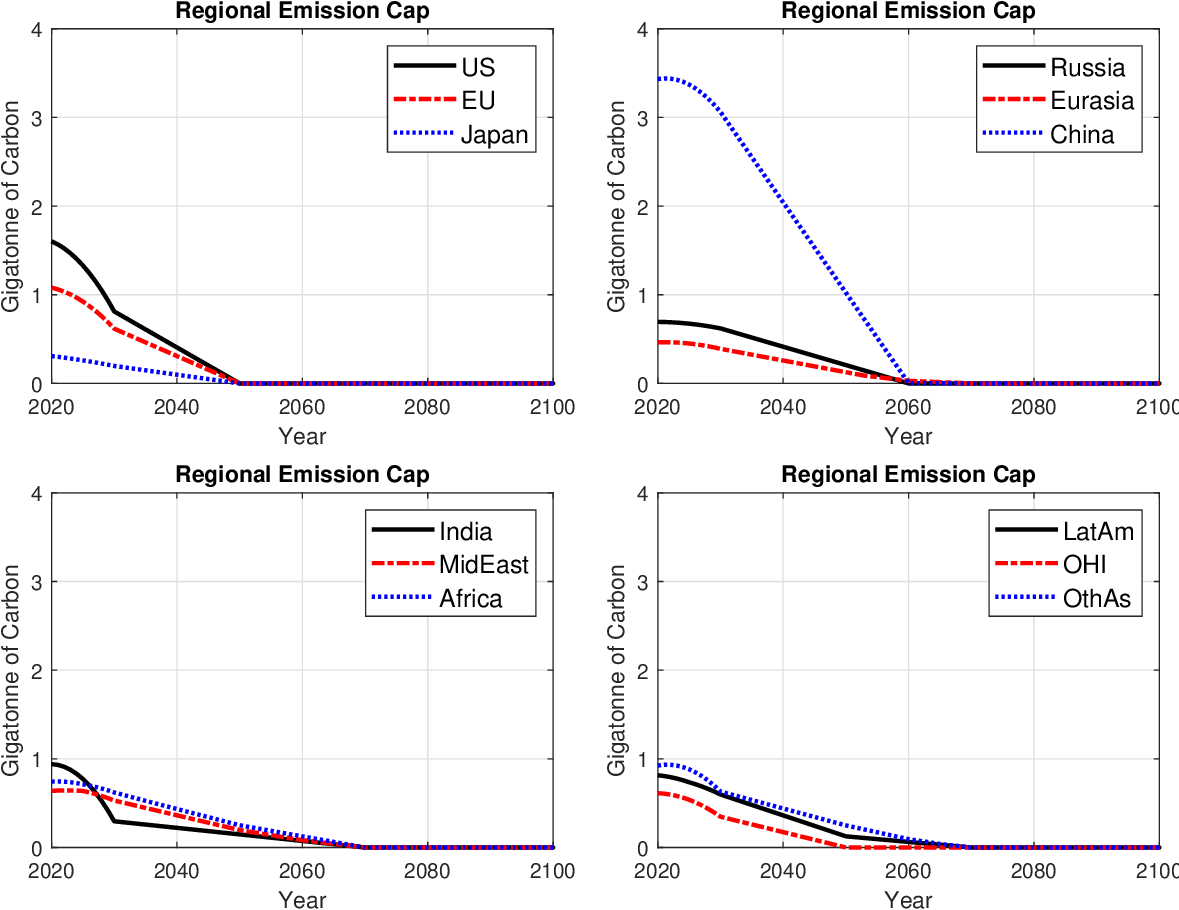}\caption{\label{fig:Emission-caps-of-regions} Regional emission cap pathways
under the baseline scenario.}
\end{figure}

Figure \ref{fig:Emission-caps-of-regions} displays the regional emission
cap pathways, measured in Gigatonne of Carbon (GtC), for the baseline
emission cap scenario, generated using the methodology described in
Section \ref{subsec:Regional-Emission-Cap}.

Table \ref{tab:Regional-emission-cap} lists the regional emission
caps for every region in five-year time steps. The regional emission
caps at annual time steps will be provided upon request.

\begin{table}[H]
\caption{\label{tab:Regional-emission-cap}Regional emission cap pathways under
the baseline scenario (unit: GtC)}

\centering{}{\small{}%
\begin{tabular}{>{\centering}p{15mm}>{\centering}p{9.5mm}>{\centering}p{9.5mm}>{\centering}p{9.5mm}>{\centering}p{9.5mm}>{\centering}p{9.5mm}>{\centering}p{9.5mm}>{\centering}p{9.5mm}>{\centering}p{9.5mm}>{\centering}p{9.5mm}>{\centering}p{9.5mm}>{\centering}p{9.5mm}}
\toprule 
 & 2020 & 2025 & 2030 & 2035 & 2040 & 2045 & 2050 & 2055 & 2060 & 2065 & 2070\tabularnewline
\midrule 
US & {\small 1.603} & {\small 1.330} & {\small 0.812} & {\small 0.609} & {\small 0.406} & {\small 0.203} & {\small 0.000} & {\small 0.000} & {\small 0.000} & {\small 0.000} & {\small 0.000}\tabularnewline
EU & {\small 1.081} & {\small 0.923} & {\small 0.617} & {\small 0.463} & {\small 0.309} & {\small 0.154} & {\small 0.000} & {\small 0.000} & {\small 0.000} & {\small 0.000} & {\small 0.000}\tabularnewline
Japan & {\small 0.308} & {\small 0.260} & {\small 0.197} & {\small 0.148} & {\small 0.099} & {\small 0.049} & {\small 0.000} & {\small 0.000} & {\small 0.000} & {\small 0.000} & {\small 0.000}\tabularnewline
Russia & {\small 0.694} & {\small 0.674} & {\small 0.621} & {\small 0.517} & {\small 0.414} & {\small 0.310} & {\small 0.207} & {\small 0.103} & {\small 0.000} & {\small 0.000} & {\small 0.000}\tabularnewline
Eurasia & {\small 0.464} & {\small 0.450} & {\small 0.390} & {\small 0.324} & {\small 0.259} & {\small 0.193} & {\small 0.126} & {\small 0.071} & {\small 0.030} & {\small 0.015} & {\small 0.000}\tabularnewline
China & {\small 3.433} & {\small 3.370} & {\small 3.061} & {\small 2.551} & {\small 2.042} & {\small 1.532} & {\small 1.023} & {\small 0.513} & {\small 0.004} & {\small 0.002} & {\small 0.000}\tabularnewline
India & {\small 0.940} & {\small 0.773} & {\small 0.295} & {\small 0.259} & {\small 0.222} & {\small 0.185} & {\small 0.148} & {\small 0.111} & {\small 0.074} & {\small 0.037} & {\small 0.000}\tabularnewline
MidEast & {\small 0.638} & {\small 0.638} & {\small 0.530} & {\small 0.446} & {\small 0.363} & {\small 0.280} & {\small 0.196} & {\small 0.140} & {\small 0.083} & {\small 0.041} & {\small 0.000}\tabularnewline
Africa & {\small 0.743} & {\small 0.712} & {\small 0.621} & {\small 0.526} & {\small 0.434} & {\small 0.342} & {\small 0.251} & {\small 0.188} & {\small 0.125} & {\small 0.063} & {\small 0.000}\tabularnewline
LatAm & {\small 0.815} & {\small 0.736} & {\small 0.598} & {\small 0.480} & {\small 0.362} & {\small 0.244} & {\small 0.126} & {\small 0.094} & {\small 0.063} & {\small 0.031} & {\small 0.000}\tabularnewline
OHI & {\small 0.613} & {\small 0.538} & {\small 0.347} & {\small 0.260} & {\small 0.173} & {\small 0.087} & {\small 0.000} & {\small 0.000} & {\small 0.000} & {\small 0.000} & {\small 0.000}\tabularnewline
OthAs & {\small 0.924} & {\small 0.882} & {\small 0.630} & {\small 0.535} & {\small 0.440} & {\small 0.345} & {\small 0.249} & {\small 0.172} & {\small 0.094} & {\small 0.040} & {\small 0.000}\tabularnewline
\bottomrule
\end{tabular}}{\small\par}
\end{table}

\begin{figure}[H]
\centering{}\includegraphics[width=0.7\textwidth]{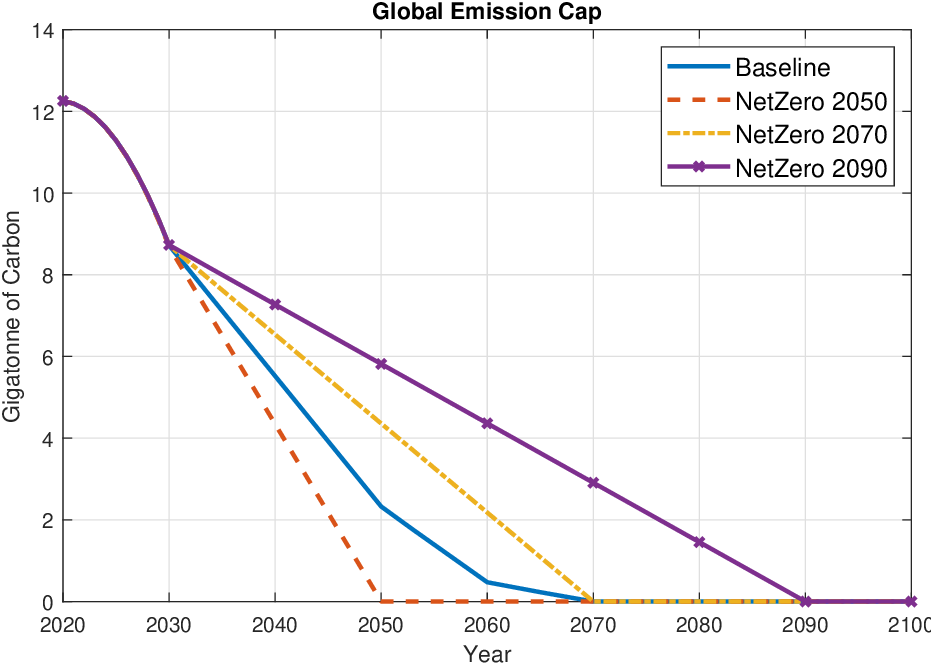}\caption{\label{fig:Global-emission-caps} Global emission cap pathways for
the different net zero scenarios.}
\end{figure}

Figure \ref{fig:Global-emission-caps} presents a comparison of global
emission cap pathways under various scenarios: the baseline scenario,
net-zero by 2050, net-zero by 2070, and net-zero by 2090. In the net-zero
scenarios, it is assumed that all countries achieve net-zero emissions
by the respective target years.

\subsection{GDP Growth Rate beyond this Century\label{sec:GDP-Growth-Rate}}

\textcolor{black}{For the GDP growth rate beyond this century, we
follow RICE to project $g_{t,i}$ for $t\geq80$. We begin by assuming
the long-run growth rate of TFP in the US is $g_{\mathrm{US},\infty}=0.0033(1-\alpha)=0.00231$
with $\alpha=0.3$. Next, we let $\widetilde{y}_{i,79}=y_{i,79}^{\mathrm{BDD}}y_{i,0}/y_{i,0}^{\mathrm{BDD}}$
be our projected per capita output in 2099, where $y_{i,0}$ is the
observed per capita output in 2020. We then assume that the TFP growth
in the US is characterized by 
\[
g_{\mathrm{US},t}=g_{\mathrm{US},\infty}+(g_{\mathrm{US},79}-g_{\mathrm{US},\infty})\exp(-0.01(t-79)),
\]
and let $\widetilde{y}_{\mathrm{US},t+1}=\widetilde{y}_{\mathrm{US},t}\exp(g_{\mathrm{US},t}/(1-\alpha))$
for $t\geq79$. For the regions other than the US, we assume their
TFP growth can be expressed in relation to the TFP growth of the US.
Specifically, we assume that, for $t\geq79$, 
\[
\begin{cases}
\widetilde{y}_{i,t+1}=\widetilde{y}_{i,t}\exp(g_{i,t}/(1-\alpha))\\
g_{i,t+1}=g_{\mathrm{US},t+1}+(1-\alpha)\chi\ln(\widetilde{y}_{\mathrm{US},t}/\widetilde{y}_{i,t})
\end{cases}
\]
where $\chi=0.005$ is chosen such that $g_{i,t}$ gradually moves
toward $g_{\mathrm{US},t}$ as $t\rightarrow\infty$. }\footnote{\textcolor{black}{Assume $\widetilde{y}_{i,t}=A_{i,t}k_{i,t}^{\alpha}$
is GDP per capita where $k_{i,t}$ is capital per capita. We have
\[
\ln\left(\frac{\widetilde{y}_{i,t+1}}{\widetilde{y}_{i,t}}\right)=g_{i,t}+\alpha\ln\left(\frac{k_{i,t+1}}{k_{i,t}}\right).
\]
If we assume the growth of $k_{t,i}$ is equal to the growth of GDP
per capita, then we have 
\[
\widetilde{y}_{i,t+1}=\widetilde{y}_{i,t}\exp(g_{i,t}/(1-\alpha)).
\]
}}

\pagebreak{}

\section{Additional Simulation Results \label{sec: Net_emissions_post_abatement}}

\subsection{Benchmark Model: Regional Emissions \label{subsec:Regional-Net-Emissions}}

\begin{figure}[h]
\begin{centering}
\includegraphics[width=0.8\textwidth]{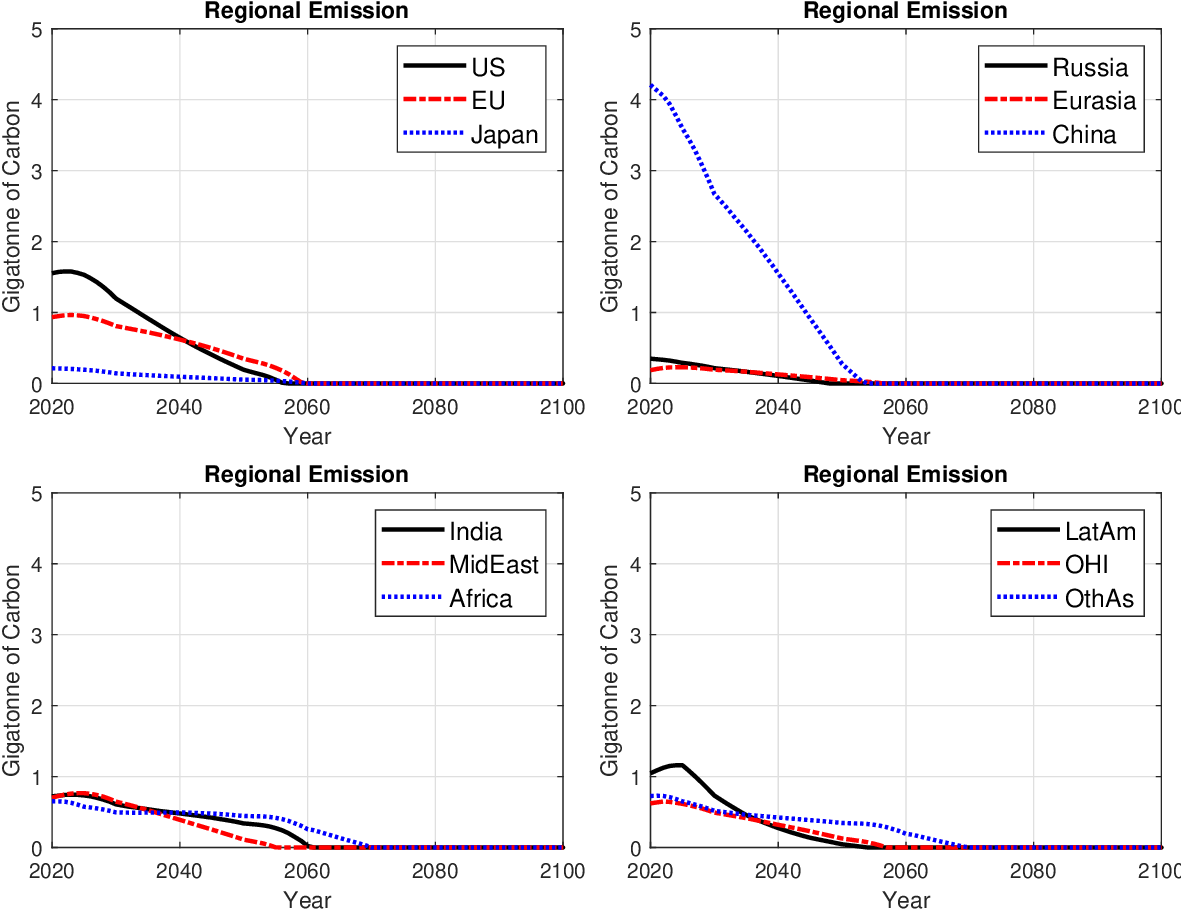}
\par\end{centering}
\centering{}\caption{\label{fig:Net_emission} Simulation results of regional emissions
under the baseline emission caps.}
\end{figure}

Figure \ref{fig:Net_emission} displays the regional emissions under
the noncooperative model with the ETS and the baseline emission cap
scenario. Russia is the first to reach net zero emissions in 2050,
followed by China and Latin America in 2056, MidEast in 2057, the
US and Eurasia in 2058, and the OHI in 2059. Then, net zero emissions
are achieved by EU in 2061, Japan and India in 2064. Finally, Africa
and non-OECD Asia achieve net zero emissions in 2070.

\pagebreak{}

\subsection{Model Comparison of ETS Implementation\label{subsec:Model_Comparison_ETS}}

\begin{figure}[H]
\centering{}\includegraphics[width=0.9\textwidth]{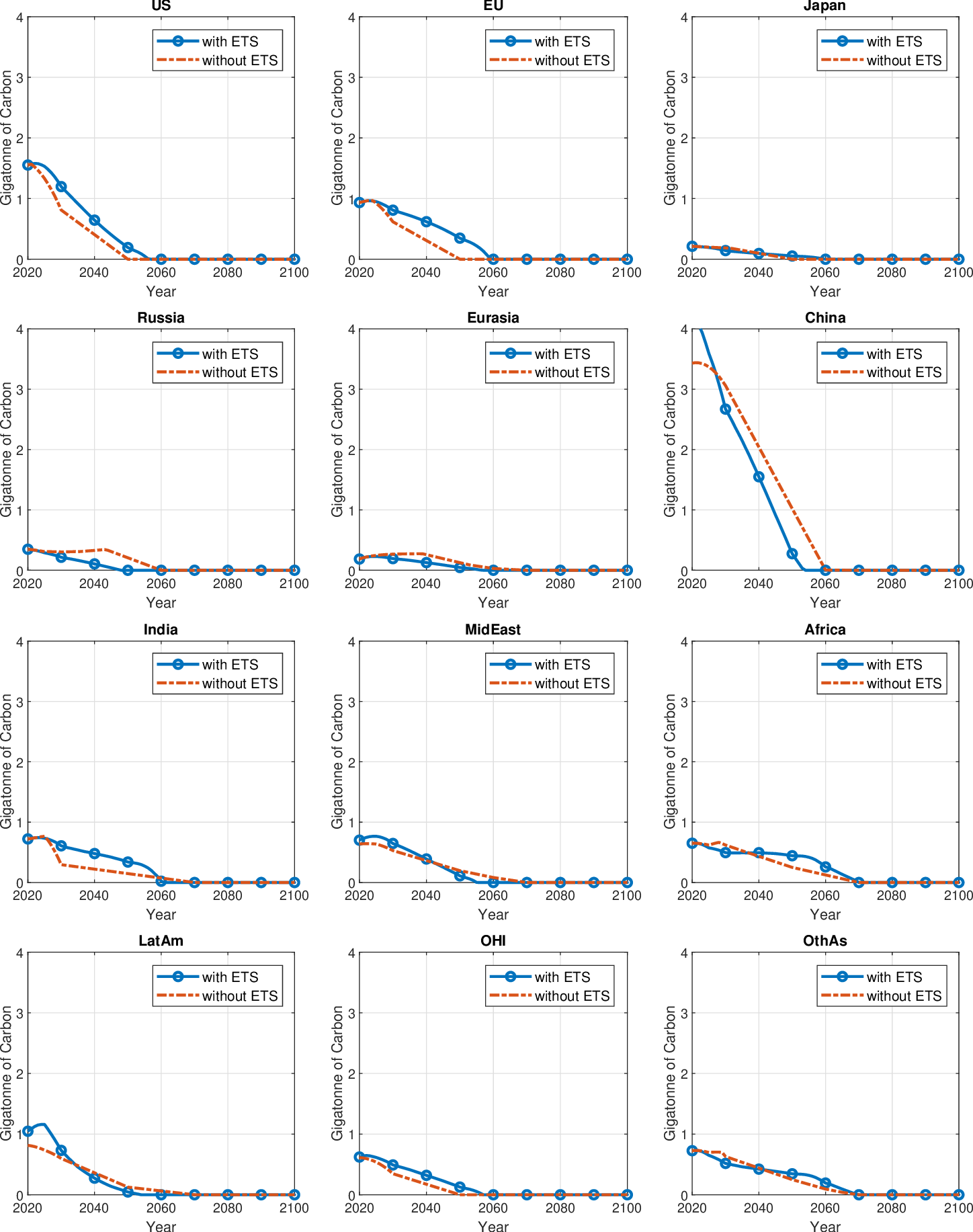}\caption{\label{fig:EE_Noncoop} Comparison of regional net emissions with
and without the ETS.}
\end{figure}

Figure \ref{fig:EE_Noncoop} compares regional net emissions between
two cases under noncooperation with the baseline emission caps: (i)
with the ETS, (ii) without the ETS.

\pagebreak{}

\begin{figure}[H]
\centering{}\includegraphics[width=0.9\textwidth]{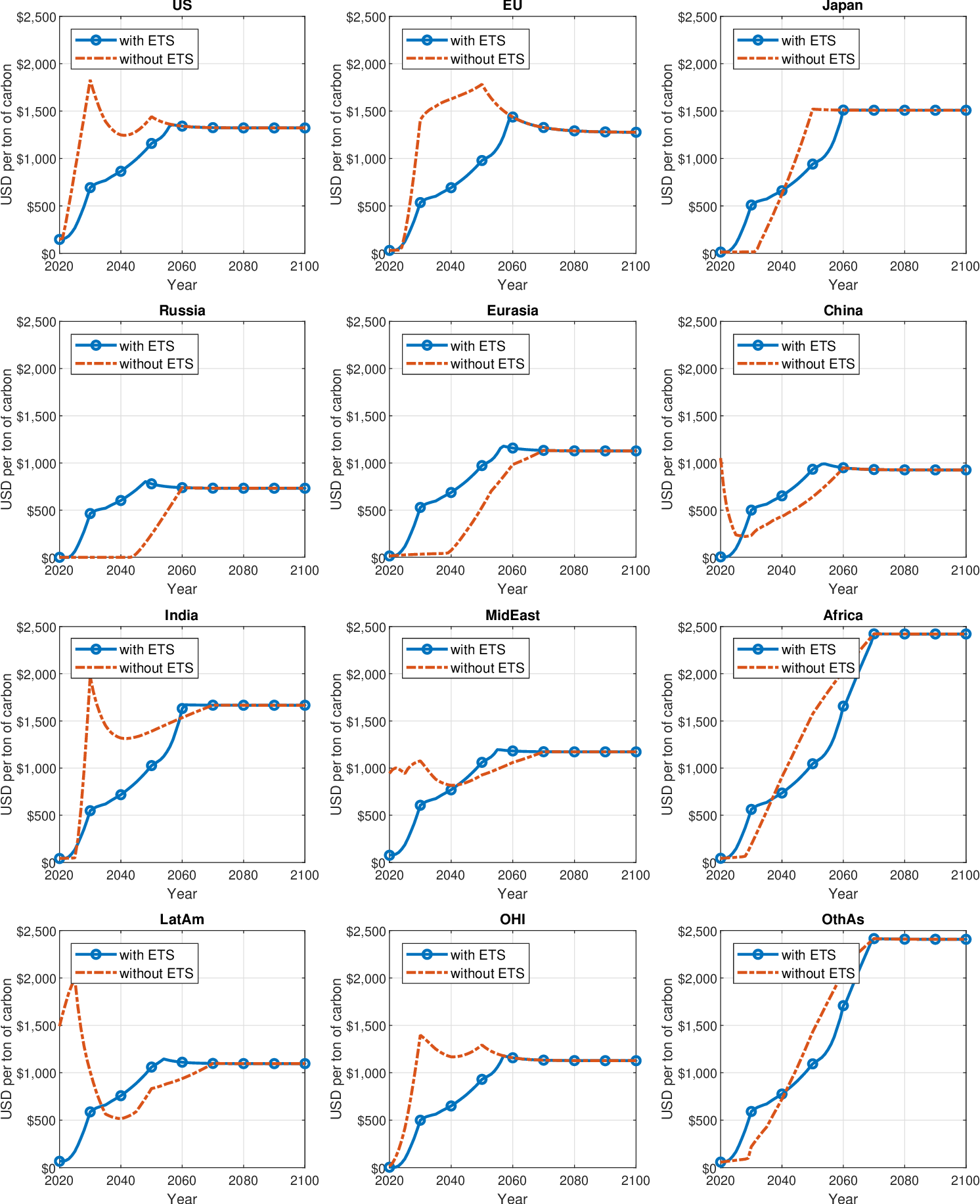}\caption{\label{fig:MAC_Noncoop}Comparison of regional MAC with and without
the ETS.}
\end{figure}

Figure \ref{fig:MAC_Noncoop} compares the regional MAC under noncooperation
with the baseline emission caps. We compare two cases: (i) with the
ETS, (ii) without the ETS.

\pagebreak{}

\begin{figure}[H]
\centering{}\includegraphics[width=0.9\textwidth]{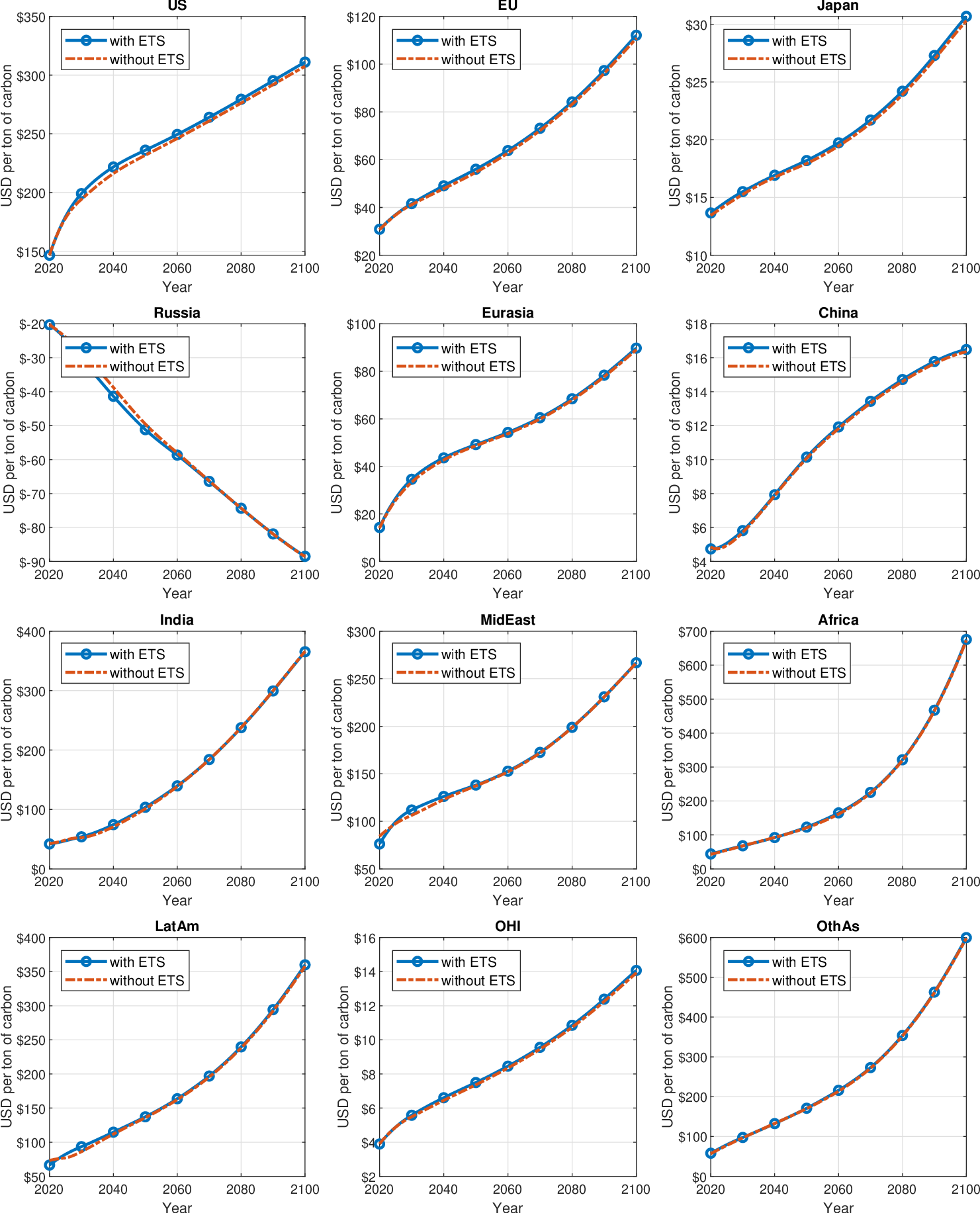}\caption{\label{fig:SCC_Noncoop} Comparison of regional SCC with and without
ETS.}
\end{figure}

Figure \ref{fig:SCC_Noncoop} compares the regional SCC under noncooperation
with the baseline emission caps, comparing two cases: (i) with the
ETS, (ii) without the ETS.

\pagebreak{}

\subsection{Alternative Policy Simulations: Partial ETS \label{subsec:sensitivity_partialETS}}

\begin{figure}[H]
\centering{}\includegraphics[width=0.9\textwidth]{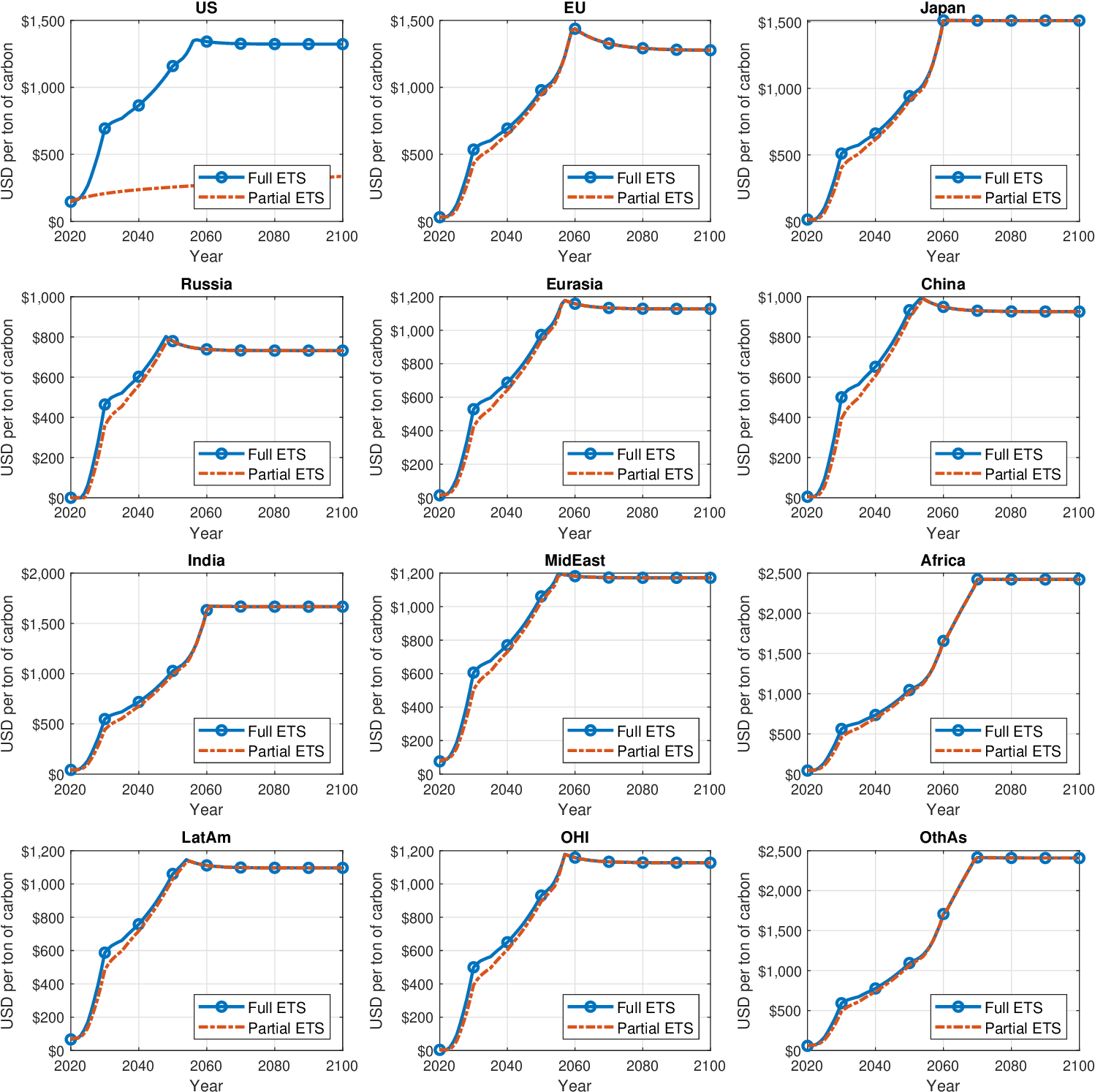}\caption{\label{fig:MAC_partialETS} Comparison of regional MAC under full
and partial ETS scenarios.}
\end{figure}

Figure \ref{fig:MAC_partialETS} compares the regional MAC under noncooperation,
comparing two cases: (i) with the full ETS, (ii) with a partial ETS.

\begin{figure}[H]
\begin{centering}
\includegraphics[width=0.9\textwidth]{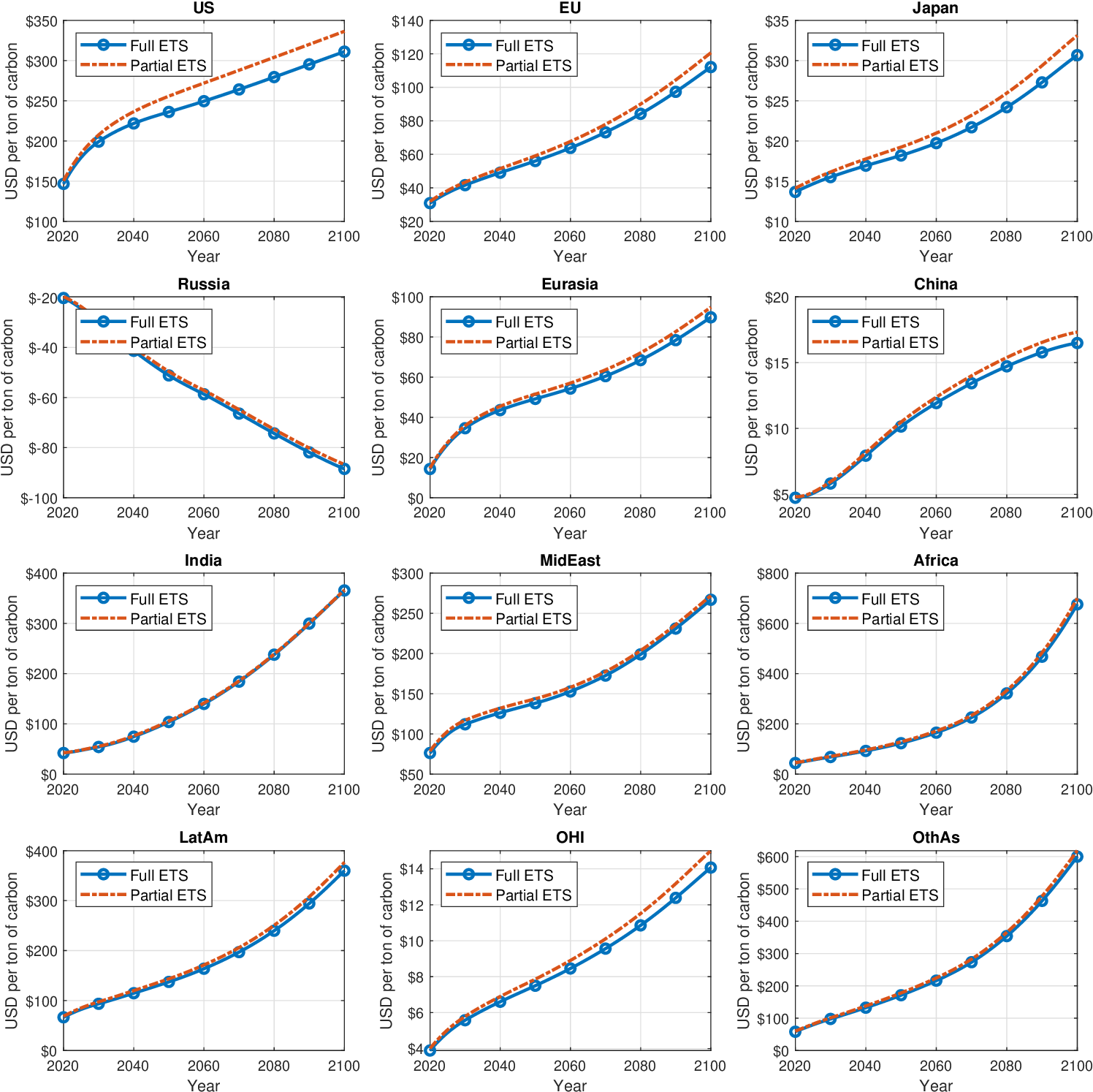}
\par\end{centering}
\centering{}\caption{\label{fig:SCC_partialETS} Comparison of regional SCC under the full
and the partial ETS scenarios.}
\end{figure}

\pagebreak Figure \ref{fig:SCC_partialETS} compares the regional
SCC under noncooperation, comparing two cases: (i) with the full ETS,
(ii) with the partial ETS without participation of the United States.

\pagebreak{}

\subsection{Alternative Policy Simulations: Net Zero Scenarios \label{subsec:sensitivity_ECAP}}

\begin{figure}[H]
\centering{}\includegraphics[width=0.9\textwidth]{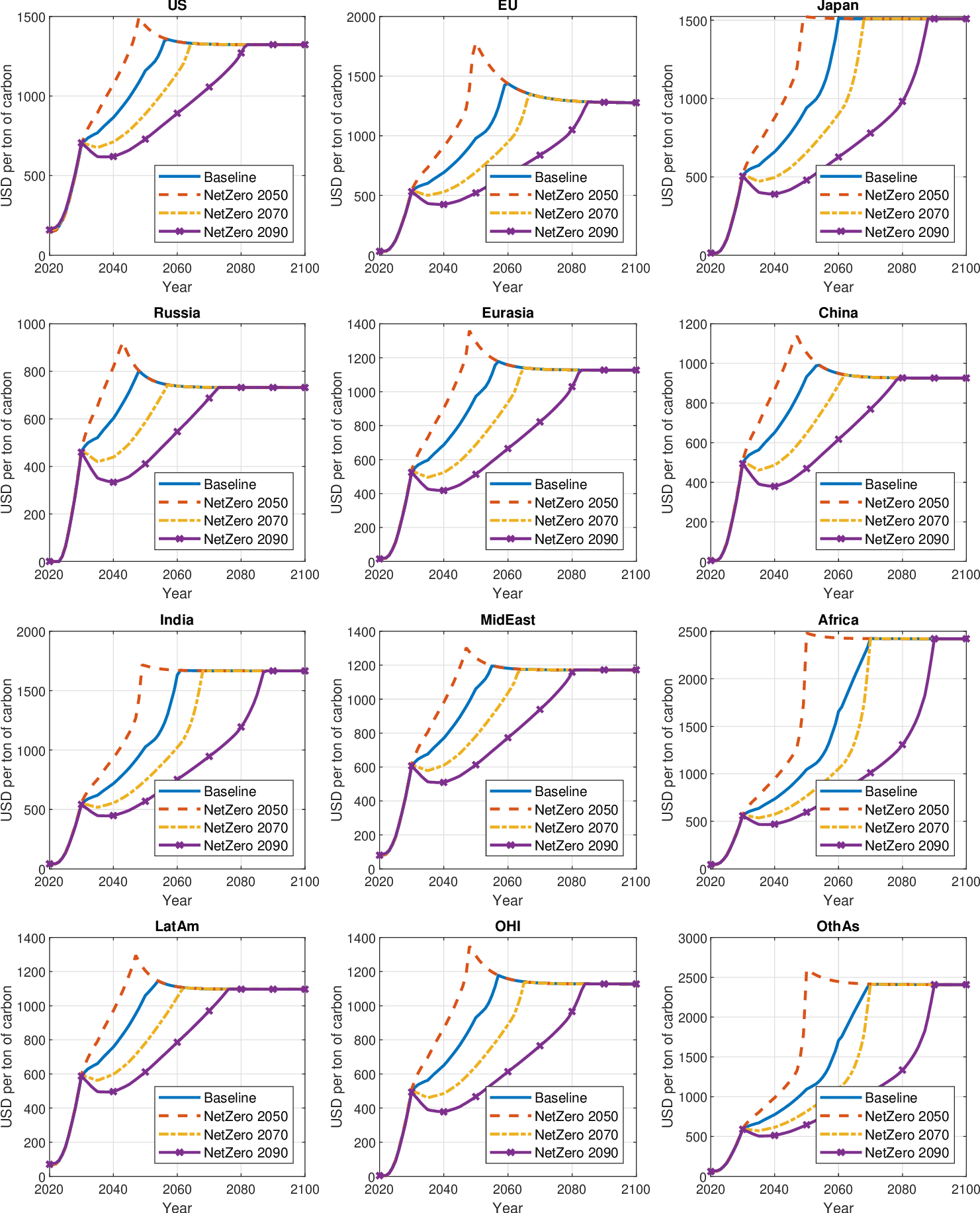}\caption{\label{fig:MAC_netzero} Comparison of regional MAC under different
emission caps.}
\end{figure}

In Figure \ref{fig:MAC_netzero}, we compare the MAC for the noncooperative
model with the ETS across alternative emission cap scenarios.

\begin{figure}[H]
\begin{centering}
\includegraphics[width=0.9\textwidth]{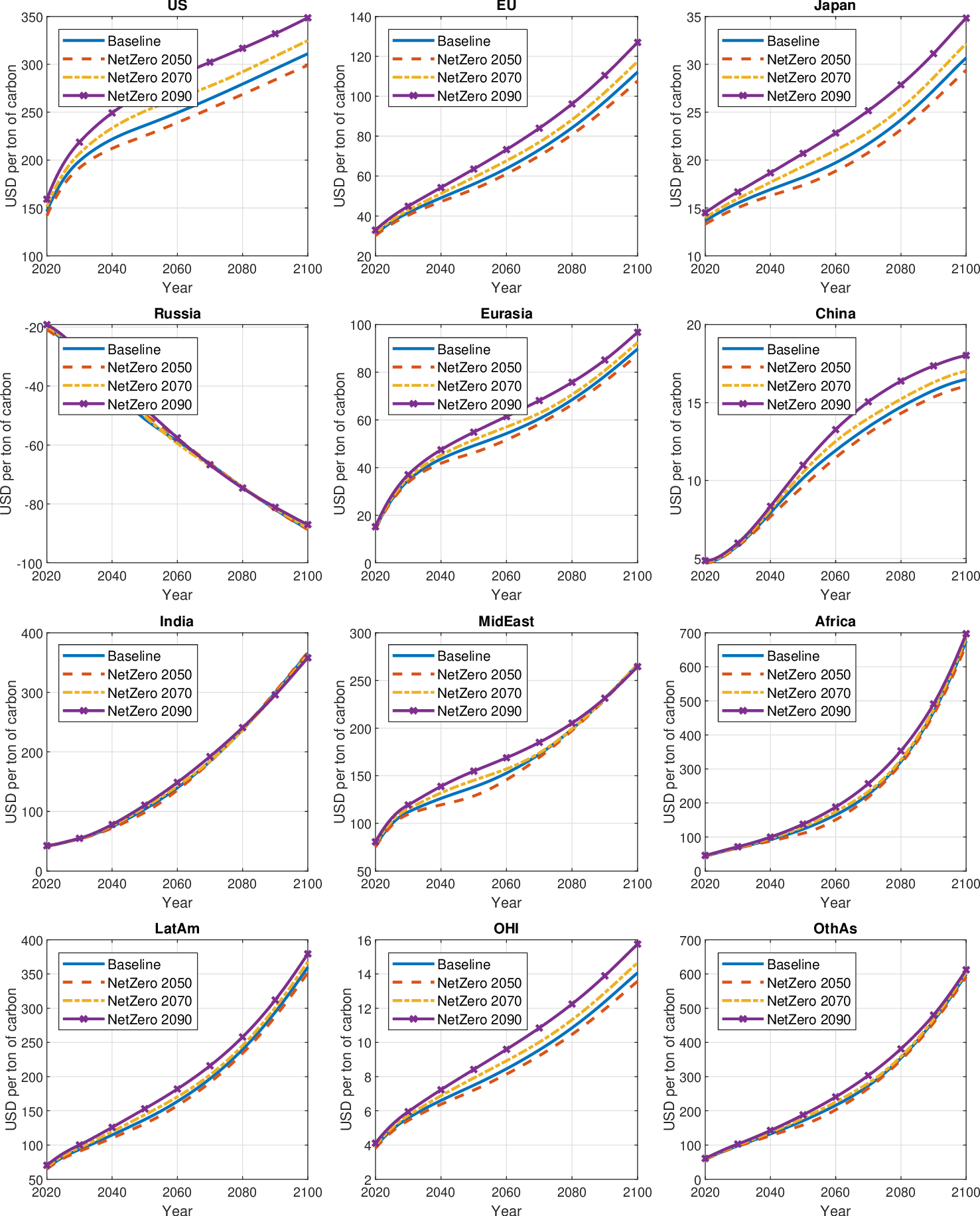}
\par\end{centering}
\centering{}\caption{\label{fig:SCC_netzero} Comparison of regional SCC under different
emission caps.}
\end{figure}

\pagebreak Similarly, Figure \ref{fig:SCC_netzero} displays the SCC
of the noncooperative model with the ETS across different emission
cap scenarios.

\pagebreak{}

\subsection{Sensitivity Analysis over Climate Damage Parameters \label{subsec:sensitivity_Damage}}

\begin{figure}[H]
\centering{}\includegraphics[width=0.9\textwidth]{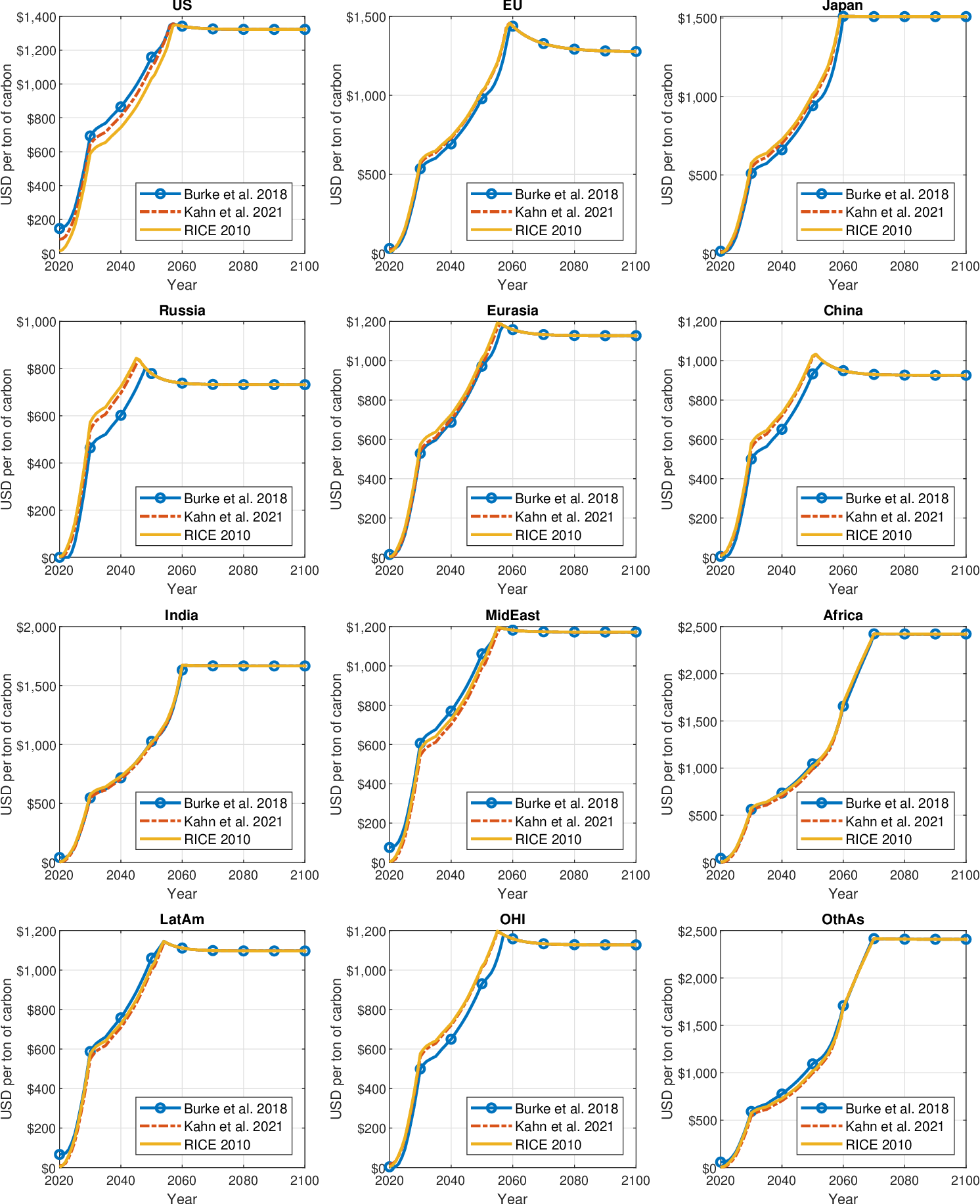}\caption{\label{fig:MAC_sensitivity_damage} Comparison of regional MAC under
different estimates of climate damages.}
\end{figure}

In Figure \ref{fig:MAC_sensitivity_damage}, we compare the MAC for
the noncooperative model with the ETS under alternative climate damage
parameters ({\small$\pi_{1,i}$, $\pi_{2,i}$}), based on projections
from \citet{kahn2021long} and \citet{nordhaus2010excel} alongside
the baseline projection from \citet{burke2018large}. 

\begin{figure}[H]
\begin{centering}
\includegraphics[width=0.9\textwidth]{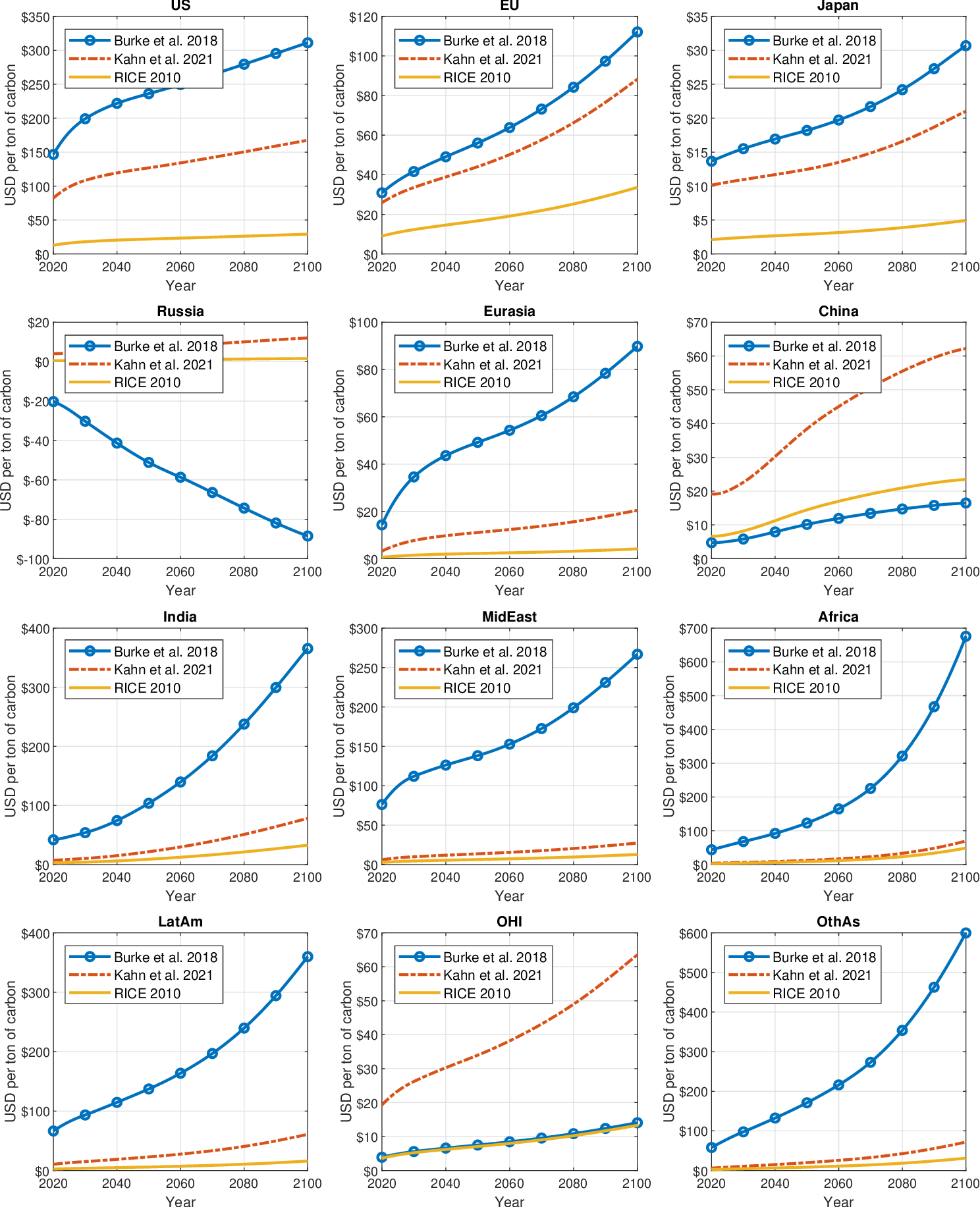}
\par\end{centering}
\centering{}\caption{\label{fig:SCC_sensitivity_damage} Comparison of regional SCC under
different estimates of climate damages.}
\end{figure}

\pagebreak Similarly, Figure \ref{fig:SCC_sensitivity_damage} displays
the SCC of the noncooperative model with the ETS under different values
of the climate damage parameters.

\pagebreak{}

\subsection{Sensitivity Analysis over Abatement Cost Parameters \label{subsec:sensitivity_abatement}}

\begin{figure}[H]
\centering{}\includegraphics[width=0.9\textwidth]{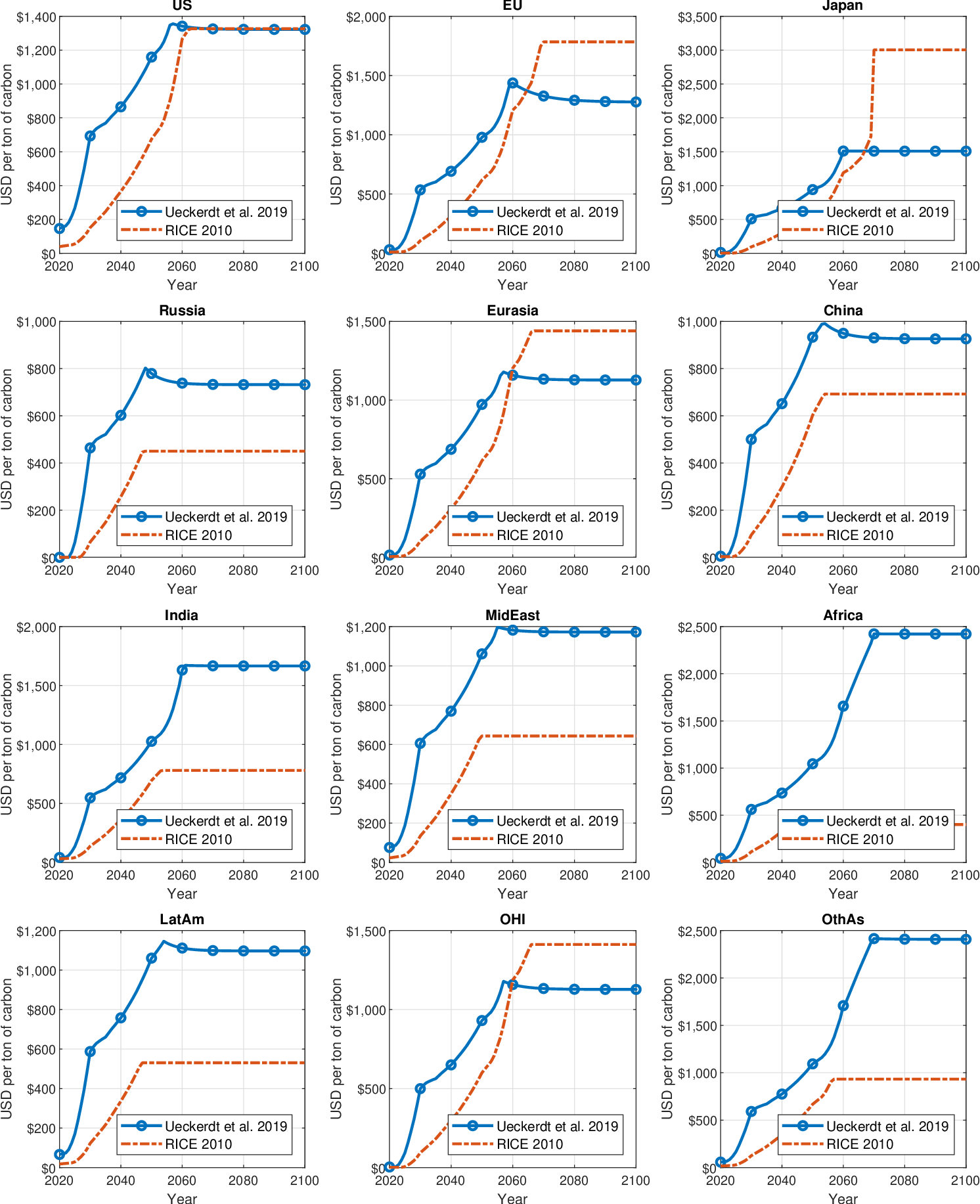}\caption{\label{fig:MAC_sensitivity_abatement} Comparison of regional MAC
under different estimates of emissions abatement cost.}
\end{figure}

In Figure \ref{fig:MAC_sensitivity_abatement}, we compare the MAC
for the noncooperative model with the ETS under alternative estimates
of the emissions abatement parameters ({\small$b_{1,i}$, $b_{2,i}$,
$b_{3,i},$ $b_{4,i}$}), calibrated from \citet{ueckerdt2019economically}
and \citet{nordhaus2010excel}.

\begin{figure}[H]
\begin{centering}
\includegraphics[width=0.9\textwidth]{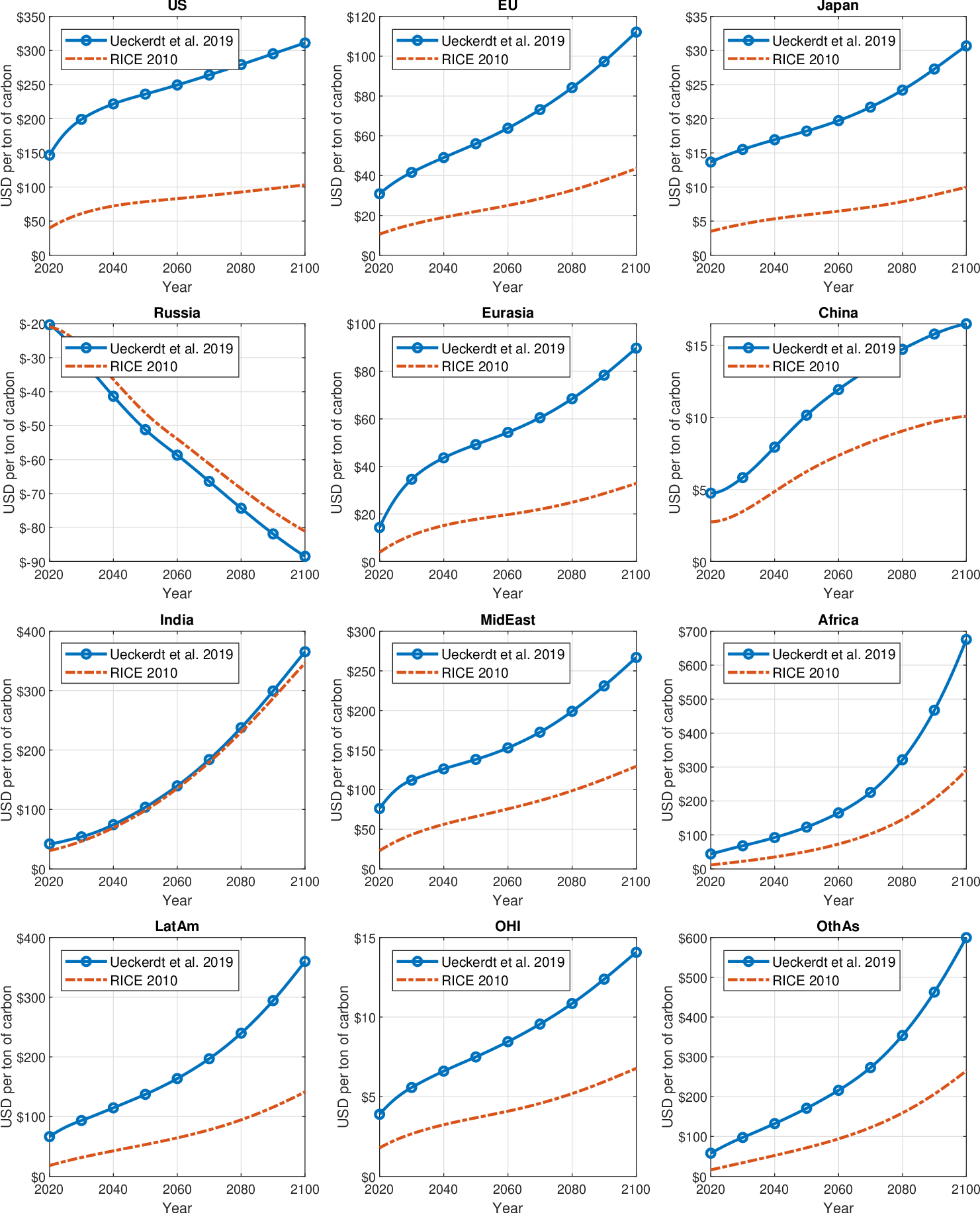}
\par\end{centering}
\centering{}\caption{\label{fig:SCC_sensitivity__abatement} Comparison of of regional
SCC under different estimates of emissions abatement cost.}
\end{figure}

\pagebreak Similarly, Figure \ref{fig:SCC_sensitivity__abatement}
displays the SCC of the noncooperative model with the ETS under different
emissions abatement cost estimates.

\pagebreak{}

\subsection{Sensitivity over TFP Growth Rates\label{subsec:Sensitivity-over-TFPgrowth}}

\begin{figure}[h]
\begin{centering}
\includegraphics[width=0.8\textwidth]{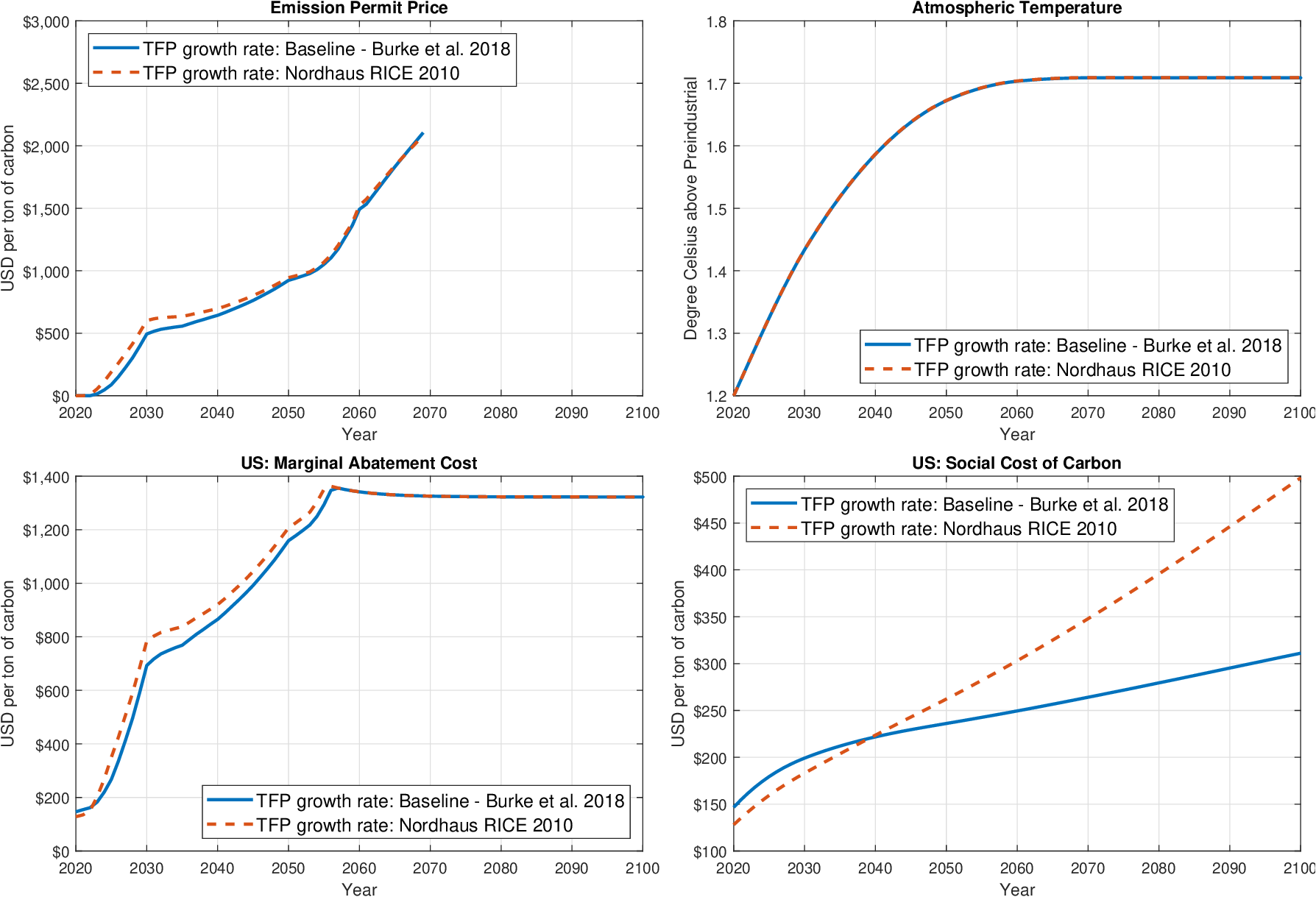}
\par\end{centering}
\centering{}\caption{\label{fig:Sensitivity_TFPgrowth} Comparison of simulation results
under different TFP growth rates.}
\end{figure}

Figure \ref{fig:Sensitivity_TFPgrowth} compares key model outcomes
under different TFP growth rates: the baseline TFP growth rates derived
from \citet{burke2018large} and the alternative rates based on \citet{nordhaus2010excel}.

\end{document}